\pdfoutput=1

\documentclass[11pt]{article}
\usepackage[margin=1in]{geometry}
\usepackage{booktabs} % For formal tables
\usepackage[ruled]{algorithm2e} % For algorithms
\usepackage{algorithmic}
\usepackage{tikz}
\usepackage{parskip}

\usepackage[sort]{natbib}
\usepackage{amsmath}
\usepackage{amsthm}
\usepackage{amssymb}
\usepackage{mathtools}
\usepackage{url}

\SetAlFnt{\small}
\SetAlCapFnt{\small}
\SetAlCapNameFnt{\small}
\SetAlCapHSkip{0pt}
\IncMargin{-\parindent}

\usepackage{csquotes}

\newcommand{\revision}[1]{{\color{black}{#1}}}

\newenvironment{remark}{\noindent \textit{{Remark.~}}}{\qed}
\newtheorem{claim}{Claim}
\newtheorem{fact}{Fact}
\newtheorem{example}{Example}
\newtheorem{lemma}{Lemma}
\newtheorem{corollary}{Corollary}

\newtheorem{theorem}{Theorem}

\newtheorem{definition}{Definition}

\newcommand{\V}{\mathcal{V}}
\newcommand{\G}{\mathcal{G}}

\newcommand{\T}{\mathbb{T}}

\newcommand{\D}{\mathbb{D}}

\title{The Complexity of Pacing for Second-Price Auctions}

\author{
    Xi Chen\\
    Columbia University\\
    \texttt{xichen@cs.columbia.edu}
    \and
    Christian Kroer\\
    Columbia University\\
    \texttt{ck2945@columbia.edu}
    \and
    Rachitesh Kumar\\
    Columbia University\\
    \texttt{rk3068@columbia.edu}
}

\begin{document}

\maketitle
\begin{abstract}
Budget constraints are ubiquitous in online advertisement auctions. To manage these constraints and smooth out the expenditure across auctions, the bidders (or the platform on behalf of them) often employ pacing: each bidder is assigned a pacing multiplier between zero and one, and her bid on each item is multiplicatively scaled down by the pacing multiplier. This naturally gives rise to a game in which each bidder strategically selects a multiplier. The appropriate notion of equilibrium in this game is known as a pacing equilibrium.
	
In this work, we show that the problem of finding an approximate pacing equilibrium is PPAD-complete for second-price auctions. This resolves an open question of \cite{conitzer2017multiplicative}. As a consequence of our hardness result, we show that the t\^atonnement-style budget-management dynamics introduced by \cite{borgs2007dynamics} are unlikely to converge efficiently for repeated second-price auctions.
This disproves a conjecture by \cite{borgs2007dynamics}, under the assumption that the complexity class PPAD is not equal to P.
Our hardness result also implies the existence of a refinement of supply-aware market equilibria which is hard to compute with simple linear utilities. 
  %$$, thereby making progress towards their open conjecture. 	
%	Furthermore, as a consequence of the relationship between pacing equilibria and supply-aware market equilibria, our work implies the existence of a refinement of supply-aware market equilibria which are hard to compute in the simple linear Fisher market utility model. 
\end{abstract}

% Paper body

\section{Introduction}

Online auctions are a mainstay of the Internet advertising industry. Whenever a user visits a webpage or searches for a keyword, interested advertisers participate in an auction to win the opportunity to promote their content to the user. Advertisers typically participate in thousands of these online ad auctions every day and are often budget constrained, which makes budget management a crucial component of online advertising. This paper is concerned with a specific method of budget management in auctions: \emph{pacing} (also known as multiplicative pacing), which has found use at platforms such as Facebook, where pacing is routinely employed as one of the ways to manage budgets on behalf of advertisers.\footnote{\url{https://www.facebook.com/business/help/1754368491258883?id=561906377587030}}

Pacing involves multiplicatively scaling down bids of advertisers in order to ensure a smooth depletion of their budgets over the entire advertising campaign, which is comprised of a large number of individual auctions. Consider the setting in which a group of buyers (advertisers) participate in a series of independent second-price auctions for a collection of items (the opportunity to display an ad to a user). If all  buyers bid their values\footnote{Bidding your value is a dominant strategy in second price auctions without budgets.} in every auction, they might all deplete their budgets before the last auction. As a remedy, pacing associates a \emph{pacing multiplier} to each buyer, which lies between zero and one, such that each buyer bids her value scaled down by her pacing multiplier. The pacing multiplier is strictly smaller than one only if the buyer would deplete her budget by bidding her true value in each auction. 

Pacing has the desirable property that, if we fix the bids of competing buyers, then pacing allows a buyer to win the items which provide the best return on investment (ratio of value to price) subject to her budget constraint. In a recent work, \citet{balseiro2019learning} exploit this property to prove the optimality of pacing for budget management: For a budget-constrained buyer who repeatedly participates in second-price auctions for which her values are drawn i.i.d. from some distribution, the optimal bidding strategy is to use pacing, both when the bids of the adversary are stochastic or adverserial. In other words, when considering the problem of bidding under budget constraints from the perspective of a single buyer, pacing is provably the best strategy to use. 

In this paper, we study the situation where \emph{every} buyer uses pacing to attempt to bid optimally in second-price auctions.
We prove that, unfortunately, if every buyer tries to bid optimally through pacing, then the resulting dynamics are \emph{not} likely to converge efficiently to an equilibrium. We do so by investigating the computational complexity of finding an equilibrium of the game in which each buyer's strategy involves selecting a pacing multiplier, called a \emph{second-price pacing game}. The natural notion of equilibrium in this game is the \emph{pacing equilibrium}
(see Definition \ref{definition_exact_pacing_equilibrium}), which was introduced and shown to always exist by %under careful tie-breaking in 
\cite{conitzer2017multiplicative}. The authors of \cite{conitzer2017multiplicative} also studied the computation of pacing equilibria by developing mixed-integer programming methods and applied them to real-world auction data, but found them plagued with poor scalability. They went on to conjecture that computing a pacing equilibrium could be PPAD-complete.\footnote{While we cite the 2021 journal version of that paper, the conjecture was made first in the 2017 arXiv version of that paper. It was also published in their 2018 conference version of the paper, which appeared at the Conference on Web and Internet Economics (WINE) that year.}

% The optimal choice of a pacing multiplier for a buyer (one which either   depletes her budget~or is equal to one) depends on pacing multipliers of other buyers. This leads to a game (see Section
%   \ref{section:model} for the model) called the \emph{second-price pacing game}, in which each buyer's strategy involves selecting a pacing multiplier. The natural notion of equilibrium in this game is the \emph{pacing equilibrium}
% (see Definition \ref{definition_exact_pacing_equilibrium}), which was introduced and shown to always exist by %under careful tie-breaking in 
% \cite{conitzer2017multiplicative}. The authors of \cite{conitzer2017multiplicative} also studied the computation of pacing equilibria by developing
%   MIP-based methods %for computing pacing equilibria 
%     and applied them to real-world auction data, 
%   but found them plagued with poor scalability. They went on to conjecture that computing
%   a pacing equilibrium could be PPAD-complete.

% state:
% \enquote{Whether an arbitrary pacing equilibrium can be computed in polynomial time is itself an interesting open problem. We believe that computing a pacing equilibrium could be a PPAD-complete problem\ldots}

Our paper resolves this open problem: it is indeed a PPAD-complete problem, and this holds~for a broad class of approximate versions of the problem as well. This provides mathematical support for the repeatedly observed empirical fact that pacing-based bidding strategies often do not converge quickly, and the resulting equilibria seem hard to compute. These two facts are evidenced by the lack of efficient dynamics and efficient algorithms for computing equilibria, despite the significant attention pacing has received for more than a decade (see Section~\ref{sec:related_work}). Through our result, we show that multi-buyer pacing is fundamentally intractable and this lack of efficient dynamics/algorithms is likely here to stay.  
 Before delving deeper, we provide a short primer on PPAD for those unfamiliar with it. %, in the hopes of making this paper and its central ideas accessible to a wider audience. 
This can be  safely skipped by any reader already familiar with the topic.%\medskip

 Like the well-known complexity class NP, PPAD (\emph{Polynomial Parity Argument in a Directed graph}, introduced by \citet{PPAD})  is a collection of computational problems.
 %which share certain run-time properties.
As~with the definition of NP-hardness and NP-completeness, a problem is said to be PPAD-hard  if it is~at least as  hard as every problem in PPAD; a problem is said to be PPAD-complete if it is contained in PPAD and is PPAD-hard. The analogy to NP extends further: the PPAD-hardness of a problem can be established by providing a polynomial-time reduction \emph{from} a problem already known to be PPAD-hard. One of the quintessential PPAD-complete problems, and the one we will employ in our reductions, is that of computing a Nash equilibrium of a bimatrix game \citep{daskalakis2009complexity,chen2006settling}. The Nash equilibrium problem has been studied extensively for decades and yet, despite much effort, no polynomial-time algorithm is known for it. Moreover, a recent spate of results showed that it is hard to solve, assuming certain strong cryptographic assumptions \citep{Crypto1,Crypto2,Crypto3,Crypto4,Crypto5}. This has motivated the conjecture that PPAD-hard problems cannot be solved efficiently. In this paper, we show that the problem of finding a pacing equilibrium is PPAD-hard. This shows that computing a pacing equilibrium is hard, unless all problems in PPAD can be solved efficiently.
 
 {On the other hand, showing that a problem is in PPAD amounts to giving a polynomial-time reduction \emph{to} a problem in PPAD. For this purpose we will avail ourselves of the fact that the algorithmic version of Sperner's lemma is known to be in PPAD \citep{PPAD,CHEN20094448}, and reduce the problem of finding a pacing equilibrium to it. 
%These two reductions together show that 
%  the problem is computationally as hard as the Nash equilibrium
%  problem (or any other PPAD-complete problem).
% Finally, a problem is said to be PPAD-complete, if it is both PPAD-hard and a member of PPAD. When hardness is captured by polynomial-time reductions, every PPAD-complete problem is as hard as any other PPAD-complete problem (see 
We refer the interested reader to \citet{goldberg2011survey} and Chapter~4 of \cite{TimSurvey} for a survey of PPAD and its complete problems.}

\subsection{Main Contributions}

We first prove that finding a pacing equilibrium is in PPAD. In particular, this implies that, when values and budgets
  of buyers are rational in the game, there always exists a pacing equilibrium in which every entry is rational and can be written using polynomially many bits. (In contrast, the existence proof of \cite{conitzer2017multiplicative} uses a convergence argument, from which it is not clear whether an equilibrium with rational entries always exists.)

\begin{theorem}\label{theo:membership}
Finding a pacing equilibrium in a second-price pacing game is in PPAD.
\end{theorem}

Next we show that the problem of finding 
  an \emph{approximate} pacing equilibrium is PPAD-hard. 
%We define a natural approximate version of pacing equilibrium (see Definition~\ref{definition_pacing_equilibrium}) and show that it is PPAD-hard to find it. 
Our notion of approximation  relaxes the definition of (exact) pacing equilibria in two ways: (i) buyers who bid \emph{close} to (but not necessarily exactly equal to) the highest bid may also win   fractions of an item; (ii) each buyer either spends \emph{most} of her budget, or her pacing multiplier is \emph{close} to one.
We use two parameters $\delta$ and $\gamma$ to capture 
  these two relaxations quantitatively and such a solution 
  is called a $(\delta,\gamma)$-\emph{approximate
  pacing equilibrium} (see Definition~\ref{definition_pacing_equilibrium}).

\begin{theorem}\label{theo:approx_hardness_informal}
For any constant $c>0$,
  finding a $(\delta,\gamma)$-approximate pacing equilibrium in a second-price pacing game with $n$ players is PPAD-hard
   when $\delta=\gamma=1/n^c$. % for some 
%. \textcolor{red}{Need to reconcile this with Theorem \ref{approximate_hardness}.}
\end{theorem}

Note that, by virtue of being a relaxation, finding an approximate pacing equilibrium is 
  in PPAD as a direct consequence of Theorem~\ref{theo:membership}. Similarly, the PPAD-hardness of finding an exact pacing equilibrium follows from Theorem~\ref{theo:approx_hardness_informal}.
%   from Theorem \ref{theo:approx_hardness_informal}. It follows from Theorem \ref{theo:membership} and \ref{theo:approx_hardness_informal}
% that 
Therefore, both problems of finding an exact and an approximate
  pacing equilibrium are complete in PPAD.
  %In other words, we show the stronger statements for both PPAD-membership and PPAD-hardness.
To the best of our knowledge, our results are the first PPAD-completeness results for budget-management in second-price auction systems such as those applied in large-scale Internet advertising.\vspace{0.05cm} \medskip

% \footnote{Note that, by virtue of being a relaxation, finding an approximate pacing equilibrium is % are always guaranteed to exist and the problem of finding them is 
%   in PPAD as a direct consequence of Theorem~\ref{theo:membership}. For the same reason, the PPAD-hardness of finding an exact pacing equilibrium follows
%   from Theorem \ref{theo:approx_hardness_informal}.}

\noindent\textbf{Implications.}
Our hardness result has implications for \cite{borgs2007dynamics}, in which the authors studied dynamic first-price and second-price auctions with budgets. They proved that pacing combined with perturbations can lead to efficient convergence of bidding dynamics under first-price auctions. They conjectured a similar convergence in the analogous second-price setting and provided experimental support for it. 
%, but leave open the problem of whether this is guaranteed to occur, and whether the convergence is efficient. 
Our definition of approximate pacing equilibria    (Definition~\ref{definition_pacing_equilibrium}) 
is able to capture their random-perturbation model, thereby bringing it under the purview of our hardness result Theorem \ref{theo:approx_hardness_informal}: if such a convergence occurs in the second-price case, then it must do so inefficiently  assuming PPAD~does not have polynomial-time algorithms (see Subsection \ref{section_approx_hardness}). Moreover, since our model also admits a stochastic interpretation, if all of the buyers employ some pacing algorithm for repeated second-price auctions with correlated value distributions and global budget constraints, then the resulting dynamics will not always converge efficiently to an equilibrium, assuming PPAD~does not have polynomial-time algorithms. In particular, this statement applies to the pacing algorithm given by \citet{balseiro2019learning}, which is an optimal bidding algorithm for a single budget-constrained buyer under both adversarial and independent-stochastic competition. Informally, the central message here is that, when multiple budget-constrained buyers bid in a way that is optimal for them individually, the resulting dynamics will not in general stabilize to an equilibrium.

Furthermore, due to connections between pacing equilibria and \emph{supply-aware} market equilibria \cite{conitzer2017multiplicative} with linear utilities, our PPAD-hardness result has novel consequences when interpreted in the language of market equilibria:
%we believe 
our result shows that a natural refinement of supply-aware market equilibria with linear utilities is PPAD-hard (finding one with prices corresponding to second-price auctions).
% we believe our result is the first to show hardness in a simple linear Fisher market utility model (albeit with a quasi-linear dependence on payments), relying on the fact that we are seeking a refinement of the general set of supply-aware market equilibria (see Subsection \ref{section_approx_hardness}).

\subsection{Techniques Used}

We prove the PPAD-hardness of finding approximate pacing equilibria (Theorem~\ref{theo:approx_hardness_informal}) by giving a reduction from the problem of finding an $\epsilon$-well-supported Nash equilibrium in win-lose bimatrix games. The second-price rule plays an important role in this reduction. Consider an item with~two interested buyers, one of which has a much higher value than the other, so much so that she always wins the good in any pacing equilibrium. Then, the payment made by this buyer on this item is determined by the bid of the lower-valued buyer, which is equal to her value times her~multiplier. This allows us to construct gadgets which capture Nash equilibria of any bimatrix game with pacing multipliers, by using the second-price rule to account for %the effect on a player's mixed strategy of the other player's 
  the expected cost of each action of a player with respect
  to the other player's
mixed strategy. A complicating factor in our proof is that pacing multipliers are always positive, whereas some actions are played with probability zero in a Nash equilibrium. To address this issue, we construct our gadgets such that they have a discontinuous behavior: there is a baseline pacing amount which corresponds to playing the corresponding action with probability zero, and only larger pacing values correspond to probabilities.

To prove the PPAD-membership of finding a pacing equilibrium (Theorem~\ref{theo:membership}), we reduce the problem  to the algorithmic version of Sperner's Lemma. 
%Directly applying Sperner's lemma to establish the existence of a pacing equilibrium 
A direct reduction proves challenging due to the discontinuous way in which the allocation of an item varies with pacing multipliers: In a pacing equilibrium, an item can only be assigned to buyers whose bids are \emph{exactly} equal to the highest bid. Similar issues were encountered in PPAD-membership proofs for market equilibrium computation~\cite{vazirani2011market}. For this reason, we start by proving the PPAD-membership of finding approximate pacing equilibria, in which items can be allocated smoothly. Then we bootstrap this result to show the PPAD-membership of  exact pacing equilibrium in two steps. The first step starts with an approximate pacing equilibrium and rounds it to obtain a pacing equilibrium in which only buyers tied for the highest bid on a good share it. We still allow %for some relaxation of the conditions that define a pacing equilibrium, by allowing 
the relaxation that each buyer can either spend \emph{most} of her budget or set her pacing multiplier \emph{close} to one. Finally, we do away with this remaining relaxation by using a LP-based technique similar to the one used in \revision{\cite{etessami2010complexity, vazirani2011market} and \cite{filos2020consensus}}, thereby showing the PPAD-membership of finding an exact pacing equilibrium. 

\subsection{Pacing in Internet Advertising}

To motivate pacing equilibrium as a solution concept, this section describes how the solution concept arises in practice as part of internet advertising platforms such as those operated by e.g. Facebook, Google, or Twitter.
As discussed previously, pacing equilibrium may arise through individual buyers optimizing their spending due to their budget constraint.
A second reason that pacing equilibrium is of practical interest is due to \emph{proxy bidders}.
When an advertiser starts a campaign, they often specify only a small set of parameters: their value for a click (or some other notion of converting an ad into value, say a video view), their budget, and their \emph{targeting criteria} which specify the subset of users they are interested in (e.g. ``people who surf'' if the ad is for surfboards).
Then, whenever an auction is run to determine which ads to show to a given user, the bid from a given advertiser is submitted by the proxy bidder acting on behalf of that advertiser. 
The proxy bidder calculates the value that advertiser $i$ has for being shown to the user in auction $j$ as $v_{ij} = v_i \cdot CTR_{ij}$, where $v_i$ is the value per click and $CTR_{ij}$ is the estimated probability that the user will click on the ad. 
If there were no budgets, then the proxy bidder should submit the bid $v_{ij}$, due to the truthfulness of the second-price auction. But in the presence of budgets, this may negatively affect the overall utility achieved by the advertiser, since they will run out of budget well before the campaign ends, and thus miss out on later strong bang-per-buck opportunities.

To address their budget constraints, the advertisers are typically offered one or more options for budget-management strategies that can be employed by the proxy bidders. Pacing as defined in this paper, via multiplicative bid scaling, is offered by Facebook by default~\citep{conitzer2017multiplicative,facebookguide}, and it is also offered on other platforms.
Intuitively speaking, the proxy bidder attempts to choose a pacing multiplier which will spend the advertiser's budget evenly across the campaign length. To ensure that this will happen, the pacing multiplier is adapted over time using a control algorithm: the algorithm will adjust the pacing multiplier up or down depending on whether the proxy bidder is currently under or overspending. 
Since we do not consider the online aspect of the problem, the pacing equilibrium solution concept that we study corresponds to the steady-state that this adaptive process would ideally arrive at (this is analogous to what was done by ~\citet{conitzer2017multiplicative,balseiro2015repeated,balseiro2017budget}).
See \citet{conitzer2017multiplicative} for a longer discussion of the pacing equilibrium model and how it relates to real-world systems.

\subsection{Additional Related Work}\label{sec:related_work}
There is a large literature on budgets in auctions, largely inspired by the Internet advertising industry. Here we survey the ones most related to our paper.
We start by surveying the literature on multi-item first or second-price auctions with budgets and the associated equilibrium issues there, since that is the setting we study. We briefly mention  some pointers to alternative approaches and models such as mechanism design or online matching.

% Most directly related to our work is that of \cite{conitzer2017multiplicative}, from which we draw our pacing equilibrium model. \cite{conitzer2017multiplicative} introduce the notion of second-price pacing equilibrium, and show that it is NP-hard to find pacing equilibria that maximize objectives such as welfare or revenue. However, the authors state:
% \enquote{Whether an arbitrary pacing equilibrium can be computed in polynomial time is itself an interesting open problem. We believe that computing a pacing equilibrium could be a PPAD-complete problem\ldots}
% Our paper resolves this open problem: it is indeed a PPAD-complete problem.

% \cite{borgs2007dynamics} study dynamic first and second-price auctions with budgets, and show that pacing combined with perturbations can lead to convergence of bidding dynamics under first-price auctions. They conjecture that this also occurs in their second-price setting and provide experimental support for it, but leave open the problem of whether this is guaranteed to occur, and whether it is efficient. Our results show that if such a convergence occurs in the second-price case, then it must do so inefficiently (see Subsection \ref{section_approx_hardness}).

\cite{balseiro2015repeated} studied budget management in second-price auctions using a \emph{fluid mean-field} model, and showed that in this model existence is guaranteed, and closed-form solutions for equilibria are derived for certain settings.
\cite{balseiro2017budget} studied several different pacing mechanisms for second-price auctions, including multiplicative pacing, and showed existence results for their setting, as well as other analytical and numerical properties.
\cite{conitzer2019pacing} studied the model of \cite{conitzer2017multiplicative}, but with each auction using a first-price rule. There, pacing equilibrium no longer constitutes best responses, but instead has a market equilibrium interpretation. In the first-price setting, pacing equilibria turn out to be easy to compute, due to a direct relationship to market equilibria. %In Section ??\ck{ref section} we elucidate this relationship by showing a general relationship between pacing equilibria and market equilibria where buyers are \emph{supply aware}.
\cite{babaioff2020non} studied non-quasi-linear agents participating in mechanisms designed for quasi-linear agents. They studied a generalization of budget constraints where agents have a concave disutility in payment, and showed that a Nash equilibria exists which employs multiplicative scaling. %\ck{note to Rachitesh: do they formally generalize our model? If so then we also show PPAD hardness of their solution concept, and we should point that out}.
Since pacing equilibrium is a special case of Nash equilibrium in the more general buyer utility model studied in \cite{babaioff2020non}, our hardness results extend to their setting.  \cite{balseiro2019learning} developed online learning methods for individual agents adapting their pacing multipliers over time, and showed that this converges to an equilibrium under certain stochastic independence assumptions. Assuming PPAD~$\neq$~P, our results can be interpreted to mean that, in the general setting which allows for correlation and discrete valuations, no dynamics can converge efficiently in the worst case (see Proposition 10 of \cite{conitzer2017multiplicative} for a formal statement connecting the stochastic and deterministic settings). %\ck{should we say something about how their result is not incompatible with ours?}
%To the best of our knowledge, our results are the first PPAD-hardness results for budget-management in second-price auction systems such as those applied in large-scale Internet advertising\ck{should maybe say this in intro instead, and also make more effort to verify it}.

% literature on mech design?
An alternative approach for handling budget constraints in multi-item settings is to design a mechanism that accounts for this explicitly, see e.g.~\cite{ashlagi2010position,goel2015polyhedral,dobzinski2012multi,dobzinski2014efficiency}. 
% literature on centralized matching
Another approach to budget-constrained allocation in online advertising is to treat the problem as an online matching problem. This research was initiated by \cite{mehta2007adwords}, see e.g. \cite{mehta2013online} for a survey.

% literature on market equilibrium (supply aware especially)
Our results are strongly related to the problem of computing market equilibria under a supply-aware model (see Subsection~\ref{section_approx_hardness} for a discussion). There have been several PPAD-completeness results for various Fisher market   models (without supply-awareness). However, these results are all for models with more complex utility functions, which give rise to the hardness.
\cite{chen2009spending} and \citet{VijayMihalis} showed that for additively-separable piecewise-linear concave utilities, finding an equilibrium in a Fisher market is PPAD-complete.
\cite{bei2016computing} showed PPAD-hardness of finding market equilibria with budget-capped utilities (this is proved using a variation on the piecewise-linear utilities proof of \cite{chen2009spending}).
In the case of indivisible goods, \cite{othman2016complexity} showed that finding  an approximate market equilibrium is hard, even one which is guaranteed to exist~\cite{budish2011combinatorial}. \revision{For the Arrow-Debreu exchange economy, \citet{chen2017complexity} showed that finding an equilibrium is PPAD-hard.}

\revision{Finally, in additional to Nash equilibrium and market equilibrium, many interesting problems have been proven to be PPAD complete in domains like auctions \citep{filos2021complexity, chen2021throttling}, fair division \citep{deng2012algorithmic, filos2020consensus} and optimization \citep{fearnley2021complexity}.}

%\textcolor{red}{We believe that our results are the first to show hardness in the case of \emph{linear} utilities, relying on the fact that we are seeking a refinement of the general set of supply-aware market equilibria with linear utilities.}

%in a simple linear Fisher market utility model (albeit with quasi-linear dependence on payments), {\color{blue}and instead rely on the fact that we are seeking a refinement of the general set of supply-aware market equilibria.}\footnote{\color{blue}Xi: This sentence
  %does not read well.}

%%% Local Variables:
%%% mode: latex
%%% TeX-master: "main"
%%% End:

\section{Model} \label{section:model}

\def\sppg{\text{second-price pacing game}}
\def\appe{approximate PE}

%We use agent, bidder and buyer interchangeably. We also use item and good interchangeably. 
We start with the definition of 
  Second-price Pacing Games.
%\begin{definition}\label{definition_pacing_game}
In a \emph{Second-price Pacing Game} (SPP game as a shorthand) $G = (n, m, (v_{ij}) , (B_i) )$,  %the following components:\vspace{0.1 cm}
%\begin{flushleft}
%\begin{itemize}
    there are $n$ buyers and $m$ (indivisible) goods. Each good is sold through independent (single slot) second-price auctions. We use $v_{ij} \geq 0$, $i\in [n]$ and $j\in [m]$,  to denote the value of good $j$ to buyer $i$, and $B_i > 0$ to denote the budget of buyer $i$. 
We will require (1) for each $j\in [m]$, 
  $v_{ij}>0$ for some $i\in [n]$, and (2) for each $i\in [n]$,
  $v_{ij}>0$ for some $j\in [m]$.
Each buyer $i$ plays the game by picking
a \emph{pacing multiplier} $\alpha_i \in [0,1]$ and then bidding 
$\alpha_i v_{ij}$ on good $j$ for each $j\in [m]$.

%for each good $j \in [m]$, at least one buyer $i\in [n]$   has a positive value $v_{ij} > 0$ for it. 
    %for some $i \in [n]$. 
%We will also require that each buyer values at least one good, i.e., for all $i \in [n]$, there exists a  $j \in [m]$ with $v_{ij}>0$.

To finish describing the game, one approach is to specify a tie-breaking rule: a rule that determines the probabilities with which a good is allocated among the highest bidders. However, \cite{conitzer2017multiplicative} showed that %, when only the highest bidders can win the good with a positive probability, 
the choice of tie-breaking rule affects equilibrium existence.
This motivated them~to introduce an equilibrium notion called
  the pacing equilibrium,
  which is not concerned with any specific tie-breaking rule, but instead includes the probability distribution used to allocate
  each good as part of the equilibrium
%which determines the tie-breaking rule as part of the equilibrium 
(see Definition~\ref{definition_exact_pacing_equilibrium}).
We will take a similar approach and work with pacing equilibrium, focusing on its computational aspects. It is worth pointing out that this only makes our
  hardness results stronger because they apply to 
   % that our hardness results use the equilibrium notion  of \cite{conitzer2017multiplicative} and thus,
  %they apply to 
  any tie-breaking rule
  (such as the one used by \cite{borgs2007dynamics}, which works
  via random perturbations; see Section~\ref{sec:implications} for a detailed discussion of the 
  implications of our hardness results).

%An alternative approach is to allow buyers \emph{close} to the highest bid to win the good with positive probability. One natural way to do this is via random perturbations, as discussed in \cite{borgs2007dynamics}. In this approach, the bids of the buyers are perturbed by a small random amount, independently and identically across buyers, and the highest perturbed bid wins the good. For the sake of generality, we will not work with any specific tie-breaking rule and instead take the approach of allowing these tie-breaking probabilities to be decided as part of the equilibrium concept. In Definition~\ref{definition_pacing_equilibrium}, we will define an equilibrium concept which contains as special cases all the equilibrium notions for second-price goods defined in \cite{borgs2007dynamics} and \cite{conitzer2017multiplicative}. We will make the standard assumption that the buyers are risk-neutral and are only concerned with the expected value and expected payment. For ease of notation, we will use fraction of good $j$ allocated to buyer $i$ to refer to the probability with which good $j$ is allocated to buyer $i$.}
    
With slight abuse of notation, we will write 
$x_{ij}\ge 0$ to denote the \emph{fraction} of good $j$ allocated to buyer $i$, which, in our indivisible goods regime, should be interpreted to mean the probability of allocating good $j$ to buyer $i$. 
Therefore, the allocation should  always satisfy $\sum_{i\in [n]} x_{ij} \leq 1$ for all $j \in [m]$. 
In addition, only buyers $i$
  with the \emph{highest} bid for good $j$ can have $x_{ij}>0$
  and they pay for good $j$ under  the \emph{second-price} rule.
  
Formally,  when the buyers use pacing multipliers $\alpha = (\alpha_1 ,\dots, \alpha_n)$,
	we let $h_j(\alpha)=\max_{i\in [n]} \alpha_iv_{ij}$ denote the \emph{highest} bid on good $j$ and 
%The price at which good $j$ is sold is denoted by 
$p_j(\alpha)$ denote the \emph{second highest bid} on good $j$, i.e., $p_j(\alpha)$ is the second largest element among $\alpha_1 v_{1j}, \dots, \alpha_n v_{nj}$ (in particular, $p_j(\alpha)=h_j(\alpha)$ when there is a tie for the highest bid).
Only buyers who bid $h_j(\alpha)$
  %(or close to $h_j(\alpha)$ %when
  %we consider approximate %equilibria in Definition %\ref{definition_pacing_equilibrium}) 
  can purchase (fractions of) good $j$ under the price $p_j(\alpha)$.
Thus, under an allocation $x=(x_{ij})$,
  the total payment of buyer $i$ is given by $\sum_{j\in [m]} x_{ij}p_j(\alpha)$, which should not exceed the budget $B_i$ of buyer $i$.
%We will use $h_j(\alpha)$ to denote the highest bid on item $j$.
%\end{definition}

Next, we define the notion of
  \emph{pacing equilibria} \cite{conitzer2017multiplicative} of SPP games.
A pacing equilibrium consists of a tuple of pacing multipliers $\alpha=(\alpha_i) $ and an allocation $x=(x_{ij}) $ of goods that satisfy the two conditions described above (i.e., only 
  buyers with the highest bid can be allocated a good and 
  their budgets are satisfied, as captured in (a) and (c) below).
In addition, we require (b) the full allocation of any good with a positive bid and (d) that there is no unnecessary pacing: if a buyer $i$ does not spend her whole budget, then her pacing multiplier should be one. Intuitively, this makes sense because if her budget is not binding, then she should participate as if each auction is a regular second-price auction.

\begin{definition}[Pacing Equilibria] \label{definition_exact_pacing_equilibrium}
Given an SPP game $G = (n, m, (v_{ij}), (B_i))$, 
%and parameters $\delta, \gamma \in [0,1)$, 
we say $(\alpha, x)$ with $\alpha=(\alpha_i)\in [0,1]^n$, $x=(x_{ij}) \in [0,1]^{nm}$
  and $\sum_{i\in [n]} x_{ij}\le 1$ for all $j\in [m]$ %with $\sum_{i\in [n]}x_{ij}\le 1$
  %for all $j\in [m]$ %which specifies the vector of pacing multipliers $\alpha$ and allocations $(x_{ij})_{i,j}$, 
  is a \emph{pacing equilibrium} if \vspace{0.15cm}
\begin{flushleft}
\begin{itemize}
    \item[(a)] Only buyers with the highest bid win the good: $x_{ij} > 0$ implies $\alpha_i v_{ij}= h_j(\alpha)$.\vspace{0.1cm}
    \item[(b)] Full allocation of each good with a positive bid:  $h_j(\alpha) > 0$  implies $\sum_{i\in [n]} x_{ij} = 1$.\vspace{0.1cm}%\footnote{Xi: Should we add $\sum_i x_{ij}\le 1$ for every good $j$?}
    \item[(c)] Budgets are satisfied: $\sum_{j\in [m]} x_{ij} p_j(\alpha) \leq B_i$.\vspace{0.1cm}
	\item[(d)] %Complementarity/
	No unnecessary pacing: $\sum_{j\in [m]} x_{ij} p_j(\alpha) < B_i$ implies $\alpha_i=1$.\vspace{0.15cm}
\end{itemize}
\end{flushleft}
%We will use \emph{PE} as a shorthand for \emph{Pacing Equilibrium}.
%When $\delta = \gamma = 0$, we will refer to it as an exact pacing equilibrium.
\end{definition}

%appropriate notion of equilibrium for pacing games. 
We will work with 
an \emph{approximate} version of pacing equilibria
  in both of our PPAD-hardness and PPAD-membership results.
  %is concerned with the computation
  %of ,
%To address the computability of equilibria of pacing games, we will make use of an approximate version of pacing equilibria, 
In an approximate pacing equilibrium, we make  two relaxations on (b) and (d); the two parameters used to capture these two relaxations are $\delta$ and $\gamma$, respectively.

%:%% (1) buyers with bids that are smaller than the highest bid of a  good can still win (fractions of) the good, as long as their bids are \emph{close} to the highest bid;
% (2) there is not too much unnecessary pacing
%when a buyer is not close to fully spending her budget.
%The two parameters used in the definition below to capture
%  these two relaxations are
%  $\delta$ and $\gamma$, respectively.
%To this end, we define this more general approximate version, which 
%
\begin{definition}[Approximate Pacing Equilibria] \label{definition_pacing_equilibrium}
Given an SPP game $G = (n, m, (v_{ij}),(B_i))$ and parameters $\delta, \gamma \in [0,1)$, we say $(\alpha, x)$, with $\alpha=(\alpha_i)\in [0,1]^n$, $x=(x_{ij}) \in [0,1]^{nm}$
  and $\sum_{i\in [n]} x_{ij}\le 1$ for all $j\in [m]$, is
a $(\delta,\gamma)$-\emph{approximate pacing equilibrium} of $G$ if  \vspace{0.1 cm}
%$$, which specifies e vector of pacing multipliers $\alpha$ and allocations $(x_{ij})_{i,j}$, is called a $\gamma$-approximate pacing equilibrium if 
\begin{flushleft}
\begin{itemize}
    \item[(a)] Only buyers \emph{close} to the highest bid win the good: %If $x_{ij}$ is the amount of good $j$ allocated to agent $i$, then 
    $x_{ij} > 0$ implies $\alpha_i v_{ij} \geq (1 - \delta) h_j(\alpha)$.\vspace{0.1 cm}
    \item[(b)] Full allocation of each good with a positive bid: $h_j(\alpha) > 0$ implies 
      $\sum_{i\in [n]} x_{ij} = 1$.\vspace{0.1 cm}
    \item[(c)] Budgets are satisfied: $\sum_{j\in [m]} x_{ij} p_j(\alpha) \leq B_i$.\vspace{0.1 cm}
	\item[(d)] \emph{Not too much} unnecessary pacing:  $\sum_{j\in [m]} x_{ij} p_j(\alpha) < (1 - \gamma)B_i$ implies $\alpha_i \geq 1 - \gamma$.\vspace{0.15cm}
\end{itemize}
\end{flushleft}
%When $\delta = \gamma = 0$, we will refer to it as an exact pacing equilibrium.
\end{definition}

For convenience we will write 
  $(\delta,\gamma)$-\emph{\appe}\ to denote $(\delta,\gamma)$-approximate pacing equilibrium,
  and write $\gamma$-\appe\ 
  to denote $(0,\gamma)$-\appe. %of a game $G$
It is clear from the definition that when $\delta=\gamma=0$,
$(\delta,\gamma)$-\appe\ captures the exact pacing equilibria of a SPP game. 
% We emphasize that $(\delta, \gamma)$-\appe\ is more general than the equilibrium notions for pacing defined in \cite{borgs2007dynamics} and \cite{conitzer2017multiplicative}, thereby incorporating them as special cases.  %\vspace{0.2cm}
%when $\delta = \gamma = 0$.
%{\color{red}For convenience we write .}

%\def\APE{\textsc{Approximate-PE}}

%We are now ready to state the computational problem that is addressed in this paper:
%\begin{flushleft}\begin{itemize}
%	\item \APE: Given a Second Price Pacing Game $G = (n, m, \{v_{ij}\}_{i,j}, \{B_i\}_i)$ and parameters $\delta, \gamma \in [0,1)$ as input (all numbers are encoded in binary), compute a $(\delta, \gamma)$-approximate pacing equilibrium for $G$.
%\end{itemize}\end{flushleft}
%Of course the computational problem \APE\ is well-defined only if the equilibrium concepts defined above are always guaranteed to exist. In the next section, we establish existence and do it in a way that places this computational problem in the class PPAD. 

\begin{remark}
     We can incorporate reserve prices in our model. Definition~\ref{definition_exact_pacing_equilibrium} can be extended in a natural way to model the presence of reserve prices (see Definition~\ref{definition_reserve_pacing_equilibrium}). All our results continue to hold with this extension. We refer the reader to Appendix~\ref{appendix_reserves} for a full discussion. 
\end{remark}

\subsection{\revision{Connections to Dynamics, Best Response and Nash Equilibrium}}\label{sec:sppe-motivation}

Before moving on to our results, we motivate the definition of pacing equilibrium by connecting it more concretely to practice and previous work. Consider a collection of $n$ buyers that participate repeatedly in $T$ second-price auctions. For each auction $t \in [T]$, the good to be sold is drawn from a collection of $m$ possible goods, with good $j$ being selected with probability $d_j > 0$. Moreover, suppose the value $v'_{ij}$ that buyer $i$ has for good $j$ is given by $\epsilon_{ij} v_{ij}/d_j$ for some $v_{ij} \geq 0$, where $\epsilon_{ij}$ is drawn independently for each buyer-good pair from some continuous distribution supported over $[1-\delta, 1]$. The $\epsilon_{ij}$ component of the value can also be thought of as a perturbation that arises from errors in estimating the click-through-rate (probability of a click) which is a crucial factor in determining the value of an advertiser in internet advertising. Finally, let $B'_i$ denote the budget of buyer $i$, which is the maximum amount she is willing to spend over all $T$ auctions.

\citet{balseiro2019learning} prove that, if we fix the bidding strategy of the other buyers, then it is optimal for a buyer to use pacing-based strategy to bid. The optimal pacing-based algorithm of \citet{balseiro2019learning} iteratively updates the pacing multiplier and satisfies the following properties: (i) If the buyer spends less than her per-period budget $B_i = B_i'/T$ in an iteration, her pacing multiplier is increased, and if the payment is greater than her per-period budget, then the multiplier is decreased; (ii) The pacing multiplier is constrained to belong to $[0,1]$ because bidding more than the value leads to negative utility. These properties are also satisfied by the algorithm proposed by \citet{borgs2007dynamics} and forms the basis of pacing algorithms used in practice which aim to smooth the expenditure of a buyer by evenly spending the budget over all auctions, i.e., aim to spend the per-period budget in each period if possible. If all of the buyers use an algorithm that satisfies these properties, the system can only stabilize when all of the buyers satisfy the no-unnecessary-pacing condition. 

The no-unnecessary-pacing condition and the optimality of pacing stem from strong duality, as argued in \citet{balseiro2015repeated} and \citet{balseiro2019learning}. We provide a brief overview of their argument here. When $T$ is large and $B'_i = \Theta(T)$, as is the case in online advertising, concentration arguments kick in and the problem of repeatedly bidding in $T$ auctions can be interpreted as repeatedly bidding in the following single-shot game: Each buyer wishes to maximize her expected utility (value $-$ payment) while keeping her expenditure below $B_i = B'_i/T$ in expectation over the randomness in the values (see \citealt{balseiro2015repeated, balseiro2019learning} for more details). This single-shot game captures the crux of the problem and its variants have been extensively studied in the literature \citep{balseiro2015repeated, balseiro2017budget, babaioff2020non, kumar2022contextual}. In fact, \citet{balseiro2019learning} show that, under some fairly stringent assumptions, their algorithm efficiently converges to an approximate pacing equilibrium of this single-shot game when all of the buyers employ it. But, these assumptions require independence of values across buyers and strong monotonicity of payments as a function of the pacing multipliers, both of which are unlikely to hold in practice.  As we show in this paper, if $\text{PPAD}\neq\text{P}$, then the convergence can no longer be efficient in the absence of these assumptions. In the rest of this subsection, we will restrict our focus to this single-shot game and connect it to SPP games and pacing equilibria.

Fix buyer $i$ and let $f_{j}$ denote the highest bid from buyers other than $i$ on good $j$. Then, the optimization problem faced by buyer $i$ in the single-shot game is given by
\begin{align*}
    \max_{b} \quad &\sum_{j=1}^m d_j \cdot \mathbb{E}_{v'_{ij}, f_j} \left[ (v'_{ij} - f_j) \mathbf{1}(b(j, v'_{ij}) \geq f_j) \right]\\
    \text{s.t.} \quad  &\sum_{j=1}^m d_j \cdot \mathbb{E}_{v'_{ij}, f_j} \left[ f_j \cdot \mathbf{1}(b(j,v'_{ij}) \geq f_j) \right] \leq B_i
\end{align*}
where $b(j, \cdot) $ denotes the bidding strategy of buyer $i$ for good $j$. Assume that the distribution of $f_j$ conditioned on $v'_{ij}$ (value of buyer $i$ for good $j$) is continuous. Then, using the strong-duality argument of \citet{balseiro2015repeated} or \citet{kumar2022contextual}, it can be shown that strong duality holds, where the dual problem is given by
\begin{align*}
    &\min_{\mu_i \geq 0}  \mu_i \cdot B +\max_b \sum_{j=1}^m d_j \cdot \mathbb{E}_{v'_{ij}, f_j} \left[ (v'_{ij} - (1 + \mu_i)f_j) \mathbf{1}(b(j, v'_{ij}) \geq f_j) \right]\\
    =& \min_{\mu_i \geq 0}  \mu_i \cdot B +  (1 + \mu_i) \max_b \sum_{j=1}^m  d_j \cdot \mathbb{E}_{v'_{ij}, f_j} \left[ \left( \frac{v'_{ij}}{1 + \mu_i} - f_j \right) \mathbf{1}(b(j, v'_{ij}) \geq f_j) \right]
\end{align*}

Therefore, if $\mu_i^* \geq 0$ is the optimal dual solution, then an optimal bidding strategy for buyer $i$ is $b(j, v_{ij}') = v_{ij}'/(1 + \mu_i^*)$ (i.e., to pace her value with the multiplier $\alpha_i = 1/(1 + \mu_i^*)$) since it is optimal for the inner Lagrangian optimization problem over $b$. Note that this argument does not require other buyers to use a pacing-based strategy. Thus, it establishes that a pacing-based best response always exists. 

Strong duality also implies that any optimal primal-dual solution pair satisfies complementary slackness: $\mu^*_i = 0$ if \begin{align*}
    \sum_{j=1}^m d_j \cdot \mathbb{E}_{v'_{ij}, f_j} \left[ f_j \cdot \mathbf{1}(v'_{ij}/(1 + \mu_i^*) \geq f_j) \right] < B_i \,. 
\end{align*}
The fixed-point argument of \citet{balseiro2015repeated} further shows that a pacing-based Nash equilibrium exists for the single-shot game where all of the buyers use pacing with multipliers $\alpha_i = 1/(1 + \mu_i)$. Moreover, if a collection of feasible dual multipliers satisfy complementary slackness and the corresponding pacing-based strategies satisfy the budget constraints, then they form a Nash equilibrium of the single-shot game described above. Now, let $\alpha_i = 1/(1 + \mu_i^*)$ be a collection of equilibrium pacing multipliers. Then, the complementary slackness condition for buyer $i$ can equivalently be written as a no-unnecessary-pacing condition: $\alpha_i = 0$ if
\begin{align*}
    &\sum_{j=1}^m d_j \cdot \mathbb{E}_{v'_{ij}, f_j} \left[ f_j \cdot \mathbf{1}(\alpha_i v'_{ij} \geq f_j) \right] < B_i
\end{align*}
As a consequence, every pacing equilibrium of this single-shot game is also a Nash equilibrium, where we define a pacing equilibrium to be any collection of pacing multipliers that satisfy the no-unnecessary-pacing condition and satisfy the budget constraint. Even if one has no interest in duality, the no-unnecessary-pacing condition is also extremely desirable in practice when the platform manages the budget of the buyer on her behalf --- it ensures that the platform bids the value of the buyer on each good unless doing so would violate her budget. Thus, as outlined above, pacing equilibrium is an important refinement of Nash equilibrium for the single-shot game in both theory and practice.

Next, we connect pacing equilibria in single-shot games to approximate pacing equilibria in SPP games. Observe that, when all of the buyers use pacing to bid, $f_j = \max_{k \neq i} \alpha_k \epsilon_{kj} v_{kj}/d_j$. Hence, the expected payment of buyer $i$ in this single-shot game can be rewritten as
\begin{align*}
    \mathbb{E}_{\{\epsilon_{ij}\}_{i,j}} \left[ \sum_{j=1}^m \left\{ \max_{k \neq i} \alpha_k \epsilon_{kj} v_{kj} \right\} \mathbf{1}\left( \epsilon_{ij}\alpha_i v_{ij} \geq \max_{k\neq i} \epsilon_{kj} \alpha_k v_{kj} \right) \right]
\end{align*}
If we ignore the perturbations $\epsilon_{ij}$, this is exactly the payment of buyer $i$ in the SPP game with values $v_{ij}$ and pacing multipliers $\alpha_i$. To account for the perturbations and connect the single-shot game to the SPP game, we can define a perturbed SPP game (like \citealt{borgs2007dynamics}) as one in which (i) the value of buyer $i$ for good $j$ is given by $\epsilon_{ij} v_{ij}$; (ii) each item is sold through second-price auction; (iii) the strategy of each buyer is her pacing multiplier $\alpha_i \in [0,1]$; (iv) $\epsilon_{ij}$ are drawn i.i.d. from some distribution with a positive density over $[1 - \delta, 1]$; (v) each buyer wishes to maximize her expected utility while satisfying her budget constraint in expectation over the perturbations ($-\infty$ utility if the budget constraint is violated). We define an approximate pacing equilibrium of this perturbed SPP game as simply a collection of budget-feasible pacing multipliers that satisfy the not-too-much-unnecessary-condition (see Appendix~\ref{appendix:perturbed_game}). Recall that approximate pacing equilibrium of SPP games allows for arbitrary allocation between all buyers close to the highest bid, and therefore includes the allocation induced by perturbations as a special case. In Appendix~\ref{appendix:perturbed_game}, we use this fact to show that computing a pacing equilibrium of perturbed SPP games is harder than computing an approximate pacing equilibrium in (unperturbed) SPP games, and therefore PPAD-hard due to Theorem~\ref{approximate_hardness}.

Finally, as we make $\delta$ smaller, this perturbed SPP game gets closer to a true SPP game. Unfortunately, the duality-based existence argument of \citet{balseiro2015repeated} and \citet{kumar2022contextual} breaks down when $\delta = 0$ because ties are no longer a zero-probability event. The following example shows that a pacing equilibrium may not exist in this case under the uniform tie-breaking rule.

\begin{example}
    Consider a setting with two buyers and one good. $v_{11}=1$, $v_{21} = v \gg 1$ and $B_1 = \infty$, $B_2 = 1/4$. Then, in any pacing equilibrium we have $\alpha_1 = 1$ because of the no-unnecessary-pacing condition. Now, if $\alpha_2 \geq 1/v$, then buyer 2 spends at least $1/2$ due to the uniform tie-breaking rule, which violates her budget. Hence, $\alpha_2 < 1/v_2$ and buyer two wins nothing and spends 0, thereby violating the no-unnecessary pacing condition.
\end{example}

\citet{conitzer2017multiplicative} show that a pacing equilibrium does exist if the ties are broken carefully, which was their motivation behind making the tie-breaking rule a part of the equilibrium concept. This equilibrium tie-breaking rule can be thought of as the limiting expected allocation in the perturbed equilibrium as $\delta$ approaches zero. They also show that, in an unperturbed SPP game, if we fix the bids of other buyers and allow a buyer to pick her bids along with the fraction of each good she wants, it is a best-response for her to use pacing to bid because it allows her to win goods that yield the highest value per unit cost---using the multiplier $\alpha_i$ ensures that a buyer wins a good if and only if $\alpha_i$ times her value is greater than the second-highest bid, i.e., if the value per unit cost is above $1/\alpha_i$. \citet{conitzer2017multiplicative} also provide a discussion on the undesirable properties of Nash equilibria in SPP games enroute to motivating pacing equilibria as a more desirable solution concept. Nevertheless, we would like to note that our hardness result can be extended to Nash equilibria: In Appendix~\ref{appendix:perturbed_game}, we prove that computing a Nash equilibrium of the perturbed SPP game is also PPAD-hard. We do so by showing that a minor modification of the game constructed in our hardness reduction for Theorem~\ref{approximate_hardness} only admits Nash equilibria that are also pacing equilibria.

\section{Hardness Results}

In this section we investigate the hardness of computing approximate pacing equilibria and show that the problem is PPAD-hard for second-price pacing games. Our most general result (Theorem~\ref{theo:approx_hardness_informal}) shows that the problem of finding a $(\delta, \gamma)$-\appe\ in a SPP game is PPAD-hard, even when $\delta$ and $\gamma$
  are polynomially small in the number of players. 
%We do so by demonstrating a reduction from the problem of finding an $\epsilon$-well-supported mixed-strategy Nash equilibrium in a 0-1 cost bimatrix game to that of finding an $(\delta, \gamma)$-approximate pacing equilibrium.
\revision{
Our result is shown by reducing the problem of computing a Nash equilibriun in a $\{0,1\}$-cost bimatrix game to that of finding a $(\delta,\gamma)$-\appe\ in a corresponding SPP game. 
Because we wish to show the result for $(\delta,\gamma)$-\appe, we must start our reduction from such approximate PE. In order to manage the resulting approximation factors, we are forced to introduce a number of additional bookkeeping gadgets, and correspondingly work with the problem of computing $\epsilon$-well-supported Nash equilibria of $\{0,1\}$-cost bimatrix games, as opposed to standard Nash equilibria.
Taken together, all these facts lead to a longer proof that may obfuscate the main ideas underlying our reduction. To better highlight the key ideas in our reduction and motivate our techniques, we are going to start by proving that finding an \emph{exact} pacing equilibrium in a SPP game is PPAD-hard, by showing a reduction from the problem of finding an exact Nash equilibrium in a $\{0,1\}$-cost bimatrix game. %This forms the content of the first subsection.
}

\subsection{Hardness of Finding Exact Pacing Equilibria} \label{section_hardness_exact}

Our reduction will be from the problem of computing a Nash equilibrium in a $\{0,1\}$-cost bimatrix game. Let $\Delta_n$ denote the set of probability distributions over $[n]$.
The input of the bimatrix problem is a pair of cost matrices $A,B \in \{0,1\}^{n\times n}$ and the goal is to find a Nash equilibrium $(x,y) \in \Delta_n \times \Delta_n$, meaning that $x$ minimizes cost given $y$, i.e. $x^TAy \leq \hat x^TAy$ for all $\hat x\in \Delta_n$, and similarly $y$ minimizes cost given $x$, i.e. $x^TBy \leq x^TB\hat y$ for all $\hat y\in \Delta_n$.
Equivalently, $(x,y)$ is a Nash equilibrium if $x_i>0$ for any $i\in [n]$
 implies that $\sum_j A_{ij} y_j\le \sum_j A_{kj}y_j$ for all $k\in [n]$, and $y_j>0$ for any $j\in [n]$ implies that $\sum_ix_iB_{ij}\le \sum_ix_iB_{ik}$ for all $k\in [n]$. 
 %, where $A_i$ denotes the $i$-th row vector of $A$ and $B_j$ denotes the $j$-th~column vector of $B$.)  
 This problem is known to be PPAD-complete \cite{WinLose}.

Given a $\{0,1\}$-cost bimatrix game $(A,B)$ with $A,B\in \{0,1\}^{n\times n}$,
 we would like to construct an SPP game $G$ in time polynomial in $n$, such that every exact PE of $G$ can be mapped back to a Nash equilibrium of the bimatrix game $(A,B)$ in polynomial time. %Then, finding a pacing equilibrium for $G$ would be at least as hard as finding a NE for the bimatrix game.
 
 Before proceeding further, we informally describe some important aspects of the construction to provide some intuition. First, in the SPP game $G$, we will encode the pair $(x,y)$ of mixed strategies in $\Delta_n$ using pacing multipliers. For each player $p \in \{1,2\}$ in the bimatrix
 game $(A,B)$ and each (pure) strategy $s \in [n]$, there will be a corresponding buyer $\mathbb{C}(p,s)$ in
the SPP game~$G$, whose pacing multiplier $\alpha(\mathbb{C}(p,s))$ will be used to encode the probability with which player $p$ plays strategy $s$ in the bimatrix game $(A,B)$.
 For now, take $x$ to be the distribution obtained by normalizing 
  $\alpha(\mathbb{C}(1,s))$, i.e., $x_t = \alpha(\mathbb{C}(1,t))/\sum_s \alpha(\mathbb{C}(1,s))$, and define $y$ similarly
  using $\alpha(\mathbb{C}(2,s))$; we will discuss the issues with this proposal
  and ways to fix them momentarily. 
 
  Second, in order to capture the best response condition of Nash equilibria, we need to~\mbox{encode} the cost borne by player $p\in \{1,2\}$ when playing a given strategy $s\in [n]$ against the mixed strategy of the other player. For simplicity, let us focus on $p =1$. We will create a set of $n$ \emph{expenditure goods} 
  $E(1,s)_1, \dots, E(1,s)_{n}$ for each pure strategy $s$ of player $1$. %consisting of $n$ goods (called , one for each opponent strategy. 
 We will set buyer $\mathbb{C}(1,s)$'s value at $1$ for each of the expenditure goods $E(1,s)_1, \dots, E(1,s)_{n}$. Additionally, each buyer $\mathbb{C}(2,t)$ will value $E(1,s)_t$ at $\nu A_{st}$, where $\nu=1/(16n)$ is set to be so small that $\mathbb{C}(1,s)$ always wins all the goods $ E(1,s)_1, \dots, E(1,s)_{n} $ under any PE of $G$. 
This means that, in any PE with multipliers $\alpha(\mathbb{C}(p,s))$, buyer $\mathbb{C}(1,s)$ pays a total of $\nu \sum_t \alpha(\mathbb{C}(2,t)) A_{st}$ for the expenditure goods $E(1,s)_1, \dots, E(1,s)_{n}$, which captures player 1's cost for playing strategy $s$ in $(A,B)$, when player $2$ uses the mixed strategy that plays each $t$ with probability defined by $\alpha(\mathbb{C}(2,t))$ after normalization. % (for fully mixed strategies; actions played with probability zero must be handled via thresholding as described next).
  
  Finally we need to make sure that the best response condition of Nash equilibria holds for a strategy pair $(x,y)$ obtained from multipliers 
  $\alpha( \mathbb{C}(p,s))$ in any PE of $G$, i.e., only  best-response strategies are played with positive probability. This poses a challenge because pacing multipliers are never zero in a pacing equilibrium, so we can't use them directly to encode probabilities in $x$ and $y$ (which need to be zero for strategies which are not best responses). To get around this issue, we will use thresholds to encode entries of $(x,y)$ using $\alpha(\mathbb{C}(p,s))$.
More formally, we add a \emph{threshold buyer} and a set of 
  \emph{threshold goods} to $G$ to make sure that 
  $\alpha(\mathbb{C}(p,s))\ge 1/2$ in any PE of $G$.
This allows us to encode $x$ by normalizing $\alpha(\mathbb{C}(1,s))- 1/2$
  and $y$ by normalizing $\alpha(\mathbb{C}(2,s))-1/2$.
The most challenging part of the construction is to have
  buyers\hspace{0.05cm}/\hspace{0.05cm}goods work together to ensure that
  both $\alpha(\mathbb{C}(1,s))-1/2$ and $\alpha(\mathbb{C}(2,s))-1/2$, $s\in [n]$, are not identically zero. We accomplish this by creating a set of \emph{normalization goods} for each buyer $\mathbb{C}(p,s)$, with the property that each buyer $\mathbb{C}(p,s)$ spends approximately $\sum_{t=1}^n \alpha(\mathbb{C}(p,t))$ on her normalization goods. This, in combination with a carefully chosen budget and the `No unnecessary pacing' condition, ensures that $\{\alpha(\mathbb{C}(p,s))-1/2\}_s$ are not identically zero.
Then, we can follow the plan described in the last paragraph
  to encode the cost of player $p$ playing $s$ using the
  expenditure of buyer $\mathbb{C}(p,s)$ on 
  $ E(p,s)_1, \dots, E(p,s)_{n} $, with careful calibration
  via the use of thresholds.
This finally helps us enforce the best response condition of Nash equilibria on $(x,y)$ in $(A,B)$ by comparing total expenditures of buyers $\mathbb{C}(p,s)$ and using implications from such comparisons. 
%, calibrated to ensure that a pacing multiplier at this threshold level corresponds to zero probability. We will define {\em threshold goods} and {\em threshold buyers} to help with this calibration.

\revision{
 We now formally define the SPP game $G$ in the next section, and then the following sections show the hardness result based on $G$.
  \subsubsection{The SPP Game}
 The game $G$ has the following set of goods:\vspace{0.15cm}
  }
\begin{itemize}
    \item Normalization goods: $n$ goods $\{N(p,s)_1, \dots,  N(p,s)_n\}$ for each $p \in \{1,2\}$ and  $s \in [n]$. \vspace{0.15cm}
    \item Expenditure goods: $n$ goods $\{E(p,s)_1, \dots, E(p,s)_{n}\}$ for each $p \in \{1,2\}$ and $s \in [n]$.\vspace{0.15cm}
    \item Threshold goods: 1 good  $T(p,s)$ for each  $p \in \{1,2\}$ and  $s \in [n]$.\vspace{0.15cm}
\end{itemize}

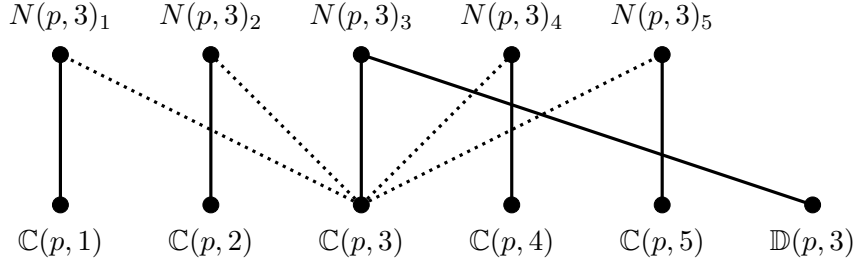
\begin{figure}
    \centering
    \begin{tikzpicture}
    %% vertices
    
    \foreach \i in {1,...,5}
    {
    	\draw[fill=black] (\i*2,0) circle (3pt);
    }
    
    \foreach \i in {1,...,5}
    {
    	\draw[fill=black] (\i*2,2) circle (3pt);
    }
    
    \foreach \i in {1,...,5}
    {
    	\node at (\i*2,-0.5) {$\mathbb{C}(p,\i)$};
    }
    
    \foreach \i in {1,...,5}
    {
    	\node at (\i*2,2.5) {$N(p,3)_{\i}$};
    }
    
    \foreach \i in {1,...,5}
    {
    	\draw[very thick] (\i*2,0) -- (\i*2,2);
    }
    
    \foreach \i in {1,2, 4,5}
    {
    	\draw[dotted, very thick] (6,0) -- (\i*2,2);
    }
    
    \draw[fill = black] (12, 0) circle (3pt);
    \node at (12, -0.5) {$\mathbb{D}(p,3)$};
    \draw[very thick] (12, 0) -- (6,2);
    
    \end{tikzpicture}
    \caption{Normalization goods for $p \in \{1,2\}$ and $s = 3$, when $n=5$. A buyer having a non-zero value for a good is represented by a line connecting the two. Solid line denotes a value of 1 and dotted line denotes a value of 2. }
    \label{fig:my_label}
\end{figure}

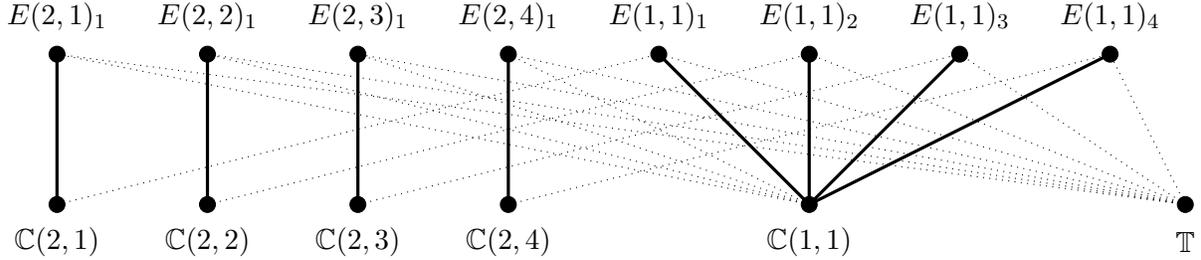
\begin{figure}
    \centering
    \begin{tikzpicture}
    %% vertices
    
    \foreach \i in {1,...,4}
    {
    	\draw[fill=black] (\i*2,0) circle (3pt);
    }
    
    \foreach \i in {1,...,8}
    {
    	\draw[fill=black] (\i*2,2) circle (3pt);
    }
    
    \foreach \i in {1,...,4}
    {
    	\node at (\i*2,-0.5) {$\mathbb{C}(2,\i)$};
    }
    
    \foreach \i in {1,...,4}
    {
    	\node at (\i*2,2.5) {$E(2,\i)_{1}$};
    }

    \node at (10,2.5) {$E(1,1)_{1}$};
    \node at (12,2.5) {$E(1,1)_{2}$};
    \node at (14,2.5) {$E(1,1)_{3}$};
    \node at (16,2.5) {$E(1,1)_{4}$};

    \foreach \i in {1,...,4}
    {
    	\draw[very thick] (\i*2,0) -- (\i*2,2);
    }

    \foreach \i in {5,...,8}
    {
    	\draw[very thick] (12,0) -- (\i*2,2);
    }
    
    \foreach \i in {1,...,4}
    {
    	\draw[dotted] (12,0) -- (\i*2,2);
    }

    \draw[dotted] (2,0) -- (10,2);
    \draw[dotted] (4,0) -- (12,2);
    \draw[dotted] (6,0) -- (14,2);
    \draw[dotted] (8,0) -- (16,2);
    
    \draw[fill = black] (12, 0) circle (3pt);
    \node at (12, -0.5) {$\mathbb{C}(1,1)$};

    \draw[fill = black] (17, 0) circle (3pt);
    \node at (17, -0.5) {$\mathbb{T}$};

    \foreach \i in {1,...,8}
    {
    	\draw[dotted] (17,0) -- (\i*2,2);
    }
    
    \end{tikzpicture}
    \caption{\revision{All the expenditure goods for which buyer $\mathbb{C}(1,1)$ has a non-zero value, when $n=4$. A buyer having a non-zero value for a good is represented by a line connecting the two. Solid lines denote a value of 1 and dotted lines denote values which are smaller than $\nu = 1/(16n)$.}}
\end{figure}

Set $\nu = 1/(16n)$. The set of buyers in $G$ is defined as follows, where we write $V(\cdot,\cdot)$ to denote the value of 
  a good (the second component) to a buyer (the first component):\vspace{0.15cm}
\begin{itemize}
    \item Buyer $\mathbb{C}(p,s)$, $p \in \{1,2\}$ and $s \in [n]$: $\mathbb{C}(p,s)$ has positive values for the following goods:\vspace{0.2cm}
    \begin{itemize}
        \item Normalization goods:
        %other than $N(p,s)_s$ have value $2$, i.e., 
        $V(\mathbb{C}(p,s), N(p,s)_i) = 2$ for all $i \in [n]\setminus{\{s\}}$;\\
%        \item The $s$'th $(p,s)$-normalization good has value $1$, i.e., 
$V(\mathbb{C}(p,s), N(p,s)_s) = 1$; and
%        \item The $s$'th $(p,t)$-normalization good at value $1$ for all $t \neq s$, i.e., 
$V(\mathbb{C}(p,s), N(p,t)_s) = 1$ for all $t \in [n]\setminus \{s\}$.\vspace{0.15cm}
        
        \item Threshold good $T(p,s)$:   $V(\mathbb{C}(p,s), T(p,s)) = 2n^4$.\vspace{0.15cm}
        
        \item Expenditure goods: % $\{E(p,s)_1, \dots, E(p,s)_n\}$ have value $1$, i.e., 
        $V(\mathbb{C}(p,s), E(p,s)_i) = 1$ for all $i \in [n]$.\\
        For $p = 1$: %$\mathbb{B}(1,s)$ values the $s$'th $(2,t)$-expenditure good at $\nu B_{st}$ for all $t \in [n]$, i.e., 
        $V(\mathbb{C}(1,s), E(2,t)_s) = \nu B_{st}$ for all $t \in [n]$.
        
        For $p = 2$: %$\mathbb{B}(2,s)$ values the $s$'th $(1,t)$-expenditure good at $\nu A_{ts}$ for all $t \in [n]$, i.e., 
        $V(\mathbb{C}(2,s), E(1,t)_s) = \nu A_{ts}$ for all $t \in [n]$.\vspace{0.2cm}
    \end{itemize}
    For $p=1$, the budget of $\mathbb{C}(1,s)$ is $n/2 + n^4 + 1/4 - \nu + \sum_{t\in [n]} \nu A_{st}/2$;\\ For $p=2$, the budget of $\mathbb{C}(2,s)$ is $n/2 + n^4 + 1/4 - \nu + \sum_{t\in [n]} \nu B_{ts}/2$.
    \vspace{0.25cm}
    \item Threshold Buyer $\mathbb{T}$: $\mathbb{T}$ has positive values only for the following goods:\vspace{0.15cm}
    \begin{itemize}
        \item Threshold goods: $V(\mathbb{T}, T(p,s)) = n^4$ for each $p \in \{1,2\}$ and  $s \in [n]$.\vspace{0.15cm}
        
        \item Expenditure goods:  %For $p=1$: The expenditure good 
        $V(\mathbb{T},E(1,s)_t)= \nu A_{st}/2$ 
        and $V(\mathbb{T},E(2,s)_t)=\nu B_{ts}/2$ for all $s,t \in [n]$.%\\
%        
%        For $p=2$: The expenditure good $E(2,s)_t$ has value $\nu B_{ts}/2$ for all $s,t \in [n]$.
\vspace{0.2cm}
    \end{itemize}
    
    $\mathbb T$ has budget $n^7$ (high enough \revision{so that $\alpha(\mathbb T) = 1$} in any PE).\vspace{0.25cm}
    
    \item Dummy buyers $\mathbb{D}(p,s)$, $p\in \{1,2\}$ and $s\in [n]$: %For each player $p \in \{1,2\}$ and strategy $s \in [n]$, there is a buyer $\mathbb{D}(p,s)$ with budget $\nu$, 
    The budget of $\mathbb{D}(p,s)$ is $\nu$ and she only values
      the normalization good $N(p,s)_s$ at $V(\mathbb{D}(p,s), N(p,s)_s)=1$.\vspace{0.15cm}
%    who values the $s$'th $(p,s)$-normalization good at 1 and every other good at 0.\vspace{0.15cm} % Moreover, $\D(p,s)$ has a  for all $p \in \{1,2\}$ and $s \in [n]$.
\end{itemize}
It is clear from the definition of $G$ that it can be constructed
  from $(A,B)$ in polynomial time.

\revision{
\subsubsection{Structure of Pacing Equilibria of $G$}
With the definition of $G$ in place, we start by showing some auxiliary structural results on the PE of $G$; these will be used to construct strategies for the bimatrix game.
}
Let $\mathcal{E}$ be a PE of the SPP game $G$.  We will use $\alpha(b)$ to denote the pacing multiplier of buyer $b$ in $\mathcal{E}$.  
Observe that, from the definition of pacing equilibria, we can conclude that $\alpha(\mathbb{T})=1$ in $\mathcal{E}$;
  otherwise $\mathbb{T}$ needs to spend all her budget of $n^7$,
  which is impossible given that no buyer has value more than $2n^4$ for any good.
The following lemma establishes bounds on 
  $\alpha(\mathbb{C}(p,s))$ in $\mathcal{E}$.

\begin{lemma}\label{mutlipliers_bounds}
    For each $p \in \{1,2\}$ and $s \in [n]$, $1/2 \leq \alpha(\mathbb{C}(p,s)) <1$ and $\alpha(\D(p,s)) = \alpha(\mathbb{C}(p,s))$. 
\end{lemma}
\def\BB{\mathbb{C}}
\begin{proof}
    Suppose for some $p \in \{1,2\}$ and $s \in [n]$, we have $\alpha(\BB(p,s)) < 1/2$. Then $\mathbb{C}(p,s)$ doesn't win any part of threshold good $T(p,s)$. Observe that she has value at most $2$ for every other good. 
Given that there are only $O(n^2)$ goods in $G$,
  %and the sum of her bids on all items other than $T(p,s)$ is at most $2 n^3$. Hence, 
  she cannot possibly spend all her budget (which is $\Omega(n^4)$). \revision{Here, we have used the fact that the payment is smaller than her bid on every item that she wins because of the second-price auction format, which in turn is always smaller than her value.}
This contradicts the assumption that $\mathcal{E}$ is a PE of $G$.
%definition of a pacing equilibrium because her budget is strictly greater than $n^4$. 
Therefore, $\alpha(\BB(p,s)) \geq 1/2$.

  Next we prove $\alpha(\mathbb{D}(p,s))=\alpha(\mathbb{C}(p,s))$.
Suppose $\alpha(\D(p,s)) > \alpha(\BB(p,s))$ for some $p \in \{1,2\}$ and $s \in [n]$. Then, buyer $\D(p,s)$ wins all of good $N(p,s)_s$ at price $\alpha(\BB(p,s))\ge 1/2$ \revision{because $\D(p,s)$ and $\mathbb{C}(p,s)$ both value $N(p,s)_s$ at 1, and the rest of the buyers have zero value for it}. This violates her budget constraint and leads to a contradiction. Therefore, $\alpha(\D(p,s)) \leq \alpha(\BB(p,s))$. Moreover, if $\alpha(\D(p,s)) < \alpha(\BB(p,s))$ (which implies $\alpha(\D(p,s))<1$) then her expenditure is zero. This violates the no unnecessary pacing condition. Hence, $\alpha(\D(p,s)) = \alpha(\BB(p,s))$ must hold. Observe that, in particular, this means that the price of 
  $N(p,s)_s$ is $\alpha(\BB(p,s))$.
    
Finally suppose $\alpha(\BB(p,s)) = 1$ for some $p \in \{1,2\}$, $s \in [n]$. Then she wins the following goods:\vspace{0.1 cm}
\begin{flushleft}    \begin{itemize}
        \item 
        All of 
        normalization goods $N(p,s)_t$ for each $t\ne s$ \revision{because $\mathbb{C}(p,s)$ has the higher value for them,} and she spends at least $1/2$ on each of them because $\alpha(\BB(p,t)) \geq 1/2$ 
        %for all $p \in \{1,2\}$, $r \in [n]$ 
        by the first part of the proof.\vspace{0.15cm}
        
        \item Part of normalization good $N(p,s)_s$ by spending at least $1 - \nu$. This is because $N(p,s)_s$ has\\ price $1$, \revision{she shares it with $\D(p,s)$,} and
        buyer $\D(p,s)$ only has budget $\nu$.
%        (Recall that $\D(p,s)$ has a budget of $\nu$).
\vspace{0.15cm}
        
        \item All of threshold good $T(p,s)$ by spending $n^4$ \revision{because she has the higher value.}\vspace{0.15cm}
        
        \item All of expenditure good $E(p,s)_t$, for each $t\in [n]$, by spending at least $\nu A_{st}/2$ if $p=1$\\ and $\nu B_{ts}/2$ if $p=2$ \revision{because she has the higher value.} \vspace{0.1 cm}
    \end{itemize}\end{flushleft}
    Hence, the total expenditure of $\BB(p,s)$ is at least $(n-1)/2 + 1 - \nu + n^4 + \sum_t \nu A_{st}/2$ if $p=1$ and at least $(n-1)/2 + 1 - \nu + n^4 + \sum_t \nu B_{ts}/2$ if $p=2$. In both cases, the budget constraint is violated, leading to a contradiction. Therefore, the lemma holds.
\end{proof}
The above lemma implies that every $\BB(p,s)$ is paced in $\mathcal{E}$ (i.e. $\alpha(\BB(p,s))<1$), thereby implying that their total expenditures must exactly equal their budgets. Additionally, we have the following corollary which will be used in the proof of Lemma~\ref{lem:exact pe is ne}.
\begin{corollary}
	For each $p \in \{1,2\}$ and  $s \in [n]$, 
	$\BB(p,s)$ spends exactly $\alpha(\BB(p,s)) - \nu$ on 
	  $N(p,s)_s$.
\end{corollary}

Next let $x_s' = \alpha(\BB(1,s)) - 1/2$ and $y_s' = \alpha(\BB(2,s)) - 1/2$ for each $s\in [n]$.
The following lemma will allow us to normalize $x'$ and $y'$ to obtain   probability distributions $x$ and $y$. 
%(which we will argue
%  form a Nash equilibrium of $(A,B)$).

\begin{lemma} \label{lem:exact pe multiplier_positive_sum}
    The following inequalities hold: $\sum_{s\in [n]} x_s' > 0$ 
    and $\sum_{s\in [n]} y_s' > 0$.
\end{lemma}

\begin{proof}
    We show $\sum_s x_s' > 0$. The proof of $\sum_s y_s' > 0$ is completely analogous. Suppose $\sum_s x_s' = 0$. Then,  $\alpha(\BB(1,s))=1/2$ for all $s \in [n]$ \revision{because $\alpha(\BB(1,s))=1/2$ by Lemma~\ref{mutlipliers_bounds}.}
We argue below that $\BB(1,1)$ violates the no-unnecessary-pacing condition.

    %By the previous lemma, this implies that $\alpha(\N(1,s)) = 1/2$ for all $s\in [n]$. 
To see this, observe  $\mathbb{C}(1,1)$ only wins a non-zero fraction of the following goods, and spends:\vspace{0.15cm}
    \begin{itemize}
        \item At most $1/2$ on each normalization good $N(1,1)_t$, $t \in [n]$, \revision{because the highest competing bid is 1/2 on these goods}.\vspace{0.1 cm}
        
        \item At most $n^4$ on the threshold good $T(1,1)$ \revision{because that is the highest competing bid}.\vspace{0.1 cm}
        
        \item At most $\nu A_{1t}$ on each expenditure good $E(1,1)_t$,  $t \in [n]$, \revision{because that is the highest possible competing bid}.\vspace{0.15cm}
    \end{itemize}
    Hence, the total expenditure of $\mathbb{C}(1,1)$ is at most $n/2 + n^4 + \sum_t \nu A_{1t}$, which is strictly less than her budget of $n/2 + n^4 + 1/4 - \nu + \sum_t \nu A_{1t}/2$, a contradiction.
\end{proof}

\revision{
\subsubsection{Extracting Bimatrix Game Equilibria from $G$}
}
Now, we are ready to define the mixed strategies $(x,y)$  for the bimatrix game $(A,B)$. Set player 1's mixed strategy $x$ to be $x_s= x_s'/ \sum_i x_i'$ and player 2's mixed strategy $y$ to be $y_s=y_s'/ \sum_i y_i'$. These are valid mixed strategies because of Lemma~\ref{mutlipliers_bounds} and Lemma~\ref{lem:exact pe multiplier_positive_sum}. 
The next lemma shows that $(x,y)$ is indeed a Nash equilibrium of $(A,B)$.
%To complete the proof of PPAD hardness, we need to show that the mixed strategies defined in this way form a NE for the bimatrix game, which is the subject of the next lemma.

\begin{lemma}
  \label{lem:exact pe is ne}
	$(x,y)$ is a Nash equilibrium for the bimatrix game   $(A,B)$.
\end{lemma}

\begin{proof}
Suppose there are $s,s^*\in [n]$ such that 
  $x_s>0$ but $\sum_t A_{st}y_t>\sum_t A_{s^*t}y_t$ (the proof 
  for $y$ is analogous).
%	Consider $s \in [n]$ such that $x_s > 0$. Then, 
Using $x_s>0$, buyer $\mathbb{C}(1,s)$ 
  spends non-zero amounts  on the following goods:\vspace{0.15cm}
\begin{itemize}
    \item $\alpha(\BB(1,t))$ on the normalization good $N(1,s)_t$ for each $t\ne s$ \revision{because $\mathbb{C}(1,s)$ has a bid strictly greater than 1, which is the value and an upper bound on the bid of $\mathbb{C}(1,t)$}.\vspace{0.1 cm}
    
    \item $\alpha(\BB(1,s)) - \nu$ on the normalization good $N(1,s)_s$ \revision{because she shares the good with $\D(1,s)$ who has a budget of $\nu$}.\vspace{0.1 cm}
    
    \item $n^4$ on the threshold good $T(1,s)$ \revision{because her bid is strictly greater than $n^4$}.\vspace{0.1 cm}
    
    \item $ \alpha(\BB(2,t))\cdot \nu A_{st}$ on the expenditure good $E(1,s)_t$ for each  $t \in [n]$.\vspace{0.15cm}
\end{itemize}

Therefore, the total expenditure of buyer $\mathbb{C}(1,s)$  is given by 
\begin{align*}
    \sum_{t\in [n]} \alpha(\BB(1,t)) &+ n^4 - \nu + \sum_{t\in [n]} \alpha(\BB(2,t))\cdot \nu A_{st}\\  &= \sum_{t\in [n]} \alpha(\BB(1,t)) + n^4 - \nu  + \sum_{t\in [n]} \nu A_{st}/2 + \sum_{t\in [n]} y_t \nu A_{st}
\end{align*}
Note that the RHS above after replacing $s$ with $s^*$:
$$\sum_{t\in [n]} \alpha(\BB(1,t)) + n^4 - \nu  + \sum_{t\in [n]} \nu A_{s^*t}/2 + \sum_{t\in [n]} y_t \nu A_{s^*t}$$
is an upper bound for the total expenditure of buyer $\BB(1,s^*)$
  (no matter whether $x_{s^*}>0$ or not).

As a result, the total expenditure of $\BB(1,s)$ minus that of 
  $\BB(1,s^*)$ is at least
$$
\left( \sum_{t\in [n]} \nu A_{st}/2 + \sum_{t\in [n]} y_t \nu A_{st}\right)
-\left(\sum_{t\in [n]} \nu A_{s^*t}/2 + \sum_{t\in [n]} y_t \nu A_{s^*t}\right)>\sum_{t\in [n]} \nu A_{st}/2-\sum_{t\in [n]} \nu A_{s^*t}/2
$$
using the assumption that $\sum_t A_{st}y_t>\sum_t A_{s^*t} y_t$.
On the other hand, the budget of $\BB(1,s)$ minus that of $\BB(1,s^*)$
  is equal to the RHS above. This is a contradiction because both
  buyers should have their total expenditures equal to their budgets.
This finishes the proof of the lemma.
\end{proof}

Thus, given a $\{0,1\}$-cost bimatrix game $(A,B)$, we have defined an SPP game $G$  which satisfies the following properties: (i) $G$ can be constructed in polynomial time; (ii) any PE $\mathcal{E}$ of $G$ can be used to construct a Nash equilibrium $(x,y)$ of $(A,B)$ in polynomial time. 
As a result, the problem of finding an exact pacing equilibrium in
  a second-price pacing game is PPAD-hard.
%Hence, if we can find  a polynomial-time algorithm for finding PE in general pacing games, then we can compose it with our reduction to find a NE in polynomial time for general 0-1 cost bimatrix games, a problem which is known to be PPAD-hard \cite{chen2007approximation}.

%\begin{theorem}\label{exact_hardness}%
%	Computing exact pacing equilibria in Second Price Pacing Games is PPAD hard. 	
%\end{theorem}

\subsection{Hardness of Finding Approximate Pacing Equilibria} \label{section_approx_hardness}

We next state our main hardness result, which extends the PPAD-hardness of finding pacing equilibria to the  approximate case of finding $(\delta, \gamma)$-approximate pacing equilibria.   

\begin{theorem} \label{approximate_hardness}
	The problem of computing a $(\delta,\gamma)$-\appe\ of an SPP game $G = (n, m, (v_{ij}),$ $(B_i))$ with $\delta=\gamma=1/n^7$ is PPAD-hard.  
\end{theorem}

The proof is relegated to Appendix~\ref{appendix_hardness}.
It uses similar ideas but entails more involved bookkeeping
%follows from a modification of the reduction given in Section \ref{section_hardness_exact} which 
  to incorporate approximations introduced in % the definition of 
  $(\delta, \gamma)$-\appe. %It ,
%, which doesn't add much to the reader's intuition, 
%and hence,
%The proof .
Theorem \ref{theo:approx_hardness_informal} follows from
  Theorem \ref{approximate_hardness} by standard padding 
  arguments (i.e., adding dummy buyers to the game).

\subsection{Implications of The Hardness Result}\label{sec:implications}

Before concluding this section, we discuss some implications of our hardness results. In \cite{borgs2007dynamics}, the authors introduced a natural bidding heuristic for optimizing the utility of budget-constrained agents who repeatedly participate in day-long auction campaigns for $m$ items, where the set of agents and items remains the same every day. The heuristic maintains a pacing multiplier for each agent, which is increased by a small amount if the buyer ran out of her daily budget before the end of the previous day, and decreased otherwise. They use random perturbation to avoid instabilities, which gives an agent who bids close to the highest bid a fraction of the item in expectation. If we ignore the intra-day temporal aspects of their model, their setting can be thought of as repeatedly playing the perturbed SPP game from Section~\ref{sec:sppe-motivation} every day. In Theorem~1 of \cite{borgs2007dynamics}, they prove that their heuristic efficiently converges for first-price auctions. Furthermore, they conjecture the convergence of the heuristic for second-price auctions to pacing multipliers which satisfy the following conditions: (i) Every agent runs out of her daily-budget close to the end of the day; (ii) Every agent either spends most of her daily budget or has a pacing multiplier close to one. \revision{In Theorem~\ref{thm:approx-perturbed} of Appendix~\ref{appendix:perturbed_game}, we show that Theorem~\ref{approximate_hardness} implies that computing an approximate pacing equilibrium of the perturbed SPP game is also PPAD hard. As a consequence, if PPAD~$\neq$~P, then ALGORITHM 1 of \cite{borgs2007dynamics} does not always converge efficiently for second-price auctions, i.e., the number of days/time-steps required for convergence cannot scale as a polynomial function of the input size and $(1/\delta, 1/\gamma)$ in the worst-case.} In other words, we have shown that Theorem 1 of \cite{borgs2007dynamics} cannot be extended to second-price auctions in any way that maintains efficient convergence unless PPAD~$=$~P, thereby making progress towards their open conjecture.

% It can be shown that this is equivalent to convergence to an $(\delta, \gamma)$-\appe\ for the corresponding pacing game, where $\delta$ comes from the definition of the random perturbations in their model, and $\gamma$ is the approximation factor used to measure the extent of the convergence. Thus, Theorem~\ref{approximate_hardness} implies that, if PPAD~$\neq$~P, then the heuristic of \cite{borgs2007dynamics} cannot converge efficiently, i.e., the number of days/time-steps required for convergence cannot scale as a polynomial function of the input size and $(1/\delta, 1/\gamma)$. In other words, we have shown that Theorem 1 of \cite{borgs2007dynamics} cannot be extended to second-price auctions in any way that maintains efficient convergence unless PPAD~$=$~P, thereby making progress towards their open conjecture.

\revision{Moreover, recall from Section~\ref{sec:sppe-motivation} that if all of the buyers employ pacing algorithms, like the one proposed by \citet{balseiro2019learning}, and the resulting dynamics converge, then they will converge to an approximate pacing equilibrium. Our hardness result (Theorem~\ref{approximate_hardness} and Theorem~\ref{thm:approx-perturbed}) implies that there exists a (correlated) value distribution such that the algorithm of \citet{balseiro2019learning}, which is optimal for a single buyer against an adversarial/stochastic competition, does not converge efficiently to an equilibrium when employed by all the buyers, unless PPAD$=$P.}

% Moreover, pacing games admit a stochastic interpretation as a one-shot second-price auction with buyers who want to satisfy their budget constraints in expectation (see Proposition~10 of \citet{conitzer2017multiplicative}). This implies that if all of the buyers employ some pacing algorithm for repeated second-price auctions with correlated value distributions under global budget constraints, then the resulting dynamics will not always converge efficiently to an equilibrium, assuming PPAD~does not have polynomial-time algorithms. In particular, this statement applies to the pacing algorithm given by \citet{balseiro2019learning}, which is an optimal bidding algorithm for a single budget-constrained buyer under both adversarial and independent-stochastic competition. 

Our hardness results are also pertinent to the relationship between pacing equilibria and market equilibria. In Proposition 5 of \cite{conitzer2017multiplicative}, the authors show that every pacing equilibrium in a second-price pacing game has an equivalent supply-aware market equilibrium with linear utilities, where supply-aware means that the buyers are aware of the supplies of each item and choose their demand set accordingly. 
%The concept of supply-aware market equilibria is an extension of the standard notion of market equilibria, in which the buyers select their demand set with no regard for the supply (but the supply constraints need to be satisfied nonetheless). %In \cite{conitzer2019pacing}, it is shown that pacing equilibria for first-price auctions correspond to the standard notion of market equilibrium and hence, can be computed efficiently using the Eisenberg-Gale convex program (see \cite{cole2017convex}). 
Thus, the relationship between pacing equilibria and market equilibria, in combination with Theorem \ref{approximate_hardness}, implies that there exists a refinement of the set of supply-aware market equilibria with linear utilities which is PPAD-hard to compute.

%, even though, in general, a supply-aware market equilibrium can be found in polynomial time using the Eisenberg-Gale program.
\section{Existence of Pacing Equilibria and Membership in PPAD}

We prove Theorem \ref{theo:membership} in this section, i.e.,
  the problem of finding a pacing equilibrium of an SPP game is in PPAD.
One consequence of this result is that every SPP game 
  with rational values $ v_{ij} $ and budgets $ B_i$ has a pacing equilibrium $(\alpha,x)$ with rational entries.
%  the existence of a pacing equilibrium with rational entries
%  in 
%In this section,
%we establish the existence of approximate pacing equilibria. \textcolor{red}{rk: After adding the LP arguement, we will be establishing the existence of exact pacing equilibria, right?}

Our plan is as follows. 
We first introduce a restricted version of approximate pacing equilibria 
  called \emph{smooth $(\delta,\gamma)$-\appe} (see 
  Definition \ref{definition_smooth_equilibrium}),
  which will only be used in Section \ref{section_existence_approx_pacing}.
We prove in Section \ref{section_existence_approx_pacing} that the problem of finding a smooth
  $(\delta,\gamma)$-\appe\ (when $\delta$ and $\gamma$ are input
  parameters encoded in binary) is in PPAD.
Given that the smooth version (Definition \ref{definition_smooth_equilibrium}) is a restriction of $(\delta,\gamma)$-\appe\ (Definition \ref{definition_pacing_equilibrium}), this implies that the problem of computing a $(\delta,\gamma)$-\appe\ is in PPAD.
  
Next we give in Section \ref{section_rounding_alg} an efficient algorithm that 
  can round any $(\delta,\gamma/2)$-approximate PE into a $\gamma$-\appe\ 
  when $\delta$ is sufficiently small.
This, combined with the PPAD-membership of $(\delta,\gamma)$-\appe,
 shows that the problem of computing $\gamma$-\appe\ is also in PPAD.
  
Finally we show in Section \ref{sec:exact} that, when $\gamma$
  is sufficiently small, any $\gamma$-\appe\ of $G$ 
  can be used to build a linear program
  which can then be solved to obtain an \emph{exact} pacing equilibrium of $G$.
It follows that the problem of computing an exact pacing equilibrium is in PPAD.
%The PPAD membership of PE follows by combining 
%  the PPAD membership of \APE\ and algorithms presented in Sections \ref{} and 
%  \ref{} (see the proof of Theorem \ref{} in Section \ref{}).

\subsection{PPAD Membership of Computing $(\delta,\gamma)$-Approximate Equilibria} \label{section_existence_approx_pacing}

We start with the definition of \emph{smooth  $(\delta,\gamma)$-\appe.} It is a refinement of $(\delta, \gamma)$-\appe\ in which the pacing multipliers $(\alpha_i)$ \emph{fully determine} the allocations $(x_{ij})$. Note that this is not the case for  $(\delta, \gamma)$-\appe\ in general: potentially there can be $(\delta, \gamma)$-\appe\ with identical multipliers but different allocations. The smooth version we consider below, on the other hand, specifies the allocations as continuous functions of multipliers.
%, while maintaining properties (a) and (b) of Definition \ref{definition_pacing_equilibrium}. 

\begin{definition}[Smooth Approximate Pacing Equilibria] \label{definition_smooth_equilibrium}
Given an SPP game $G = (n, m, (v_{ij}), (B_i))$ and two parameters $\delta \revision{\in (0,1)}, \gamma \in [0,1)$, %a tuple $(\alpha, (x_{ij})_{i,j}) \in [0,1]^{n + nm}$, 
%a pair $(\alpha, x)$ with $\alpha=(\alpha_i)\in [0,1]^n$ and $x=(x_{ij})_{i,j}\in [0,1]^{nm}$
%which specifies the vector of pacing multipliers $\alpha$, 
we say that $(\alpha, x)$ with $\alpha=(\alpha_i)\in [0,1]^n$, $x=(x_{ij}) \in [0,1]^{nm}$
  and $\sum_{i\in [n]} x_{ij}\le 1$ for all $j\in [m]$ %with $\sum_{i\in [n]}x_{ij}\le 1$
  %for all $j\in [m]$ %which specifies the vector of pacing multipliers $\alpha$ and allocations $(x_{ij})_{i,j}$, 
  %is a \emph{pacing equilibrium} if \vspace{0.15cm}
is a \emph{smooth $(\delta, \gamma)$-\appe}\ of $G$ if \vspace{0.15cm}
%\textcolor{red}{(What should we name this smooth version?)} if 
\begin{flushleft}
\begin{itemize}
    \item[(a)] Only buyers close to the highest bid win the good and the allocation $x$ is completely specified by $\alpha$: For each $i\in [n]$ and $j\in [m]$,
    $x_{ij}$ (as a function of $\alpha$) is given by
    %For pacing multipliers $\alpha \in [0,1]^n$, the fraction of good $j$ allocated to agent $i$, $x_{ij}$, is a function of $\alpha$ and is given by
			\begin{align*}
	    		x_{ij} (\alpha) := \frac{[\alpha_i v_{ij} - (1-\delta)
	    		%\max_{k} \alpha_k v_{kj}
	    		h_j(\alpha)]^+}{\sum_{r\in [n]} [\alpha_r v_{rj} - (1-\delta)
	    		%\max_{k} \alpha_k v_{kj}
	    		h_j(\alpha)]^+}
			\end{align*}
			where $[y]^+$ is $y$ if $y \geq 0$ and $0$ otherwise. (We assume by default
			that $0/0=0$.)\vspace{0.15cm}
			%Assume that the above expression evaluates to zero if the denominator $\sum_i [\alpha_i v_{ij} - (1-\delta)\max_{i} \alpha_i v_{ij}]^+ = 0$.
    \item[(b)] Budgets are satisfied: $\sum_{j\in [m]} x_{ij}(\alpha) p_j(\alpha) \leq B_i$.\vspace{0.15cm}
	\item[(c)]  Not too much unnecessary pacing:   $\sum_{j\in [m]} x_{ij}(\alpha) p_j(\alpha) < (1 - \gamma)B_i$ implies $\alpha_i \geq 1 - \gamma$.\vspace{0.15cm}
\end{itemize}
\end{flushleft}
\end{definition}

Observe from the definition that, if $(\alpha,x)$ is a smooth  $(\delta, \gamma)$-\appe\ of
  an SPP game~$G$, %for the pacing game $G = (n, m, \{v_{ij}\}_{i,j}, \{B_i\}_i)$, 
  then it must be a $(\delta,\gamma)$-\appe\ of $G$ as well.
%  is a $(\delta,\gamma)$-approximate pacing equilibrium for the pacing game $G = (n, m, \{v_{ij}\}_{i,j}, \{B_i\}_i)$. 
%This is because the allocations defined in Definition \ref{definition_smooth_equilibrium} satisfy conditions (a)-(b) of Definition \ref{definition_pacing_equilibrium} and moreover, conditions (c)-(d) are identical in both definitions. 
Therefore, the PPAD membership of computing a smooth $(\delta, \gamma)$-\appe\ 
  in an SPP game implies directly the PPAD membership for $(\delta,\gamma)$-\appe.
A similar statement holds for establishing their existence.
%  the existence of a smooth $(\delta,\gamma)$-approximate pacing equilibrium in $G$. A similar statement holds for establishing membership in PPAD.

% The goal of this subsection is to establish the existence of a smooth $(\delta,\gamma)$-approximate pacing equilibrium for $G$ and perform the closely related task of placing the problem of computing $(\delta,\gamma)$-\appe\ in PPAD. 
The main tools we will use are Sperner's Lemma 
  and the search problem it defines.%\medskip

\noindent\textbf{High-dimensional Sperner's Lemma.}  
We review Sperner's lemma.
%Consider the $(n-1)$-simplex $S = \{\beta = (\beta_1,\dots,\beta_n) \mid \sum_i \beta_i = 1, \beta_i \geq 0\ \forall\ i\}$ and its standard $\omega$-fine discretization,
%  for some $\omega>0$ with $1/\omega$ being an integer:
%$S_\omega$ consists of all points in $S$ whose coordinates are multiples of $\omega$.
%Assume that $G$ has at most one bidder with an infinite budget. The following lemma shows that this assumption is without loss of generality.
%
%\begin{lemma}
%    Consider a  $\delta$-smoothed pacing game $G' = (n + k, m, \{v'_{ij}\}_{i,j}, \{B'_i\}_i)$ in which bidders $\{1, \dots, n-1\}$ have finite budgets and the rest of the bidders have infinite budget, for some $k \in \mathbb{Z}_+$. Define a $\delta$-smoothed pacing game $G = (n, m, \{v_{ij}\}_{i,j}, \{B_i\}_i)$ as
%    \begin{itemize}
%        \item $v_{ij} = v'_{ij}$ for all $i \in [n-1], j \in [m]$ and $B_i = B'_i$ for all $i \in [n-1]$.
%        \item $v_{nj} = \max_{i \geq n} v'_{ij}$ for all $j \in [m]$ and $B_n = \infty$.
%    \end{itemize}
%    If $G$ has a $\gamma$-approximate pacing equilibrium, then $G'$ has a $\gamma$-approximate pacing equilibrium.
%\end{lemma}
%
%\begin{proof}
%    Consider a $\gamma$-approximate pacing equilibrium $\alpha \in [0,1]^n$ of $G$. Set $\alpha'_i = \alpha_i$ for all $i \in [n-1]$ and $\alpha_i = 1$ for all $i \geq n$. It is straightforward to check that $\alpha'$ is a $\gamma$-approximate pacing equilibrium of $G'$.
%\end{proof}
Consider a $(n-1)$-dimen\-sional simplex $S= \{\sum_{i=1}^{n} \alpha_i v_i \hspace{0.06cm}|\hspace{0.06cm} \alpha_i \geq 0, \sum_{i=1}^{n} \alpha_i = 1\}$, where $v_1,\dots,v_n$ are $n$ vertices of $S$. A triangulation of $S$ is a partition of $S$ into smaller subsimplices such that any two subsimplices  either are disjoint or share a full face of a certain dimension. A \emph{Sperner coloring} $T$ of a triangulation of~$S$ is then an assignment of $n$ colors $\{1,\ldots,n\}$ to  vertices of the triangulation (union of the vertices of subsimplices that make up the triangulation) such that\vspace{0.15cm}
\begin{flushleft}\begin{itemize}
    \item Vertices of the original simplex $S$ each receive a different color: $T(v_i)=i$ for each $i\in [n]$.\vspace{0.1cm}
    \item Vertices on each face of $S$ are colored using only the colors of the vertices defining that face: For any vertex $u=\sum_i\beta_iv_i$ in the triangulation,
      we have $T(u)\ne j$ if $\beta_j=0$.\vspace{0.15cm}
\end{itemize}\end{flushleft}
A \emph{panchromatic} subsimplex of $T$ is one in the triangulation whose vertices have all the $n$ colors.%\medskip

\textbf{Sperner's Lemma:} Every Sperner coloring
  $T$ of any triangulation of $S$ has a panchromatic\\ subsimplex.%\medskip

Before proceeding with the formal proof of PPAD membership (with its added burden of rigorously attending to complexity-theoretic details), we provide an informal argument for the existence of smooth $(\delta, \gamma)$-\appe\ which forms the basis of its PPAD membership proof.  Let $G$ be an SPP game and $S$ be the standard simplex
$S = \{\beta = (\beta_1,\dots,\beta_n) \hspace{0.06cm}|\hspace{0.06cm} \beta_i\ge 0,\sum_i \beta_i = 1\}$ from now on.
We~will assign a color to each point $\beta\in S$ (informally) as follows: Construct a pacing multiplier $\alpha_i (t) = t \beta_i$ for each $i\in [n]$, where $t$ is a scalar. Increase $t$, starting at $0$, and instruct each buyer $i\in [n]$ to say ``Stop'' when either $\alpha_i(t) =1$ or $\sum_j x_{ij}(\alpha(t))p_j(\alpha(t)) = B_i$ happens. Color $\beta$ with $k$ if buyer $k$ is the first to say ``Stop'' (with tie breaking done arbitrarily, e.g., taking the smallest such $k$).

Let $t^*(\beta)$ be  the value of $t$ at which some buyer says ``Stop'' for the first time. Then the buyer that says ``Stop'' first is either spending her budget or is not paced, i.e. she satisfies both the budget constraint (b) and the `No unnecessary pacing' condition (c) (see Definition \ref{definition_exact_pacing_equilibrium}). 
%Moreover, at $t=t^*(\beta)$, all other buyers satisfy their budget constraint and have a pacing multiplier that is smaller than $1$. 
Now, by taking a triangulation of $S$, 
  it is easy to verify that the coloring described above
  induces a Sperner coloring and thus, Sperner's lemma implies the existence of a 
  panchromatic subsimplex $Q$.
It follows from our coloring that every buyer says ``Stop'' at one of the
  vertices of $Q$ and hence, every buyer satisfies (b) and (c) of Definition \ref{definition_exact_pacing_equilibrium} at one of its vertices.
%  is precisely the one for which all the buyers say "Stop" at one of its vertices and hence, all the agents satisfy the `No unnecessary pacing' condition at one of its vertices. 
%We establish the existence of this panchromatic subsimplex by invoking Sperner's Lemma. Moreover, 
By proving the Lipschitzness of $t^*(\beta)$ and the total
  expenditures of buyers, both as functions of $\beta$,
  we show that when the triangulation is \emph{fine} enough,  
  any point $\beta$ in a panchromatic subsimplex yields a 
  $(\delta, \gamma)$-\appe\ of $G$.

With the blueprint of the proof in place, we now proceed with the formal proof that places the problem of computing smooth $(\delta,\gamma)$-\appe\ in PPAD. Let $S$ be the standard simplex as above,
and we consider \emph{Kuhn's triangulation} of $S$~\citep{kuhn1960some,deng2012algorithmic}.
%:
%  $S = \{\beta = (\beta_1,\dots,\beta_n) \mid \sum_i \beta_i = 1, \beta_i \geq 0\ \forall\ i\}$.
Given any $\omega>0$ with 
  $1/\omega$ being an integer,
  Kuhn's triangulation uses
  $S_\omega$ as its vertices, where
  $S_\omega$ 
%We consider the standard $\omega$-\emph{fine discretization} $S_\omega$ of $S$,
  %for some $\omega>0$ with $1/\omega$ being an integer:
  consists of all points $\beta\in S$ whose coordinates $\beta_i$ are integer multiples of $\omega$.
Kuhn's triangulation  also has the property that any two vertices of a  subsimplex of the triangulation has $\ell_\infty$-distance at most $2\omega$. 
%$\|q-q'\|\le 2\omega$.

\revision{A proof of the following PPAD membership result 
  can be found in \cite{etessami2010complexity} (see the proof of 
  item 2 of Proposition 2.2; note that on page 2548 they reduce
  the problem they are interested in to the problem of finding 
  a panchromatic subsimplex in a Sperner coloring over Kuhn's 
  triangulation and then show the latter is in PPAD):}
  
%A \emph{unit subsimplex} of $S_w$ is a set of $n$ (distinct) points $Q=\{q_1,\ldots,q_n\}$ 
%  in $S_w$ such that $\|q_i-q_j\|_\infty=\omega$ for all $i\ne j$.
%It follows from Sperner's lemma that every Sperner   coloring $T:S_\omega\rightarrow [n]$
%  has a panchromatic unit subsimplex $P$ (by taking e.g. the standard Kuhn's triangulation with $S_\omega$ as its vertices \cite{} and noting that every of its subsimplex is a unit one%  using $S_\omega$ \cite{}
%  ). Indeed%
% the following computational problem is known to be in PPAD \cite{}:

\begin{theorem}\label{theo1}
Given a Boolean circuit\footnote{The circuit has $O(n\log (1/\omega))$ input 
  variables to encode a point of $S_\omega$ and has $\lceil \log n\rceil$ 
  output gates to encode the output of the Sperner coloring $T$.} that encodes a Sperner coloring
  $T:S_\omega\rightarrow [n]$ of Kuhn's triangulation for some $\omega$ and $n$, the problem
  of finding a  panchromatic subsimplex is in PPAD.
\end{theorem}
  
%The rest of this subsection concerns itself with the application of Sperner's Lemma to establish the existence of a $\gamma$-approximate pacing equilibrium of $G$. To this end, consider an $(n-1)$-simplex, $S = \{\beta = (\beta_1,\dots,\beta_n) \mid \sum_i \beta_i = 1, \beta_i \geq 0\ \forall\ i\}$. Let $T$ be a $\omega$-fine triangulation of $S$, i.e. if $S_0$ is a sub-simplex part of $T$, then $\max_{x,y \in S_0} \|x -y\| < \omega$. 

We prove the PPAD membership of the problem of finding a smooth $(\delta,\gamma)$-\appe\ by giving a polynomial-time reduction to the problem described in Theorem 
 \ref{theo1}. 
 %to show that the problem of finding a smooth
 % $(\delta,\gamma)$-\appe\  is also in PPAD.
Given an SPP game $G=(n,m,(v_{ij}),(B_i))$ and parameters $\delta$ and $\gamma$ (which we assume without loss of generality that $\delta,\gamma<1/4$), 
  we set the parameter $\omega$ to be %(We will use $B_{\min}$ and $B_{\max}$ to denote the smallest and the largest budgets respectively.)
$$
\omega= \frac{\min(B_{\min},1)}{\left(2^{|G|}/{\delta} \right)^{10,000}}\cdot \frac{\gamma}{2}
$$
where $B_{\min}:=\min_{i\in [n]} B_i$ and $|G|$ denotes the number of bits needed to represent $G$. %and
%$$Q\left(2^{|G|}, 1/
%\delta\right) = \left(2^{|G|}/\delta\right)^{10,000}$$
%is a large enough polynomial of $2^{|G|}/\delta$.
We define a coloring $T: S_\omega\rightarrow [n]$, following ideas described in the sketch of existence above, and prove that $T$ satisfies the following properties:

\begin{lemma}\label{membershiplem}
\begin{flushleft}\begin{enumerate}
    \item $T$ is a Sperner coloring;
    \item Every panchromatic  subsimplex of $T$ in the triangulation can be used to compute 
      a smooth $(\delta,\gamma)$-\appe\ of the SPP game $G$ in polynomial time.  
\item There is a polynomial-time algorithm that
outputs $T(\beta)$ on inputs $G$, $\omega$, $\delta$ and $\beta\in S_\omega$.
\end{enumerate}\end{flushleft}
\end{lemma}

%It follows from the definition $T$ that one can compute a Boolean 
%  circuit that encodes $T$ in polynomial time (in the number of bits needed
  %to encode $G$, $\delta$ and $\gamma$).
The PPAD membership of computing a smooth $(\delta,\gamma)$-\appe\  in an SPP game follows directly by combining  Theorem \ref{theo1} and Lemma \ref{membershiplem}.

We now give the definition of the coloring $T:S_\omega\rightarrow [n]$.
Let $\beta=(\beta_1,\ldots,\beta_n)$ be a vertex of $S_\omega$.
%We will define a labelling of $\beta$ as follows: 
Set $\alpha_i(t) = t \beta_i$, where $t$ is a positive scalar.
As discussed earlier, we set the color $T(\beta)$ of $\beta$ by 
 increasing $t$, starting at $0$, and instructing each buyer $i$ to say ``Stop'' when either $\alpha_i(t) =1$ or $$\sum_{j\in [m]} x_{ij}(\alpha(t))\cdot p_j(\alpha(t)) = B_i.$$ 
The color $T(\beta)$ of $\beta$ is set to be $k\in [n]$ if buyer $k$ is the first buyer to say ``Stop''
(with arbitrary tie breaking, e.g., by taking the smallest such $k$). 

More formally, recall that for $t>0$,
\begin{align*} 
    x_{ij} (\alpha(t)) = \frac{[t \beta_i v_{ij} - (1-\delta)\max_{k} t \beta_k v_{kj}]^+}{\sum_r [t \beta_r v_{rj} - (1-\delta)\max_{k} t\beta_k v_{kj}]^+} = \frac{[\beta_i v_{ij} - (1-\delta)\max_{k} \beta_k v_{kj}]^+}{\sum_r [\beta_r v_{rj} - (1-\delta)\max_{k} \beta_k v_{kj}]^+}=x_{ij}(\beta),
\end{align*}
which does not depend on $t$.
%Therefore, $x_{ij}(\alpha(t))$ is a constant and does not depend on $t$ for $t>0$. For convenience we denote the RHS of (\ref{eq:x1}) by $x_{ij}(\beta)$. 
%\footnote{If $\sum_r [\beta_r v_{rj} - (1-\delta)\max_{k} \beta_k v_{kj}]^+ = 0$, then take $x_{ij}(\beta)$ to be zero}. 
Also, for $t \geq 0$, $p_j(\alpha(t)) = t p_j(\beta)$, where we write $p_j(\beta)$ to denote the second largest element among $\beta_1 v_{1j}, \dots, \beta_n v_{nj}$. For each buyer $i\in [n]$, define %\footnote{From our standing assumption of $B_i > 0$ for all bidders $i$, we get that $t_i(\beta) \geq \min\left\{1, \frac{B_i}{\sum_j v_{ij}}\right\}>0$ for all $i$.} 
$$t_i(\beta) = \min\left\{\frac{1}{\beta_i}, \frac{B_i}{\sum_j x_{ij}(\beta) p_j(\beta)}\right\},$$ where the first term is $+\infty$ if $\beta_i=0$ and the second term is $+\infty$ if $\sum_j x_{ij}(\beta) p_j(\beta) = 0$.
Note that $t_i(\beta)$ is exactly the value of $t$ at which buyer $i$ would say ``Stop'' in the informal coloring procedure described earlier. 
Given our assumption of $B_i>0$, we have $t_i(\beta)>0$ for all $i\in [n]$.
Additionally, define $t^*(\beta) = \min_{i\in [n]} t_i(\beta)$.
Given that $\beta_i$'s sum to $1$, we have that $t^*(\beta)\le n$   because $\beta_i\ge 1/n$ for some $i\in [n]$. 
We record the discussion as the following lemma:

\begin{lemma}\label{triviallem}
For every $\beta\in S_\omega$ we have $0<t^*(\beta)\le n$.
%and $t^*(\beta)\ne \infty$ 
\end{lemma}

Finally, the color $T(\beta)$ of $\beta\in S_\omega$ is set to be the 
  smallest $i\in [n]$ such that $t_i(\beta)=t^*(\beta)$.
We are now ready to prove Lemma \ref{membershiplem}.

\begin{proof}[Proof of Lemma \ref{membershiplem}]
Part (3) of Lemma \ref{membershiplem} follows from the description of $T$. %above that a Boolean circuit
%  that encodes $T$ can be %computed in time polynomial in the number of bits needed
%  to encode $G$, $\delta$ and $\gamma$.
To prove part (1) ($T$ is a Sperner coloring), consider a vertex $\beta\in S_\omega$ on the facet of $S$ opposite to the vertex $e_i$, \revision{i.e., $\beta_i = 0$}. Hence, $t_i(\beta) = \infty$, which by Lemma \ref{triviallem} implies that $T(\beta)\ne i$ given that $t^*(\beta)\le n$.  
%we show cannot be $+\infty$. To this end, using the definition of $t^*$, we get
%\begin{align*}
%    \alpha_i(t^*(\beta)) \leq 1 \textrm{    and     }\sum_j x_{ij}(\alpha(t^*(\beta)))p_j(\alpha(t^*(\beta))) \leq B_i
%\end{align*}
%for all bidders $i$. Observe that the above procedure can be used to label any point $\beta \in S$ (not necessarily a vertex of $T$) and hence, $t^*$ can be defined for every point in $S$.

%Therefore, we set the label of $\beta$ to be any index in $\textrm{arg}\min_i t_i(\beta)$. 

%All that is left is to show that any panchromatic unit subsimplex 
%  of $T$ can be used to compute a smooth $(\delta,\gamma)$-\appe\ in polynomial time.

%\begin{lemma}\label{lastlemma}
%
To prove part (2), we show that   if $q$ is a vertex of any panchromatic subsimplex of $T$, then
  $(\alpha,x)$ must be a smooth $(\delta,\gamma)$-\appe\ of $G$ where $\alpha= t^*(q)\cdot q$
  and $x=(x_{ij})$ has $x_{ij}=x_{ij}(q)$.
  %$P=\{p_1,\ldots,p_n\}$ of $T$ with $T(p_i)=i$,
%  any vertex $\beta\in P$ satisfies that $t^*(\beta)\cdot \beta$ is a smooth $(\delta,\gamma)$-\appe\ of $G$.
%\end{lemma}
%\begin{proof}
%Let $p=p_1$ without loss of generality.

First %it follows 
%Let $t^*=t^*(\beta)$, which by Lemma \ref{triviallem} is positive and is not $\infty$.
%Let $\alpha$ and $x$ be such that $\alpha_i=t^*\cdot \beta_i$ and 
%  $x_{ij}=x_{ij}(\beta)$.
it follows from the definition of $t^*(\beta)$ and $x_{ij}(\beta)$ that
  $\alpha_i\in [0,1]$ and $x_{ij}\in [0,1]$.
Conditions (a) and (b) of Definition \ref{definition_smooth_equilibrium} also trivially \revision{hold for all vertices of the triangulation}. 
It suffices to prove (c) for all $i\in [n]$, which means
  the complementarity condition that
  either  $\alpha_i\ge 1-\gamma$ or the expenditure of buyer $i$ is at least $(1-\gamma)B_i$. \revision{Fix an arbitrary $i \in [n]$.}

For this purpose we note that
  given the subsimplex is panchromatic, it has a vertex $q'$ such that $T(q')=i$, which 
  implies that if we used $q'$ to define $\alpha'$ and $x'$
  (i.e. $\smash{\alpha'=t^*(q')\cdot q'}$ and $\smash{x'_{ij}=x_{ij}(q')}$), then they would satisfy
  the above complementarity condition for buyer $\smash{i}$ with $\smash{\gamma=0}$.
The following claim shows that both the multiplier $ t^*(\beta)\cdot \beta_i$  and 
the total expenditure of buyer $i$: $$
\sum_{j\in [m]} x_{ij}(\beta)\cdot p_j\big(t^*(\beta)\cdot \beta\big)
=
t^*(\beta)\sum_{j\in [m]} x_{ij}(\beta)\cdot p_j(\beta)
$$
are smooth as functions of $\beta$. 
Intuitively this allows us to use the complementarity condition for buyer $i$
  at $q'$ to show that the same condition holds at $q$ \emph{approximately} given that $\|q-q'\|_\infty\le 2\omega$ (as a property of subsimplices in Kuhn's triangulation).

\begin{claim}\label{lipschitz}
	Let $L = (2^{|G|}/\delta)^{10,000}$. Then for any panchromatic subsimplex $S_0$ of $T$ and buyer $i \in [n]$, the following Lipschitz conditions hold for all $\beta, \beta' \in S_0$:
\begin{align*}
	%\big| t^*(\beta)\cdot \beta -  t^*(\beta')\cdot 
	%\beta'| = 
	\Big|\hspace{0.04cm}t^*(\beta)\cdot \beta_i - t^*(\beta')\cdot \beta_i'\hspace{0.04cm}\Big| &\leq L\cdot  \|\beta - \beta'\|_\infty\quad\text{and}\\
	\left|\hspace{0.04cm} t^*(\beta)\sum_{j\in [m]} x_{ij}(\beta)\cdot p_j(\beta) - t^*(\beta')\sum_{j\in [m]} x_{ij}(\beta') \cdot p_j(\beta') \hspace{0.04cm}\right| &\leq L\cdot \|\beta - \beta'\|_\infty\\[-2.8ex]
\end{align*}
\end{claim}

\revision{
We use Claim \ref{lipschitz} to finish the proof of the lemma and consign the claim's proof to Appendix \ref{appendix_lipschitz}.  
Given $T(q')=i$, one of the following two cases holds:
\begin{itemize}
    \item $t^*(q')\cdot q_i' =1$, which by Claim~\ref{lipschitz} and our chocie of $\omega$ implies
    \begin{align*}
        \alpha_i=  t^*(q)\cdot q_i \ge 1- 2L\omega \ge 1-\gamma
    \end{align*}
    \item $t^*(q')\sum_j x_{ij}(q') p_j(q')=B_i$, which in combination with Claim~\ref{lipschitz} and our choice of $\omega$ implies that the expenditure of buyer $i$ exceeds $(1 - \gamma)B_i$:
    \begin{align*}
        t^*(q)\sum_{j\in [m]} x_{ij}(q) \cdot p_j(q)\ge B_i-2L\omega
\ge B_i-B_{\min}\gamma \ge (1-\gamma)B_i.
    \end{align*}
\end{itemize}
% we have either $t^*(q')\cdot q_i' =1$, which by Claim \ref{lipschitz}
%   implies that 
% $$
% \alpha_i=  t^*(q)\cdot q_i \ge 1- 2L\omega \ge 1-\gamma 
% $$
% by our choice of $\omega$,
% or $t^*(q')\sum_j x_{ij}(q') p_j(q')=B_i$ which implies that 
% $$
% t^*(q)\sum_{j\in [m]} x_{ij}(q) \cdot p_j(q)\ge B_i-2L\omega
% \ge B_i-B_{\min}\gamma \ge (1-\gamma)B_i. 
% $$
Since $i \in [n]$ was arbitrary, this finishes the proof that $(\alpha,x)$ is a smooth $(\delta,\gamma)$-approximate
  \appe.}
%Finally, consider a fully colored subsimplex $S_0$ of $T$ and a point $\beta^* \in S_0$. As $S_0$ is fully-colored, for each $i \in [n]$, there exists a vertex $\beta^{(i)}$ of $S_0$ such that $\beta^{(i)}$ is labelled $i$, i.e., there exists a vertex $\beta^{(i)}$ of $S_0$ for which either $\alpha_i(t^*(\beta^{(i)})) = 1$ or $\sum_j x_{ij}(\alpha(t^*(\beta^{(i)})))p_j(\alpha(t^*(\beta^{(i)})) = B_i$. Therefore, if we pick $\omega \leq \min\left\{\gamma B_{\min}/C_1, \gamma B_{\min}/C_2\right\}$, then Lemma \ref{lipschitz} implies that $\alpha(t^*(\beta^*))$ is a $\gamma$-approximate equilibrium of $G$. %\textcolor{red}{Citation needed:} This implies that $(\delta, \gamma)$-PE is in PPAD.
\end{proof}

\subsection{PPAD Membership of Computing $ \gamma $-\appe}\label{section_rounding_alg}
%Fix $\gamma \in (0,1)$. 
Consider an SPP game $G = (n, m, \{v_{ij}\}_{i,j}, \{B_i\}_i)$. As before, we will use $|G|$ to denote the number of bits required to represent $G$. The main result of this subsection shows that (informally) 
  when $\delta$ is small enough, any $(\delta,\gamma/2)$-\appe\ of $G$
  can be efficiently rounded to a $ \gamma $-\appe.
It follows from the PPAD membership of $(\delta,\gamma)$-\appe\ established in the previous
  subsection that the problem of computing a  $\gamma $-\appe\ 
  is in PPAD as well.

%establishes the existence of a $\gamma$-approximate pacing equilibrium for $G$. We will do so by showing that, for $\delta, \gamma'$ small enough, we can round a $(\delta, \gamma')$-PE to a $\gamma$-PE. Combining this with the existence result of the previous subsection gives the existence of a $\gamma$-PE for $G$. Moreover, our rounding algorithm runs in polynomial time, thereby allowing us to show the membership of $\gamma$-PE in PPAD as a direct consequence of the PPAD membership of $(\delta, \gamma')$-PE established earlier.\\

Before presenting the rounding algorithm, we motivate the main idea behind it. Observe that the major difference between $(\delta, \gamma)$-\appe\  and $\gamma$-\appe\ is the ability of buyers that don't have the highest bid to win the good in the former. In order to round a $(\delta, \gamma')$-\appe\ $(\alpha^*,x^*)$ to obtain a $ \gamma $-\appe\ $(\alpha',x')$ of $G$ (where $\gamma'=\gamma/2$ in the rest of this subsection), 
  we set $x'=x^*$ and need to round $\alpha^*$ 
  to $\alpha'$
  to ensure that all the winners are tied for the highest bid and at the same time, the multiplier and total expenditure of each 
  buyer changes only slightly. 
  
We now present an informal argument that demonstrates how this is achieved in our rounding algorithm when there are only two buyers ($n=2$). Define the set of all valuation ratios
$$\tilde{\V} = \left\{\frac{v_{ar}}{v_{br}} : a, b \in [n], r \in [m] \text{ such that } v_{ar}, v_{br} >0\right\}.$$
Set $\delta$ to be small enough: for all $y, z \in \V$ with $y z>1$, we have $(1 - \delta)^2 y z > 1$. Consider~a~$(\delta,\gamma')$-\appe\ $(\alpha^*,x^*)$.
Assume without loss of generality that there is a good $j$ such that 
  $\alpha_1^* v_{1j}$   $ = c_j \alpha_2^* v_{2j}$ and 
  $ 1 - \delta \leq c_j \leq 1/(1- \delta)$. \revision{If no such
  good $j$ exists then every good is fully allocated 
  to the buyer with the highest bid \revision{because only bidders with bids greater than $(1 - \delta)$ times the highest bid can win the item in a $(\delta, \gamma')$-\appe}, and  thus,
  $(\alpha^*,x^*)$
  is already a $\gamma'$-\appe.}
%in which buyer $1$ and buyer $2$ share some good $j$, i.e., $x_{1j},x_{2j}>0$. Even though buyers $1$ and $2$ might not have identical bids, by definition of $(\delta, \gamma)$-\appe, there exists a constant $(1 - \delta)< c_j < 1/(1- \delta)$ such that $\alpha_1 v_{1j} = c_j \alpha_2 v_{2j}$. 
We show that after 
  scaling the pacing multiplier of buyer 2 from $\alpha_2^*$ to $c_j \alpha_2^*$ (and letting $\alpha'=(\alpha_1^*,c_j\alpha_2^*)$ be the new multipliers),
  $(\alpha',x^*)$ satisfies the property that $x_{i\ell}^*>0$ for
  any $i$ and $\ell$ implies
  buyer $i$ has the highest bid for good $\ell$. 
This is trivially true for good $\ell=j$ given that the two 
  buyers are now tied on good $j$. \revision{The remaining goods can be divided into two categories and we argue about each one separately:
\begin{itemize}
    \item Consider good $\ell$ such that $\alpha_1^* v_{1\ell} = c_\ell \alpha_2^* v_{2 \ell}$
  and $c_\ell$ satisfies either $c_\ell < 1-\delta$ or 
  $c_\ell > 1/(1-\delta)$. Given that 
  we only changed the multiplier of buyer 2 by a factor of $  1-\delta \leq c_j \leq 1/(1-\delta)$, the highest bidder does not change. Moreover, the highest bidder won the entire good in the $(\delta, \gamma')$-approximate PE because $  1-\delta \leq c_j \leq 1/(1-\delta)$ and continues to do so in the $(\delta, \gamma')$-approximate PE because the allocation does not change.

  \item Consider a good $\ell$ such that $\alpha_1^* v_{1\ell} = c_\ell \alpha_2^* v_{2 \ell}$ and $c_\ell$ satisfies $(1 - \delta) \leq c_\ell \leq 1/(1 -\delta)$. Then, we can write $\alpha_1^*/ \alpha^*_2 = c_j(v_{2j}/ v_{1j}) = c_\ell (v_{2\ell}/ v_{1\ell})$, which implies $(c_j / c_\ell) (v_{2j}/ v_{1j}) (v_{1\ell}/ v_{2\ell}) = 1$. Observe that $c_j/c_\ell \in [(1 - \delta)^2, 1/(1 - \delta)^2]$. Thus, by our choice of $\delta$, we get $(v_{2j}/ v_{1j}) (v_{1\ell}/ v_{2\ell}) = 1$, which implies  $c_j = c_\ell$. Hence, both buyers are tied in good $\ell$.
\end{itemize}}
To finish the proof that $(\alpha',x^*)$ is a $\gamma$-\appe,
  it suffices to show that the budget constraint 
  and the not too much unnecessary pacing condition still
  hold approximately after the small scaling of $\alpha_2^*$.
In the rest of this subsection, we extend  the aforementioned line of reasoning to design a rounding algorithm for the general setting, and prove its correctness.

Building on $\tilde{\V}$ defined above, we can define the set of valuation ratio products
\begin{align*}
	\V = \Big\{ y_1 y_2 \dots y_{k}: k\in [2n]\ \text{and}\ y_i \in \tilde{\V}  \ \text{for each}\ i \in [k]\Big\},
\end{align*}
i.e., $\V$ consists of all products of no more than $2n$ numbers from $\tilde{\V}$.
Given $G$ and $\gamma\in [0,1)$ (with $\gamma'=\gamma/2$),
  we choose $\delta\in [0,1)$ to be small enough to satisfy the following two conditions:\vspace{0.1cm}
%Let $\gamma'$ be such that $(1 - \gamma')^2 > (1 - \gamma)$. Moreover, pick $\delta$ small enough such that:
\begin{enumerate}
	\item[] $(1 - \delta)^{2^{n }} > (1 - \gamma')$ \ and \ $(1 - \delta)^{2^{n }} z > 1$ for all  $z \in \V$ such that $z > 1$. \vspace{0.1cm} %, then 
	%\item If $z \in \V$ such that $z > 1$, then $(1 - \delta)^{2^{n + 2}} > z$.
\end{enumerate}
It suffices to set $\delta$ to be $1/2^N$ where $N$ is polynomial
  in $|G|$ and $\log (1/\gamma)$.

%is easy to verify that $\delta$ can be computed in polynomial time
%  in $|G|$ and the number of bits needed to encode $\gamma$. 

Let $(\alpha^*,x^*)$ be a $(\delta,\gamma')$-\appe\ of $G=(n, m, (v_{ij}), (B_i))$, where $\gamma'=\gamma/2$ and $\delta$ satisfies the two conditions above.
We will use $W_j$ to denote the winners of the good $j$ under $x^*$: $W_j$ consists of buyers $i$ with $x_{ij}^* > 0$. Moreover, recall that $h_j(\alpha )$ denotes the highest bid on good $j$ when the pacing multipliers are given by $\alpha$. 
Our rounding algorithm is presented in Algorithm~\ref{alg:rounding}.
The polynomial reduction then follows from the following performance guarantee
  of the rounding algorithm, which we prove in the rest of the subsection:
  
\begin{lemma}[Correctness]\label{alg:correct}
The rounding algorithm takes $(\alpha^*,x^*)$, $\delta$ and $G$ as input and runs in polynomial time.
Let $\alpha'$ be the tuple of multipliers returned by the rounding algorithm.
Then $(\alpha',x^*)$ is a $\gamma$-\appe\ of $G$.
\end{lemma}  
  
The rounding algorithm maintains an undirected graph $\G$ over vertices
  $[n]$ as buyers. \revision{$\G$ starting out with an empty edge set and edges are added according to Algorithm~\ref{alg:rounding} to keep track of the rounding-updates performed on $\alpha$.}
We use 
$C_\G(i)$ to denote the connected component of $i$ in the graph $\G$.
The algorithm also maintains an edge labeling $I(\cdot)$ that maps each
  edge of the graph $\G$ to a good $j\in [m]$ (which intuitively is the good that
  caused the creation of this edge).
We remark that the labeling $I(\cdot)$  
  is only relevant for the analysis of the algorithm below.
  %and can be safely omitted from the pseudo-code.
%that the algorithm constructs:
Now, we proceed to prove Lemma \ref{alg:correct}. 

% \begin{minipage}{0.9 \textwidth}
\begin{figure}[t!]
\begin{algorithm}[H] 
   \caption{Rounding Algorithm}
   \label{alg:rounding}
   %\textbf{Round}$(\alpha^*,x^*)$
    \begin{algorithmic}\vspace{0.08cm}
            \item[\textbf{Initialize:}] Graph $\G = (V,E)$ with $V = [n]$ and $E = \emptyset$; $\alpha = \alpha^*$
            \item[\textbf{While}] there exists a good $j \in [m]$
            and a buyer $i\in W_j$ such that $\alpha_i v_{ij}
            <h_j(\alpha)$, \revision{i.e., $i$ does not have the highest bid on $j$ but wins a positive fraction of it}:\vspace{0.15cm}% \text{ for some buyer}\ i \in W_j$
            \begin{enumerate}
                \item Pick $k,i \in [n]$ and $j \in [m]$ such that 
                $i \in W_j$ and $\alpha_i v_{ij}< \alpha_k v_{kj}=h_j(\alpha) $\vspace{0.1cm}  
                \item Set $\alpha_a \leftarrow ({h_j(\alpha)}/{\alpha_i v_{ij}})\cdot \alpha_a$ for every buyer $a \in C_\G(i)$\vspace{0.1cm}  
                \item Set $E \leftarrow E \cup \{\{i,k\}\}$ and $I(\{i,k\}) = j$\vspace{0.15cm}
            \end{enumerate}
            \item[\textbf{Return:}] $\alpha' \coloneqq (1 - \delta)^{2^{n} }\alpha$
    \end{algorithmic}
 \end{algorithm}%~\captionof{figure}{The rounding algorithm for $\gamma$-\appe.}\label{fig:alg}
% \medskip\medskip\medskip
\vspace{-4mm}
\end{figure}

% the correctness of the rounding algorithm. % We show that if the algorithm outputs $\alpha'$, then $(\alpha', (x_{ij}(\alpha^*))_{i,j})$ is a $(\delta, \gamma)$-approximate pacing equilibrium for $G = (n, m, \{v_{ij}\}_{i,j}, \{B_i\}_i)$.\\

\revision{
\begin{lemma}\label{lemma:ratio-of-mult}
    Suppose in the $t_0$ iteration of the while loop,  $\{i,k\}$ is the edge that was just added to $\G$ 
   with $I(\{i,k\})=j$, then at the end of this iteration we have
   %in iteration $t_0$ of the while loop, then, at the end of iteration $t_0$, we have 
   $C_\G(i) = C_\G(k)$ and
\begin{align*}
    \frac{\alpha_{i}}{\alpha_{k}} = \frac{v_{kj}}{v_{ij}} \,. \tag{\#}
\end{align*}
Moreover, (\#) holds for all iterations $t \geq t_0$. 
\end{lemma}
\begin{proof}
    We prove the lemma using induction on the iterations on the while loop. For the base case $t = t_0$, note that (\#) holds at the end of the iteration due to Step 2 of Algorithm~\ref{alg:rounding}. Moreover, since edge $\{i,k\}$ is added to  $\G$ in Step 3, we also have $C_\G(i) = C_\G(k)$ at the end of iteration $t_0$. Moreover, since no edges are removed during the run of Algorithm~\ref{alg:rounding}, $\{i,k\} \in E$ for iterations after $t_0$, and hence $C_\G(i) = C_\G(k)$ at the end of all iterations $t \geq t_0$. Suppose (\#) holds at the end of iteration $t-1$ for some $t-1 \geq t_0$. Then, either both $\alpha_i$ and $\alpha_k$ will both be updated identically or neither of them will be updated because $C_\G(i) = C_\G(k)$, thereby maintaining (\#). This completes the induction and establishes the lemma.
\end{proof}}
% First, we observe that, if $\{i,k\}$ is the edge that was just added to $\G$ 
%    with $I(\{i,k\})=j$, then at the end of this iteration we have
%    %in iteration $t_0$ of the while loop, then, at the end of iteration $t_0$, we have 
%    $C_\G(i) = C_\G(k)$ and
% \begin{align*}
%     \frac{\alpha_{i}}{\alpha_{k}} = \frac{v_{kj}}{v_{ij}} \tag{\#}
% \end{align*}
% Given that we raise the multipliers of a whole component in each while loop, it can be shown by induction that (\#) continues to hold for all iterations after the edge $\{i,k\}$ is added to $\G$. 

Next we prove that at the end of each iteration, bids for the same good from buyers in the same component of $\G$ are either tied or not very close.

\begin{lemma}\label{rounding_small_delta}
	After each iteration of the while loop, and for each good $j\in [m]$, all buyers from the same connected component of $\G$ are either tied for $j$, or their bids for $j$ are multiplicatively separated by a factor \revision{larger} than $(1 - \delta)^{2^n}$.
\end{lemma}

\begin{proof}
%	Suppose, for contradiction, that this statement doesn't hold after some iteration:
Let $\G$ be the current graph and $a,b \in [n]$ be two buyers in the same connected component of $\G$.
Assuming $\alpha_a v_{aj} > \alpha_b v_{bj}$ for some $j\in [m]$, we show below that  $ (1 - \delta)^{2^n}\alpha_a v_{aj}> \alpha_b v_{bj}$ from which the lemma follows.
Given that $a$ and $b$ are connected in $\G$, we write  $\{a,i_1\}, \{i_1, i_2\}, \dots,$ $\{i_L, b\}$ to denote a path from $a$ to $b$ in $\G$ with $L<n$. 
%let $(1 - \delta)^{2^n} \leq d \leq 1$ be such that $d \alpha_a v_{aj} = \alpha_b v_{bj}$. 
Then, using Lemma~\ref{lemma:ratio-of-mult}, we can write
\begin{align*}
    1> \frac{\alpha_b v_{bj}}{\alpha_a v_{aj}} =\frac{v_{bj}}{v_{aj}}\cdot \frac{\alpha_{i_1}}{\alpha_a} \frac{\alpha_{i_2}}{\alpha_{i_1}}\frac{\alpha_{i_3}}{\alpha_{i_2}} \dots \frac{\alpha_{b}}{\alpha_{i_L}} = \frac{v_{bj}}{v_{aj}}\cdot  \frac{v_{aI(\{a,i_1\})}}{v_{i_1 I(\{a,i_1\})}}\frac{v_{i_1 I(\{i_1,i_2\})}}{v_{i_2 I(\{i_1,i_2\})}} \dots \frac{v_{i_L I(\{i_L,b\})}}{v_{b I(\{i_L, b\})}}%\\
    %\implies& 1 = \frac{1}{d} \cdot \frac{v_{aI(\{a,i_1\})}}{v_{i_1 I(\{a,i_1\})}}\frac{v_{i_1 I(\{i_1,i_2\})}}{v_{i_2 I(\{i_1,i_2\})}} \dots \frac{v_{i_L I(\{i_L,b\})}}{v_{b I(\{i_L, b\})}} \frac{v_{bj}}{v_{aj}}\\
\end{align*}
\revision{Hence, $\alpha_a v_{aj}/\alpha_a v_{bj} \in \V$ and $\alpha_a v_{aj}/\alpha_b v_{bj} > 1$. Therefore,  our choice of $\delta$ implies that
\begin{align*}
    (1 - \delta)^{2^n} \cdot \frac{\alpha_a v_{aj}}{\alpha_b v_{bj}} > 1 \,,
\end{align*}
as required.}
%This contradicts the assumption that $\alpha_a v_{aj} > \alpha_b v_{bj}$. Hence, the lemma holds.
\end{proof}

Initially (in $\alpha^*$)
  we have every $i\in W_j$ has
  $\alpha_i^*v_{ij}\ge (1-\delta)h_j(\alpha^*)$
  (given that $(\alpha^*,x^*)$ is a $(\delta,\gamma')$-\appe).
The next lemma shows that, at the end of each iteration,
  $\alpha_i v_{ij}$ of every
  $i\in W_j$ (note that $W_j$ is always defined using the original allocation $x^*$)
  remains not far from
  $h_j(\alpha )$.
\begin{lemma} \label{rounding_progress}
    After $t$ iterations of the while loop, every $j\in [m]$ and $i \in W_j$ satisfy $$\alpha_i v_{ij} \geq (1 - \delta)^{2^t}\cdot h_j(\alpha).$$
\end{lemma}

\begin{proof}
    The proof follows from induction. The base case of $t=0$ follows from 
    definition. 
    
    Suppose the statement holds after $(t - 1)$ iterations, and let's focus on some $j\in [m]$ and $i\in W_j$ during the $t$-th iteration.
    By our inductive hypothesis, we have $${\alpha_i v_{ij}} \ge  (1 - \delta)^{2^{t-1}}\cdot  {h_j(\alpha)}$$ before the start of the $t$-th iteration. 
    On the other hand, note that all changes to $\alpha$ occur in step 2 of the while loop, and moreover, all such changes result in an increase of some entries of $\alpha$. 
It also follows from the inductive hypothesis and the choices of $k,i,j$ in step 1 of the while loop that 
  entries of $\alpha$ can only
  go up by a multiplicative factor of 
%    Therefore, for all items $j \in [n]$, the highest bid increases by a multiplicative factor of 
at most 
$ \smash{{1}/{(1 - \delta)^{2^{t-1}}}}$. Therefore, after the $t$-th iteration, we have $${\alpha_i v_{ij}} \ge  (1 - \delta)^{2^{t-1}} \cdot (1 - \delta)^{2^{t-1}} \cdot h_j(\alpha)  =  (1 - \delta)^{2^{t}}\cdot 
h_j(\alpha).$$ This completes the induction step.
\end{proof}

 Lemmas \ref{rounding_small_delta} and  \ref{rounding_progress}  imply that, in each of the first 
$n$ iterations of the while loop, buyers $i$ and~$k$ picked in step 1 must belong to different connected components of $\G$. As a result, there are at most $n-1$ iterations of the while loop given that we merge two connected components in each loop.
On the one hand, this implies that the rounding algorithm
  terminates in polynomial time.
On the other hand, at the termination of the while loop, 
  for every good $j \in [m]$, we have $\alpha_i v_{ij} = h_j(\alpha)$ for all $i \in W_j$, i.e., every winner of $j$ under $x^*$ has the highest bid for $j$.

The next lemma shows that the $\alpha'$
  returned by the rounding algorithm is close 
  to $\alpha^*$.
%Before proving Lemma \ref{

\begin{lemma} \label{change_in_alpha}
Let $\alpha'$ be the tuple of multipliers returned by the rounding algorithm.
Then $$(1 - \delta)^{2^{n} } \alpha^* \leq \alpha' \leq \alpha^*.$$	
\end{lemma}

\begin{proof}
	By Lemma \ref{rounding_progress}, in iteration $t$ of the while loop, each entry of $\alpha$ either stays the same or increases multiplicatively by a factor of at most $\smash{1/(1 - \delta)^{2^{t-1}}}$. As there are at most $n-1$ iterations of the while loop, we have for every $i\in [n]$:
	\begin{align*}
		(1-\delta)^{2^n }\cdot \alpha_i^*\le \alpha_i' \coloneqq (1-\delta)^{2^n }\cdot \alpha_i\le (1 - \delta)^{2^{n} } \prod_{t=1}^{n-1} \frac{1}{(1 - \delta)^{2^{t-1}}} \cdot \alpha_i^* %= (1 - \delta)^{2^{n}} \frac{1}{(1 - \delta)^{\sum_{t=1}^{n-1} 2^{t-1}}} \alpha^* 
		\leq \alpha_i^*.
	\end{align*}
	This finishes the proof of the lemma.
%	Moreover, note that no component of $\alpha$ decreases in any step of the while loop. Hence, we also have $(1 - \delta)^{2^{n}} \alpha^* \leq \alpha'$.
\end{proof}

%As $\alpha^*$ is a $(\delta, \gamma')$-\appe\ of a $G$, the following conditions are satisfied:
% \begin{itemize}	     
%	\item $\alpha^*_i = 1$ whenever $B_i = \infty$.
%	
%	\item Not too much unnecessary pacing: For all $i$, $\sum_j x_{ij}(\alpha^*) p_j(\alpha^*) \leq B_i$. In addition,\\ if $\sum_j x_{ij}(\alpha^*) p_j(\alpha^*) < (1 - \gamma')B_i$, then $\alpha^*_i \geq 1 - \gamma'$.
%\end{itemize}

We are now ready to prove Lemma \ref{alg:correct}.

\begin{proof}[Proof of Lemma \ref{alg:correct}]
We have already shown that the algorithm runs in polynomial time. 
Assuming that $(\alpha^*,x^*)$ is a $(\delta,\gamma')$-\appe\ of $\G$, we show that $(\alpha',x^*)$ is a $\gamma$-\appe\ of $\G$ \revision{by establishing conditions (a)-(d) of Definition~\ref{definition_pacing_equilibrium}}.
Using Lemma \ref{change_in_alpha}, we have $\alpha'\in [0,1]^n$.
Condition (a) has already been established earlier using 
  Lemmas \ref{rounding_small_delta} and  \ref{rounding_progress}.
Condition (b) holds because we kept the same allocation $x^*$ and 
  given how we obtain $\alpha'$ from $\alpha^*$, the set of goods $j$ with $h_j(\alpha^*)>0$ is the same as that in $\alpha'$.
Condition (c) follows easily from Lemma \ref{change_in_alpha}.
So it suffices to verify that (d) holds with $\gamma$.
%it suffices to prove (a) (with $\delta=0$) and (d) (with $\gamma=2\gamma'$) in Definition \ref{}.

%First,  let us check that 
%  every buyer $i\in W_j$ for some $j\in [m]$ has the highest bid: $\alpha_i'v_{ij}=h_j(\alpha')$,
%  which is equivalent to 
%  $\alpha_i v_{ij}=h_j(\alpha)$.

%check that $(\alpha', (x_{ij}(\alpha^*))_{i,j})$ are a valid pacing parameter-allocation pair for $G$. As we observed before, at the termination of the while loop, for every item $j \in [m]$, we have $\alpha_i v_{ij} = h_j(\alpha)$ for all $i \in W_j$, i.e., the winners of $j$ under $\alpha^*$ have the highest bid for $j$. Therefore, by definition of $\alpha'$, we get $\alpha'_i v_{ij} = h_j(\alpha')$ for all $j \in [m]$, $i \in W_j$. Moreover, $\sum_j x_{ij}(\alpha^*) p_j(\alpha^*) \leq B_i$ implies that $\sum_j x_{ij}(\alpha^*) p_j(\alpha') \leq B_i$, because $\alpha' \leq \alpha^*$. Hence, we have satisfied the validity of $(\alpha', (x_{ij}(\alpha^*))_{i,j})$. Also, unless $p_j(\alpha^*) = 0$, we have $\sum_i x_{ij}(\alpha^*) = 1$. 

To see this we have for each buyer $i\in [n]$ that either 
  $\alpha^*_i\ge 1-\gamma'$ or $\sum_{j} x^*_{ij}p_j(\alpha^*)
  \ge (1-\gamma')B_i$.
For the former case, we have from  Lemma \ref{change_in_alpha}
  that $$\alpha_i'\ge (1-\delta)^{2^n}\cdot (1-\gamma')
  > (1-\gamma')^2\ge 1-\gamma$$
using $(1-\delta)^{2^n}>  1-\gamma' $ from the choice of $\delta$
  and that $\gamma=2\gamma'$.
For the latter case, it follows from  
  Lemma \ref{change_in_alpha} and our choice of $\delta$  that
	\begin{align*}
	p_j(\alpha')\ge (1-\delta)^{2^n}\cdot p_j(\alpha^*)
	  >
		(1 - \gamma')\cdot p_j(\alpha^*)
		%< (1 - \delta)^{2^{n}} p_j(\alpha^*) \leq 
	\end{align*}
 	for all $j \in [m]$. \revision{Here we have used the fact that $p_j((1 - \delta)^{2^n} \alpha^*) = (1 - \delta)^{2^n} p_j(\alpha^*)$ and $p_j(\alpha) \geq p_j(\tilde \alpha)$ whenever $\alpha \geq \tilde \alpha$.}
 	%Using this, together with $(1 - \gamma')^2 > (1 - \gamma)$ (using $\gamma'=\gamma/2$) and the no unnecessary pacing property of $\alpha^*$, yields the following chain of implications for each $i \in [n]$,
As a result, the total expenditure of buyer $i$ in $(\alpha',x^*)$ 
  is 
 	\begin{align*}
 		\sum_{j\in [m]} x_{ij}^*\cdot   p_j(\alpha') 
 		>(1-\gamma')\sum_{j\in [m]} x_{ij}^*\cdot p_j(\alpha^*)
 		\ge (1-\gamma')^2 B_i
 		\ge (1-\gamma) B_i.
% 		\\
% 		< (1 - \gamma)B_i \Rightarrow  (1 -\gamma') \sum_{j\in [m]} x_{ij}^*  p_j(\alpha^*) < (1 - \gamma) B_i \Rightarrow \sum_{j\in [m]} x_{ij}^*  p_j(\alpha^*) < (1 - \gamma')B_i
 		%\\
% 		&\implies \\
 %		&\implies \alpha'_i \geq (1 - \gamma')^2\\
 %		&\implies \alpha'_i > (1 - \gamma)
 	\end{align*}
%Twhich implies that 
%  $\alpha^*_i \geq 1 - \gamma'$ and thus,
%  $\alpha'_i >1 - \gamma.$
Therefore, we have shown that $(\alpha', x^*)$ is a $\gamma$-\appe\ of $G$.
%This finishes the proof of Lemma \ref{alg:correct}.
%of $G = (n, m, \{v_{ij}\}_{i,j}, \{B_i\}_i)$.
\end{proof}

\subsection{PPAD Membership of Computing Exact Pacing Equilibria}\label{sec:exact}

In the last subsection we showed that the problem of finding a  $\gamma$-\appe\ of a second-price pacing game $G$ is in PPAD.
Finally we show in this subsection that the problem of finding an exact equilibrium of a pacing game is also in PPAD.
To this end, we show that when $\gamma$ is small enough (though with bit length polynomial in $|G|$), any $\gamma$-\appe $(\alpha',x')$ of $G$ can be ``rounded'' into an exact equilibrium by solving a linear program defined using support information extracted from $(\alpha',x')$. \revision{This technique is similar to the one used in \cite{etessami2010complexity, vazirani2011market} and \cite{filos2020consensus}}.
For this purpose we recall the following fact about linear programs:

\def\LP{\mathsf{LP}}

\begin{fact}\label{fact:LP}
There is a polynomial $r(\cdot)$ with the following property. 
Let $\LP$ be a linear program that minimizes a non-negative variable $\gamma$.
Then an optimal solution of $\LP$ has either $\gamma=0$ or
  $\gamma\ge 1/2^{r(|\LP|)}$, where $|\LP|$ denotes the number of bits needed to represent $\LP.$
\end{fact}
Given a $\gamma$-\appe\ $(\alpha',x')$ of $G=(n,m,(v_{ij}),(B_i))$
  (for some sufficiently small $\gamma$ to be specified later),
  we extract from $(\alpha',x')$ the following support information:\vspace{0.15cm}
\revision{
\begin{flushleft}\begin{enumerate}
    \item $I'\subseteq [n]$ consists of buyers $i\in [n]$ who are almost unpaced, i.e.,
      $\alpha_i'\ge 1-\gamma$. Given that $(\alpha',x')$ is a $\gamma$-\appe, \revision{condition (d) of Definition~\ref{definition_pacing_equilibrium} implies that} 
$$
      \sum_{j\in [m]} x'_{ij} p_j(\alpha')\ge (1-\gamma)B_i,\quad
      \text{for all $i\notin I$}.
$$
\item For each $j\in [m]$,  
    $W_j'$ is the set of buyers $i\in [n]$ with
    $x_{ij}'p_j(\alpha')>0$ (which implies   $\alpha_i' v_{ij}=h_j(\alpha')$). These are buyers who win good $j$ and pay a positive amount for it.\vspace{0.1cm}
  \item For each $j\in [m]$, let $s_j\in [n]$ be the smallest index $i$
    such that $\alpha_i'v_{ij}=h_j(\alpha')$, i.e., $s_j$ is the smallest index among the buyers who have the highest bid in good $j$.\vspace{0.1cm}
\item For each $j\in [m]$,
    let $t_j\in [n]$ be the smallest index $i\ne s_j$
    such that $\alpha_i'v_{ij}=\max_{k\ne s_j} \alpha_k'v_{kj}$
    (so we have that $\alpha_{t_j}' v_{t_j j}=p_j(\alpha')$).\vspace{0.15cm}
    %\item $J'\subseteq [m]$ consists of $j\in [m]$ with
    %  $p_j(\alpha')>0$ (so $p_j(\alpha')=0$ for all $j\notin J$).
      %and thus,
      %$x_{ij}'p_j(\alpha')=0$ for all $i\in [n]$ and $j\notin J$).
    
    %\item For each $j\in J$, we use $\ell_j'\in [n]$ to denote the smallest
    %  $i\in [n]$ such that $\alpha_i'v_{ij}=p_j(\alpha')$.
\end{enumerate}\end{flushleft}}

On the other hand, given any $I\subseteq [n]$, 
$W=(W_j\subseteq [n] :j\in [n])$, $s=(s_j\in [n]:j\in [m])$,~and
  $t=(t_j\in [n]:j\in [m])$, % it is easy to verify that 
%Let $I\subseteq [n]$ (which captures the set of bidders with multipliers close to $1$), $J\subseteq [m]$ (where $\overline{J}$ captures goods for which all bids are zero), $W_j\subseteq [n]$ for each $j\in J$ (where $W_j$, as in the previous subsection, captures bidders with the highest bid for good $j$), and $\ell_j\in [n]$ for each $j\in J$.
%We write $W_j^-$ to denote $W_j$ after deleting the smallest integer in $W_j$.
  we use $\LP(I,W,s,t)$ to denote the following linear program 
  on $n+nm+1$
  variables $\alpha=(\alpha_i:i\in [n])$, $q=(q_{ij}:i\in [n],j\in [m])$
  and $\tau$ (where each variable $q_{ij}$ captures the amount buyer $i$ pays for good $j$):
\begin{align*}
&\text{minimize}\ \tau\\    
&\tau\ge 0, \alpha_i\in [0,1], q_{ij}\ge 0\ \text{for all $i\in [n]$ and $j\in [m]$}\\
&q_{ij}=0\ \text{for all $j\in [m]$ and $i\notin W_j$}\\
%p_j\ge \alpha_iv_{ij}\ \text{for all $i\in [n]$ and $j\in [m]$}\\
&\alpha_{s_j} v_{s_j j}\ge \alpha_k v_{kj}\ \text{for all $j\in [m]$ 
  and $k\in [n]$}\\
&\alpha_{t_j} v_{t_j j}\ge \alpha_k v_{kj}\ \text{for all $j\in [m]$
  and $k\ne s_j \in [n]$}\\
(a)\ \ &\text{$\alpha_iv_{ij}\ge \alpha_{s_j} v_{s_j j}$ 
for all $j\in [m]$ and $i\in W_j$ }\\
(b)\ \ &\text{\(\textstyle\sum_{k\in [n]} q_{kj}=\alpha_{t_j}v_{t_j j}\) for all $j\in [m]$ }\\
%\ \ &\text{$\alpha_{\ell_j}v_{\ell_j j}\ge \alpha_k v_{kj}$ for all $j\in J$ and $k\in[n]$ with
%  $k\ne \min(W_j)$.}\\
(c)\ \  &\text{$\textstyle\sum_{j\in [m]} q_{ij}\le B_i$ for all $i\in [n] $}\\
(d)\ \  &\text{$\alpha_i\ge 1-\tau$ for all $i\in I$ and $\textstyle\sum_{j\in [m]} q_{ij}\ge (1-\tau)B_i$ for all $i\notin I$}
\end{align*}

\revision{Here, $(a)$ ensures that the buyers in $W_j$ have the highest bid on good $j$; $(b)$ ensures that the total payment of all buyers for good $j$ is equal to the second highest bid; $(c)$ ensures that the budgets are satisfied; and $(d)$ ensures that the not-too-much-unnecessary-pacing condition is satisfied.} The lemma below follows directly from the definition of
  $\gamma$-\appe\ and the way $I',W',$ $s'$ and $t'$ are extracted from
  $(\alpha',x')$.
 
\begin{lemma}\label{easy1}
Suppose $(\alpha',x')$ is a $\gamma$-\appe\ of $G$.
Then $(\alpha',q',\gamma)$ is a feasible solution to the linear program $\LP(I',W',s',t')$,
  where $q'=(q_{ij}')$ with $q_{ij}'=x_{ij}' p_j(\alpha')$.
\end{lemma}

On the other hand, the next lemma shows that if 
$\LP(I,W,s,t)$ has a feasible solution $(\alpha,q,0)$~for some 
  $I,W,s$ and $t$, %the linear program 
  then $(\alpha,x)$ is an exact pacing equilibrium, where 
  $x=(x_{ij})$ and $x_{ij}=q_{ij}/p_j(\alpha)$ if $p_j(\alpha)>0$;
  when  $p_j(\alpha)=0$ we set $x_{s_j j}=1$ and $x_{ij}=0$
  for all other $i$.
%  $x_{ij}=0$ if .

%The following lemma shows that if $\LP(I,J,W_j)$ has a solution $(\alpha,q,\gamma)$ with $\gamma=0$, then $(\alpha,x)$
%is a pacing equilibrium of $G$, where $x_{ij}=q_{ij}/(\alpha_iv_{ij})$ for each $i\in [n]$ and $j\in [m]$ (with $0/0=0$ by default).

\begin{lemma}\label{easy2}
If $(\alpha,q,0)$ is a feasible solution to $\LP(I,W,s,t)$, then $(\alpha,x)$ is an exact equilibrium.
\end{lemma}
\begin{proof}
%\textcolor{red}{rk:Xi, I added the proof, take a look when you get the chance} 
Let $(\alpha,q,0)$ be a feasible solution to $\LP(I,W,s,t)$. Set $\alpha$ to be the pacing multipliers of buyers in $G$ 
  and define the allocation $x=(x_{ij})$ as above.
  %for which $(\alpha, x)$ forms a pacing equilibrium of $G$. If $p_j(\alpha) = 0$ for good $j \in [m]$, then set $x_{s_jj} = 1$. Otherwise, set $x_{ij} = q_{ij}/p_j(\alpha)$.
Then, the LP constraints imply that the highest bid on good $j$ is $h_j(\alpha) = \alpha_{s_j} v_{s_j j}$ and the second highest bid is $p_j(\alpha) = \alpha_{t_j} v_{t_jj}$. Next we note that, in the latter case, the constraints of the LP force the set of winners $\{i \mid x_{ij} > 0\}$ of good $j \in [m]$ to be a subset of $W_j$. \revision{This is because $x_{ij} > 0$ implies $q_{ij} > 0$ and $q_{ij} = 0$ for all $i \notin W_j$.} Now, it is straightforward to see that constraints (a)-(d), in combination with $\tau = 0$, imply that $(\alpha, x)$ satisfies the corresponding conditions (a)-(d) of Definition~\ref{definition_exact_pacing_equilibrium}. 
\end{proof}
%\end{fact}

Given the definition of 
  $\LP(I,W,s,t)$, there is a polynomial $r'(\cdot)$ such that $$\max_{I,W,s,t} \big|\LP(I,W,s,t)\big|\le r'(|G|).$$
Now we can set $\gamma$ to be
  smaller than $1/2^{r(r'(|G|))}$ (with bit length still polynomial in $|G|$).
To finish the proof of Theorem \ref{theo:membership}, we let $(\alpha',x')$ be a $\gamma$-\appe\ of $G$.
It follows from Lemma \ref{easy1} that $(\alpha',q',\gamma)$
  is a feasible solution to $\LP(I',W',s',t')$.
%Given $(\alpha',x')$ we let
%\begin{enumerate}
%    \item $I$ be the set of $i\in [n]$ with $\alpha_i\ge 1-\gamma$;
%    \item $J$ be the set of $j\in [m]$ with $h_j(\alpha')>0$; and
%    \item $W_j$, $j\in J$, be the set of $i\in [n]$ with %$\alpha_iv_{ij}=h_j(\alpha')$.
%\end{enumerate}
Next it follows from Fact \ref{fact:LP} that this linear program  has a feasible solution $(\alpha,q,0)$ and the latter can be computed in polynomial time.
Lemma \ref{easy2} shows that $(\alpha,x)$,
  which can be computed in polynomial time, is a pacing equilibrium of $G$.

\section{Conclusion}

We studied the computational complexity of pacing equilibria in  second-price pacing games with multiplicative pacing. Our results show that finding a pacing equilibrium, whether exact or approximate, is a PPAD-complete problem. As discussed previously, these results close the open problem from \cite{conitzer2017multiplicative} on the   complexity of pacing equilibria, and make progress towards resolving the conjecture of \cite{borgs2007dynamics} by showing that their dynamics is unlikely to converge efficiently in second-price auctions.
More generally, our results show that algorithms for budget-smoothing in auctions, an important problem for Internet advertising, cannot be expected to efficiently find even approximate pacing equilibria in the worst case. 

There are several interesting future questions and implications to investigate based on our work. Perhaps most importantly, we would like to understand exactly when budget-smoothing becomes hard. As discussed in the literature review, \cite{balseiro2019learning} developed regret minimization algorithms for the case of i.i.d. and continuous stochastic valuations. Yet our results imply that for general correlated valuations convergence cannot occur efficiently. The question is now which types of correlated stochastic valuations admit efficient algorithms, and which types are hard.
It would also be interesting to understand whether other methods of budget smoothing (such as those discussed by \cite{balseiro2017budget}) lead to PPAD-complete equilibrium problems as well.

In the direction of positive results, our PPAD membership proof  suggests that \mbox{complementary} pivoting may be a fruitful research direction for computing pacing equilibria. This is especially pertinent because approaches based on mixed-integer programming seem to scale poorly~\cite{conitzer2017multiplicative}.

\newpage

%%% Local Variables:
%%% mode: latex
%%% TeX-master: "main"
%%% End:

\bibliographystyle{plainnat}
\bibliography{refs}

\begin{thebibliography}{41}
\providecommand{\natexlab}[1]{#1}
\providecommand{\url}[1]{\texttt{#1}}
\expandafter\ifx\csname urlstyle\endcsname\relax
  \providecommand{\doi}[1]{doi: #1}\else
  \providecommand{\doi}{doi: \begingroup \urlstyle{rm}\Url}\fi

\bibitem[Ashlagi et~al.(2010)Ashlagi, Braverman, Hassidim, Lavi, and
  Tennenholtz]{ashlagi2010position}
Itai Ashlagi, Mark Braverman, Avinatan Hassidim, Ron Lavi, and Moshe
  Tennenholtz.
\newblock Position auctions with budgets: Existence and uniqueness.
\newblock \emph{The BE Journal of Theoretical Economics}, 10\penalty0 (1),
  2010.

\bibitem[Babaioff et~al.(2020)Babaioff, Cole, Hartline, Immorlica, and
  Lucier]{babaioff2020non}
Moshe Babaioff, Richard Cole, Jason Hartline, Nicole Immorlica, and Brendan
  Lucier.
\newblock Non-quasi-linear agents in quasi-linear mechanisms.
\newblock \emph{arXiv preprint arXiv:2012.02893}, 2020.

\bibitem[Balseiro et~al.(2017)Balseiro, Kim, Mahdian, and
  Mirrokni]{balseiro2017budget}
Santiago Balseiro, Anthony Kim, Mohammad Mahdian, and Vahab Mirrokni.
\newblock Budget management strategies in repeated auctions.
\newblock In \emph{Proceedings of the 26th International Conference on World
  Wide Web}, pages 15--23, 2017.

\bibitem[Balseiro et~al.(2022)Balseiro, Kroer, and Kumar]{kumar2022contextual}
Santiago Balseiro, Christian Kroer, and Rachitesh Kumar.
\newblock Contextual standard auctions with budgets: Revenue equivalence and
  efficiency guarantees.
\newblock In \emph{EC}, 2022.
\newblock URL \url{https://arxiv.org/abs/2102.10476}.

\bibitem[Balseiro and Gur(2019)]{balseiro2019learning}
Santiago~R Balseiro and Yonatan Gur.
\newblock Learning in repeated auctions with budgets: Regret minimization and
  equilibrium.
\newblock \emph{Management Science}, 65\penalty0 (9):\penalty0 3952--3968,
  2019.

\bibitem[Balseiro et~al.(2015)Balseiro, Besbes, and
  Weintraub]{balseiro2015repeated}
Santiago~R Balseiro, Omar Besbes, and Gabriel~Y Weintraub.
\newblock Repeated auctions with budgets in ad exchanges: Approximations and
  design.
\newblock \emph{Management Science}, 61\penalty0 (4):\penalty0 864--884, 2015.

\bibitem[Bei et~al.(2016)Bei, Garg, Hoefer, and Mehlhorn]{bei2016computing}
Xiaohui Bei, Jugal Garg, Martin Hoefer, and Kurt Mehlhorn.
\newblock Computing equilibria in markets with budget-additive utilities.
\newblock In \emph{24th Annual European Symposium on Algorithms, ESA 2016},
  page~8. Schloss Dagstuhl-Leibniz-Zentrum fur Informatik GmbH, Dagstuhl
  Publishing, 2016.

\bibitem[Bitansky et~al.(2015)Bitansky, Paneth, and Rosen]{Crypto1}
N.~Bitansky, O.~Paneth, and A.~Rosen.
\newblock On the cryptographic hardness of finding a nash equilibrium.
\newblock In \emph{Proceedings of the 56th Annual Symposium on Foundations of
  Computer Science}, pages 1480--1498, 2015.

\bibitem[Borgs et~al.(2007)Borgs, Chayes, Immorlica, Jain, Etesami, and
  Mahdian]{borgs2007dynamics}
Christian Borgs, Jennifer Chayes, Nicole Immorlica, Kamal Jain, Omid Etesami,
  and Mohammad Mahdian.
\newblock Dynamics of bid optimization in online advertisement auctions.
\newblock In \emph{Proceedings of the 16th international conference on World
  Wide Web}, pages 531--540, 2007.

\bibitem[Budish(2011)]{budish2011combinatorial}
Eric Budish.
\newblock The combinatorial assignment problem: Approximate competitive
  equilibrium from equal incomes.
\newblock \emph{Journal of Political Economy}, 119\penalty0 (6):\penalty0
  1061--1103, 2011.

\bibitem[Chen and Deng(2006)]{chen2006settling}
Xi~Chen and Xiaotie Deng.
\newblock Settling the complexity of two-player nash equilibrium.
\newblock In \emph{2006 47th Annual IEEE Symposium on Foundations of Computer
  Science (FOCS'06)}, pages 261--272. IEEE, 2006.

\bibitem[Chen and Deng(2009)]{CHEN20094448}
Xi~Chen and Xiaotie Deng.
\newblock On the complexity of 2d discrete fixed point problem.
\newblock \emph{Theoretical Computer Science}, 410\penalty0 (44):\penalty0
  4448--4456, 2009.

\bibitem[Chen and Teng(2009)]{chen2009spending}
Xi~Chen and Shang-Hua Teng.
\newblock Spending is not easier than trading: on the computational equivalence
  of fisher and arrow-debreu equilibria.
\newblock In \emph{International Symposium on Algorithms and Computation},
  pages 647--656. Springer, 2009.

\bibitem[Chen et~al.(2007)Chen, Teng, and Valiant]{WinLose}
Xi~Chen, Shang-Hua Teng, and Paul Valiant.
\newblock The approximation complexity of win-lose games.
\newblock In \emph{Proceedings of the 18th annual ACM-SIAM symposium on
  Discrete algorithms}, pages 159--168, 2007.

\bibitem[Chen et~al.(2017)Chen, Paparas, and Yannakakis]{chen2017complexity}
Xi~Chen, Dimitris Paparas, and Mihalis Yannakakis.
\newblock The complexity of non-monotone markets.
\newblock \emph{Journal of the ACM (JACM)}, 64\penalty0 (3):\penalty0 1--56,
  2017.

\bibitem[Chen et~al.(2021)Chen, Kroer, and Kumar]{chen2021throttling}
Xi~Chen, Christian Kroer, and Rachitesh Kumar.
\newblock Throttling equilibria in auction markets.
\newblock In \emph{WINE}, 2021.
\newblock URL \url{https://arxiv.org/abs/2107.10923}.

\bibitem[Choudhuri et~al.(2019)Choudhuri, Hubacek, Kamath, Pietrzak, Rosen, and
  Rothblum]{Crypto5}
A.R. Choudhuri, P.~Hubacek, C.~Kamath, K.~Pietrzak, A.~Rosen, and G.N.
  Rothblum.
\newblock Finding a nash equilibrium is no easier than breaking {Fiat-Shamir}.
\newblock In \emph{Proceedings of the 51st Annual ACM Symposium on Theory of
  Computing}, pages 1103--1114, 2019.

\bibitem[Conitzer et~al.(2019)Conitzer, Kroer, Panigrahi, Schrijvers, Sodomka,
  Stier-Moses, and Wilkens]{conitzer2019pacing}
Vincent Conitzer, Christian Kroer, Debmalya Panigrahi, Okke Schrijvers, Eric
  Sodomka, Nicolas~E Stier-Moses, and Chris Wilkens.
\newblock Pacing equilibrium in first-price auction markets.
\newblock In \emph{Proceedings of the 2019 ACM Conference on Economics and
  Computation}, pages 587--587, 2019.

\bibitem[Conitzer et~al.(2021)Conitzer, Kroer, Sodomka, and
  Stier-Moses]{conitzer2017multiplicative}
Vincent Conitzer, Christian Kroer, Eric Sodomka, and Nicolas~E Stier-Moses.
\newblock Multiplicative pacing equilibria in auction markets.
\newblock \emph{Operations Research}, 2021.

\bibitem[Daskalakis et~al.(2009)Daskalakis, Goldberg, and
  Papadimitriou]{daskalakis2009complexity}
Constantinos Daskalakis, Paul~W Goldberg, and Christos~H Papadimitriou.
\newblock The complexity of computing a nash equilibrium.
\newblock \emph{SIAM Journal on Computing}, 39\penalty0 (1):\penalty0 195--259,
  2009.

\bibitem[Deng et~al.(2012)Deng, Qi, and Saberi]{deng2012algorithmic}
Xiaotie Deng, Qi~Qi, and Amin Saberi.
\newblock Algorithmic solutions for envy-free cake cutting.
\newblock \emph{Operations Research}, 60\penalty0 (6):\penalty0 1461--1476,
  2012.

\bibitem[Dobzinski and Leme(2014)]{dobzinski2014efficiency}
Shahar Dobzinski and Renato~Paes Leme.
\newblock Efficiency guarantees in auctions with budgets.
\newblock In \emph{International Colloquium on Automata, Languages, and
  Programming}, pages 392--404. Springer, 2014.

\bibitem[Dobzinski et~al.(2012)Dobzinski, Lavi, and Nisan]{dobzinski2012multi}
Shahar Dobzinski, Ron Lavi, and Noam Nisan.
\newblock Multi-unit auctions with budget limits.
\newblock \emph{Games and Economic Behavior}, 74\penalty0 (2):\penalty0
  486--503, 2012.

\bibitem[Etessami and Yannakakis(2010)]{etessami2010complexity}
Kousha Etessami and Mihalis Yannakakis.
\newblock On the complexity of {N}ash equilibria and other fixed points.
\newblock \emph{SIAM Journal on Computing}, 39\penalty0 (6):\penalty0
  2531--2597, 2010.

\bibitem[Facebook(2017)]{facebookguide}
Facebook.
\newblock Your guide to facebook bid strategy.
\newblock
  \url{https://www.facebook.com/gms_hub/share/biddingstrategyguide_final.pdf},
  2017.
\newblock Accessed: 2021-06-30.

\bibitem[Fearnley et~al.(2021)Fearnley, Goldberg, Hollender, and
  Savani]{fearnley2021complexity}
John Fearnley, Paul~W Goldberg, Alexandros Hollender, and Rahul Savani.
\newblock The complexity of gradient descent: Cls= ppad $\cap$ pls.
\newblock In \emph{Proceedings of the 53rd Annual ACM SIGACT Symposium on
  Theory of Computing}, pages 46--59, 2021.

\bibitem[Filos-Ratsikas et~al.(2020)Filos-Ratsikas, Hollender, Sotiraki, and
  Zampetakis]{filos2020consensus}
Aris Filos-Ratsikas, Alexandros Hollender, Katerina Sotiraki, and Manolis
  Zampetakis.
\newblock Consensus-halving: Does it ever get easier?
\newblock In \emph{Proceedings of the 21st ACM Conference on Economics and
  Computation}, pages 381--399, 2020.

\bibitem[Filos-Ratsikas et~al.(2021)Filos-Ratsikas, Giannakopoulos, Hollender,
  Lazos, and Po{\c{c}}as]{filos2021complexity}
Aris Filos-Ratsikas, Yiannis Giannakopoulos, Alexandros Hollender, Philip
  Lazos, and Diogo Po{\c{c}}as.
\newblock On the complexity of equilibrium computation in first-price auctions.
\newblock In \emph{Proceedings of the 22nd ACM Conference on Economics and
  Computation}, pages 454--476, 2021.

\bibitem[Garg et~al.(2016)Garg, Pandey, and Srinivasan]{Crypto2}
S.~Garg, O.~Pandey, and A.~Srinivasan.
\newblock Revisiting the cryptographic hardness of finding a nash equilibrium.
\newblock In \emph{Proceedings of the 36th Annual International Cryptology
  Conference on Advances in Cryptology}, pages 579--604, 2016.

\bibitem[Goel et~al.(2015)Goel, Mirrokni, and Leme]{goel2015polyhedral}
Gagan Goel, Vahab Mirrokni, and Renato~Paes Leme.
\newblock Polyhedral clinching auctions and the adwords polytope.
\newblock \emph{Journal of the ACM (JACM)}, 62\penalty0 (3):\penalty0 1--27,
  2015.

\bibitem[Goldberg(2011)]{goldberg2011survey}
Paul~W Goldberg.
\newblock A survey of ppad-completeness for computing nash equilibria.
\newblock \emph{arXiv preprint arXiv:1103.2709}, 2011.

\bibitem[Hubacek and Yogev(2017)]{Crypto4}
P.~Hubacek and E.~Yogev.
\newblock Hardness of continuous local search: Query complexity and
  cryptographic lower bounds.
\newblock In \emph{Proceedings of the 28th Annual ACM-SIAM Symposium on
  Discrete Algorithms}, pages 1352--1371, 2017.

\bibitem[Kuhn(1960)]{kuhn1960some}
Harold~W Kuhn.
\newblock Some combinatorial lemmas in topology.
\newblock \emph{IBM Journal of research and development}, 4\penalty0
  (5):\penalty0 518--524, 1960.

\bibitem[Mehta(2013)]{mehta2013online}
Aranyak Mehta.
\newblock Online matching and ad allocation.
\newblock \emph{Foundations and Trends in Theoretical Computer Science}, 8
  (4):\penalty0 265--368, 2013.
\newblock URL \url{http://dx.doi.org/10.1561/0400000057}.

\bibitem[Mehta et~al.(2007)Mehta, Saberi, Vazirani, and
  Vazirani]{mehta2007adwords}
Aranyak Mehta, Amin Saberi, Umesh Vazirani, and Vijay Vazirani.
\newblock Adwords and generalized online matching.
\newblock \emph{Journal of the ACM (JACM)}, 54\penalty0 (5):\penalty0 22--es,
  2007.

\bibitem[Othman et~al.(2016)Othman, Papadimitriou, and
  Rubinstein]{othman2016complexity}
Abraham Othman, Christos Papadimitriou, and Aviad Rubinstein.
\newblock The complexity of fairness through equilibrium.
\newblock \emph{ACM Transactions on Economics and Computation (TEAC)},
  4\penalty0 (4):\penalty0 1--19, 2016.

\bibitem[Papadimitriou(1994)]{PPAD}
Christos~H. Papadimitriou.
\newblock On the complexity of the parity argument and other inefficient proofs
  of existence.
\newblock \emph{Journal of Computer and System Sciences}, 48\penalty0
  (3):\penalty0 498--532, 1994.

\bibitem[Rosen et~al.(2017)Rosen, Segev, and Shahaf]{Crypto3}
A.~Rosen, G.~Segev, and I.~Shahaf.
\newblock Can {PPAD} hardness be based on standard cryptographic assumptions?
\newblock In \emph{Proceedings of the 15th International Conference on Theory
  of Cryptography}, pages 173--205, 2017.

\bibitem[Roughgarden(2020)]{TimSurvey}
Tim Roughgarden.
\newblock Complexity theory, game theory, and economics: The barbados lectures.
\newblock \emph{Foundations and Trends in Theoretical Computer Science},
  14\penalty0 (3–-4):\penalty0 222--407, 2020.

\bibitem[Vazirani and Yannakakis(2011{\natexlab{a}})]{VijayMihalis}
Vijay~V. Vazirani and Mihalis Yannakakis.
\newblock Market equilibrium under separable, piecewise-linear, concave
  utilities.
\newblock \emph{J. ACM}, 58\penalty0 (3), 2011{\natexlab{a}}.

\bibitem[Vazirani and Yannakakis(2011{\natexlab{b}})]{vazirani2011market}
Vijay~V Vazirani and Mihalis Yannakakis.
\newblock Market equilibrium under separable, piecewise-linear, concave
  utilities.
\newblock \emph{Journal of the ACM (JACM)}, 58\penalty0 (3):\penalty0 1--25,
  2011{\natexlab{b}}.

\end{thebibliography}

\appendix

\section{Proof of Theorem \ref{approximate_hardness}} \label{appendix_hardness}

%Consider 
\def\eps{\epsilon}
%We give a polynomial-time reduction from the problem of finding 
%  an $\eps$-well-supported Nash equilibrium (see definition below)
%  in a $\{0,1\}$-cost bimatrix game
%  to that of finding a $(\delta,\gamma)$-\appe\ in an SPP game
%  $G$

Consider a $\{0,1\}$-cost $n\times n$ bimatrix game  
  $(A,B)$ and let $\epsilon = 1/n$.
Recall that an $\eps$-well-supported Nash equilibrium  
  is a pair $(x,y)\in \Delta_n\times \Delta_n$ such that
$x_i>0$ for any $i\in [n]$ implies that $\sum_j A_{ij}y_j\le 
  \sum_j A_{kj}y_j+\eps$ for all $k$ and 
  $y_j>0$ for any $j\in [n]$ implies $\sum_i x_i B_{ij}
  \le \sum_i x_i B_{ik}+\eps$ for all $k\in [n]$.
  
In this section we show how to construct an SPP game $G$ with $4n+1$ buyers from 
  the bimatrix game $(A,B)$ in 
  time polynomial 
  in $n$  such that every $(\delta, \gamma)$-\appe\ of $G$,
  where $\delta = \gamma=\epsilon/n^6$, can be mapped back to an 
  $\eps$-well-supported Nash equilibrium of $(A,B)$ in polynomial time. 
Theorem \ref{theo:approx_hardness_informal} follows from the PPAD-completeness of 
  the problem of finding an $\eps$-well-supported Nash equilibrium
  in a $\{0,1\}$-cost bimatrix game with $\eps=1/n$ \cite{WinLose}.

The SPP game $G$ contains the following goods:\vspace{0.15cm}
%\begin{itemize}
%    \item $(p,s)$-Strategy Gadget: For each player $p \in \{1,2\}$ and strategy $s \in [n]$, there are $n$ normalization goods $\{N(p,s)_1, \dots,  N(p,s)_n\}$.
    
    %\item $(p,s)$-Threshold Good: For each player $p \in \{1,2\}$ and strategy $s \in [n]$, there is a threshold good $T(p,s)$.
    
    %\item $(p,s)$-Expenditure Gadget: For each player $p \in \{1,2\}$ and strategy $s \in [n]$, there are $n$ items $\{E(p,s)_1, \dots, E(p,s)_{n}\}$.
%\end{itemize}
\begin{itemize}
    \item Normalization goods: $n$ goods $\{N(p,s)_1, \dots,  N(p,s)_n\}$ for each $p \in \{1,2\}$ and  $s \in [n]$. \vspace{0.15cm}
    \item Expenditure goods: $n$ goods $\{E(p,s)_1, \dots, E(p,s)_{n}\}$ for each $p \in \{1,2\}$ and $s \in [n]$.\vspace{0.15cm}
    \item Threshold goods  $T(p,s)$ for each  $p \in \{1,2\}$ and  $s \in [n]$.\vspace{0.15cm}
\end{itemize}

Set $\nu = 1/(16n)$. The set of buyers in $G$ is defined as follows:\vspace{0.15cm}
\begin{flushleft}\begin{itemize}
    \item Buyer $\mathbb{C}(p,s)$, $p \in \{1,2\}$ and $s \in [n]$: $\mathbb{C}(p,s)$ has positive values for the following goods:\vspace{0.2cm}
    \begin{itemize}
        \item Normalization goods:
        %other than $N(p,s)_s$ have value $2$, i.e., 
        $V(\mathbb{C}(p,s), N(p,s)_i) = 16$ for all $i \in [n]\setminus{\{s\}}$;\\
%        \item The $s$'th $(p,s)$-normalization good has value $1$, i.e., 
$V(\mathbb{C}(p,s), N(p,s)_s) = 1$; and
%        \item The $s$'th $(p,t)$-normalization good at value $1$ for all $t \neq s$, i.e., 
$V(\mathbb{C}(p,s), N(p,t)_s) = 1$ for all $t \in [n]\setminus \{s\}$.\vspace{0.15cm}
        
        \item Threshold good $T(p,s)$:   $V(\mathbb{C}(p,s), T(p,s)) = 2n^4$.\vspace{0.15cm}
        
        \item Expenditure goods: % $\{E(p,s)_1, \dots, E(p,s)_n\}$ have value $1$, i.e., 
        $V(\mathbb{C}(p,s), E(p,s)_i) = 1$ for all $i \in [n]$.\\
        For $p = 1$: %$\mathbb{B}(1,s)$ values the $s$'th $(2,t)$-expenditure good at $\nu B_{st}$ for all $t \in [n]$, i.e., 
        $V(\mathbb{C}(1,s), E(2,t)_s) = \nu B_{st}$ for all $t \in [n]$.
        
        For $p = 2$: %$\mathbb{B}(2,s)$ values the $s$'th $(1,t)$-expenditure good at $\nu A_{ts}$ for all $t \in [n]$, i.e., 
        $V(\mathbb{C}(2,s), E(1,t)_s) = \nu A_{ts}$ for all $t \in [n]$.\vspace{0.2cm}
    \end{itemize}
    For $p=1$, the budget of buyer $\mathbb{C}(1,s)$ is $n/2 +  n^4 + 1/4 + \sum_{t\in [n]} \nu A_{st}/2$;\\ For $p=2$, the budget of buyer $\mathbb{C}(2,s)$ is $n/2 +  n^4 + 1/4 + \sum_{t\in [n]} \nu B_{ts}/2$.
    \vspace{0.25cm}
    \item Threshold Buyer $\mathbb{T}$: $\mathbb{T}$ has positive values only for the following goods:\vspace{0.15cm}
    \begin{itemize}
        \item Threshold goods: $V(\mathbb{T}, T(p,s)) = (1-\delta)n^4$ for each $p \in \{1,2\}$ and  $s \in [n]$.\vspace{0.15cm}
        
        \item Expenditure goods:  %For $p=1$: The expenditure good 
        $V(\mathbb{T},E(1,s)_t)= \nu A_{st}/2$ 
        and $V(\mathbb{T},E(2,s)_t)=\nu B_{ts}/2$ for all $s,t \in [n]$.%\\
%        
%        For $p=2$: The expenditure good $E(2,s)_t$ has value $\nu B_{ts}/2$ for all $s,t \in [n]$.
\vspace{0.2cm}
    \end{itemize}
    
    Buyer $\mathbb T$ has budget $n^7$. % (high enough so that she is never paced in any PE).
    \vspace{0.25cm}
    
    \item Dummy buyer $\mathbb{D}(p,s)$, $p\in \{1,2\}$ and $s\in [n]$: %For each player $p \in \{1,2\}$ and strategy $s \in [n]$, there is a buyer $\mathbb{D}(p,s)$ with budget $\nu$, 
    The budget of $\mathbb{D}(p,s)$ is $\nu$ and she only values
      the normalization good $N(p,s)_s$: $V(\mathbb{D}(p,s), N(p,s)_s)=1$.\vspace{0.15cm}
%    who values the $s$'th $(p,s)$-normalization good at 1 and every other good at 0.\vspace{0.15cm} % Moreover, $\D(p,s)$ has a  for all $p \in \{1,2\}$ and $s \in [n]$.
\end{itemize}\end{flushleft}

Let $\mathcal{E}$ be a $(\delta, \gamma)$-\appe\ of the game $G$. We will use $\alpha(\cdot )$ to denote pacing multipliers of buyers in $\mathcal{E}$. Observe that, from the definition of approximate pacing equilibria, we must have $\alpha(\T)\in [1 - \gamma, 1]$.  The following lemma establishes bounds on pacing multipliers of other buyers.

\begin{lemma}\label{appx:mutlipliers_bounds}
    For each $p \in \{1,2\}$ and $s \in [n]$, we have $$\frac{(1 -\delta)^2}{2} \leq \alpha(\BB(p,s)) \leq \frac{7}{8}\quad \text{and}\quad (1 - \delta)\cdot  \alpha(\BB(p,s)) \leq \alpha(\D(p,s)) \leq \frac{\alpha(\BB(p,s))}{1 - \delta}.$$ 
\end{lemma}

\begin{proof}
    Suppose for some $p \in \{1,2\}$ and $s \in [n]$, we have $\alpha(\BB(p,s)) < (1-\delta)^2/2$. Then $\mathbb{C}(p,s)$ doesn't win any part of the threshold good $T(p,s)$. Observe that she has value at most $16$ for every other good. 
Given that there are only $O(n^2)$ goods in $G$,
  %and there are at most $n^3$ goods in $G$. Therefore, the sum of her bids on all items other than $T(p,s)$ is at most $2 n^3$. Hence, 
  she can not possibly spend all her budget (which is $\Omega(n^4)$). 
This contradicts the assumption that $\mathcal{E}$ is an approximate PE.
%definition of a pacing equilibrium because her budget is strictly greater than $n^4$. 
Therefore, we have $\alpha(\BB(p,s)) \geq (1-\delta)^2/2$ for
  each $p\in \{1,2\}$ and $s\in [n]$.

  Next we prove the inequality about $\alpha(\D(p,s))$.
Suppose $(1-\delta) \alpha(\D(p,s)) > \alpha(\BB(p,s))$ for some $p \in \{1,2\}$ and $s \in [n]$. Then, $\D(p,s)$ wins all of good $N(p,s)_s$ at price $\alpha(\BB(p,s))\ge (1-\delta)^2/2$. This violates her budget constraint and leads to a contradiction. Hence $\alpha(\D(p,s)) \leq \alpha(\BB(p,s))/(1-\delta)$. Moreover, if $\alpha(\D(p,s)) < (1-\delta) \alpha(\BB(p,s))$ (which implies $\alpha(\D(p,s))<1-\delta=1-\gamma$) then her expenditure is zero. This violates the no unnecessary pacing condition. Hence the inequality about $\alpha(\D(p,s))$ must hold. Observe that, in particular, this means that the price of 
  $N(p,s)_s$ is between $(1-\delta)\alpha(\BB(p,s))$ and 
  $\alpha(\BB(p,s))/(1-\delta)$.
    
Finally suppose $\alpha(\BB(p,s)) >7/8$ for some $p \in \{1,2\}$, $s \in [n]$. Then she wins:\vspace{0.1 cm}
\begin{flushleft}    \begin{itemize}
        \item 
        All of 
        normalization good $N(p,s)_t$, for each $t\ne s$, by spending at least $(1-\delta)^2/2$ on each of them because $\alpha(\BB(p,t)) \geq (1-\delta)^2/2$ 
        %for all $p \in \{1,2\}$, $r \in [n]$ 
        by the first part of the proof.\vspace{0.15cm}
        
        \item Part of normalization good $N(p,s)_s$ by spending at least $(1-\delta)(7/8) - \nu$. This is because $N(p,s)_s$ has price 
        at least $(1-\delta)(7/8)$ and
        buyer $\D(p,s)$ only has budget $\nu$.
%        (Recall that $\D(p,s)$ has a budget of $\nu$).
\vspace{0.15cm}
        
        \item All of threshold good $T(p,s)$ by spending   $\alpha(\T)(1-\delta)n^4\ge (1-\delta)^2n^4$ (using $\gamma=\delta$).\vspace{0.15cm}
        
        \item All of expenditure good $E(p,s)_t$, for each $t\in [n]$, by spending at least $\alpha(\T)\nu A_{st}/2$ if $p=1$\\ and $\alpha(\T) \nu B_{ts}/2$ if $p=2$.\vspace{0.1 cm}
    \end{itemize}\end{flushleft}
    Hence, the total expenditure of $\BB(p,s)$ when $p=1$ is at least $$
    (1-\delta)^2\cdot \frac{n-1}{2} + (1-\delta)\cdot \frac{7}{8} - \nu + (1-\delta )^2n^4 + (1-\delta)\sum_{t\in [n]} \nu A_{st}/2$$ 
    which is strictly higher the budget (using $\delta=1/n^7$).
The same also holds for $p=2$. In both cases, the budget constraint is violated, leading to a contradiction. Therefore, the lemma holds.
\end{proof}

In particular, the above lemma implies that 
  the total expenditure of each buyer $\BB(p,s)$ is at least
  $(1-\gamma)$-fraction of her budget (and of course is also
  bounded from above by her budget).
%that the construction of $G$ ensures that all the strategy buyers are paced in $\mathcal{E}$, thereby implying that their total expenditures must exactly equal their budgets.
We also get the following corollary:

\begin{corollary}
	For each $p \in \{1,2\}$ and $s \in [n]$,  the expenditure 
	of $\BB(p,s)$ on $N(p,s)_s$ lies in the following interval $[ (1 - \delta)\alpha(\BB(p,s)) - \nu, \alpha(\BB(p,s))-(1-\delta)\nu]$
\end{corollary}

 Next, we define two vectors $x'$ and $y'$ with
   $$x_s' = \big\{\alpha(1,s) - (\alpha(\T)/2)\big\}^+\quad\text{and}\quad y_s' = \big\{\alpha(2,s) - (\alpha(\T)/2)\big\}^+$$ for each $s \in [n]$, where $a^+$ ddenotes $\max\{a,0\}$. The following lemma will allow us to normalize $x'$ and $y'$ to obtain valid probability distributions.

\begin{lemma} \label{multiplier_positive_sum}
    The following inequalities hold: $\sum_s x_s' > 1/8$ and $\sum_s y_s' > 1/8$.
\end{lemma}

\begin{proof}
    We prove $\sum_s x_s' > 1/8$. The proof of $\sum_s y_s' > 1/8$ is  analogous. Suppose that $\sum_s x_s' \leq 1/8$. Then, buyer $\mathbb{B}(1,1)$ only wins a non-zero fraction of the following goods, and spends:\vspace{0.15cm}
    \begin{itemize}
        \item At most $\alpha(\BB(1,t))$ on each
        normalization good $N(1 ,1)_t$ for each $t\in [n]$. The total
        expenditure is 
$$
\sum_{t\in [n]} \alpha(\BB(1,t)) \le  n \alpha(\T)/2 + \sum_{t\in [n]} x_t' \leq n/2 + 1/8.
$$      %, \dots, N(p,s)_n\}$.\vspace{0.1 cm}
        
        \item At most $(1 - \delta)n^4$ on the threshold good $T(1,1)$.\vspace{0.1 cm}
        
        \item At most $\nu A_{1t}$ on each expenditure good $E(1,1)_t$,  $t \in [n]$.\vspace{0.15cm}
    \end{itemize}
    Hence, the total expenditure of buyer $\mathbb{C}(1,1)$ is at most $n/2 + 1/8 + (1 - \delta)n^4 + \sum_t \nu A_{1t}$, which is strictly less than her budget, a contradiction.
\end{proof}

Now, we are ready to define the mixed strategies $(x,y)$  for the bimatrix game $(A,B)$. Set player 1's mixed strategy $x$ to be $x_s= x_s'/ \sum_i x_i'$ and player 2's mixed strategy $y$ to be $y_s=y_s'/ \sum_i y_i'$. These are valid mixed strategies because of Lemma~\ref{mutlipliers_bounds} and Lemma~\ref{lem:exact pe multiplier_positive_sum}. 
The next lemma shows that $(x,y)$ is indeed an $\eps$-well-supported  Nash equilibrium of $(A,B)$.

%Now, we are ready to define the mixed strategies for the bimatrix game corresponding to pacing equilibrium $\mathscr{E}$. Set player 1's mixed strategy to be $\{x_s/ \sum_s x_s\}_s$ and player 2's mixed strategy to be $\{y_s/ \sum_s y_s\}_s$. These are valid mixed strategies because of Lemma \ref{mutlipliers_bounds} and Lemma \ref{multiplier_positive_sum}. To complete the proof of PPAD hardness, we need to show that the mixed strategies defined in this way form an $\epsilon$-well-supported NE for the bimatrix game, which is the subject of the next lemma.

\begin{lemma}
	$(x,y)$ is an $\epsilon$-well-supported Nash equilibrium of the bimatrix game $(A,B)$.
\end{lemma}

\begin{proof}
Assume there are $s,s^*\in [n]$ such that 
  $x_s>0$ but $\sum_t A_{st}y_t>\sum_t A_{s^*t}y_t+\eps$; the proof 
  for $y$ is analogous.
%	Consider $s \in [n]$ such that $x_s > 0$. Then, 
Using $x_s>0$, buyer $\mathbb{C}(1,s)$ 
  spends non-zero amounts  on the following goods:\vspace{0.15cm}
\begin{itemize}
    \item $\alpha(\BB(1,t))$ on the normalization good $N(1,s)_t$ for each $t\ne s$.\vspace{0.1 cm}
    \item at least $(1-\delta)\cdot \alpha(\BB(1,s)) - \nu$ on the normalization good $N(1,s)_s$.\vspace{0.1 cm}
    
    \item $\alpha(\T)\cdot (1-\delta)n^4$ on the threshold good $T(1,s)$.\vspace{0.1 cm}
    
    \item $\max\{\alpha(\BB(2,t)),\alpha(\T)/2\} \cdot \nu A_{st}$ on the expenditure good $E(1,s)_t$ for each  $t \in [n]$.\vspace{0.15cm}
\end{itemize}
Therefore, the total expenditure of buyer $\mathbb{C}(1,s)$  is at least
\begin{align*}
    \sum_{t\in [n]}& \alpha(\BB(1,t))   -\delta\cdot \alpha(\BB(1,s))-\nu +\alpha(\T)\cdot (1-\delta)n^4 + \sum_{t\in [n]} \max\big\{\alpha(\BB(2,t)),\alpha(\T)/2\big\}\cdot \nu A_{st}\\  &= \sum_{t\in [n]} \alpha(\BB(1,t)) -\delta\cdot \alpha(\BB(1,s))-\nu +\alpha(\T)\cdot (1-\delta)n^4 + \alpha(\T)\sum_{t\in [n]} \nu A_{st}/2 + \nu \sum_{t\in [n]} y_t A_{st}.
\end{align*}
On the other hand, buyer $\BB(1,s^*)$ spends (without assuming 
  $x_{s^*}>0$):\vspace{0.15cm}
%	Consider $s \in [n]$ such that $x_s > 0$. Then, $\mathbb{S}(1,s)$ only wins a non-zero fraction of the following goods and spends:
\begin{itemize}
    \item $\alpha(\BB(1,t))$ on the normalization good $N(1,s^*)_t$ for each $t\ne s^*$.\vspace{0.1 cm}
    \item at most $\alpha(\BB(1,s^*))-(1-\delta)\nu$ on the normalization good $N(1,s^*)_{s^*}$.\vspace{0.1 cm}
    
    \item at most $\alpha(\T)\cdot (1-\delta)n^4$ on the threshold good $T(1,s^*)$.\vspace{0.1 cm}
    
    \item $\max\{\alpha(\BB(2,t)),\alpha(\T)/2\} \cdot \nu A_{s^*t}$ on the expenditure good $E(1,s^*)_t$ for each  $t \in [n]$.\vspace{0.15cm}
\end{itemize}
Therefore, the total expenditure of buyer $\BB(1,s^*)$ is at most
$$
\sum_{t\in [n]} \alpha(\BB(1,t))-(1-\delta)\nu+\alpha(\T)\cdot (1-\delta)n^4 + \alpha(\T)\sum_{t\in [n]} \nu A_{s^*t}/2 + \nu \sum_{t\in [n]} y_t A_{s^*t}.
$$

Using the assumption that $\sum_t A_{st}y_t > \sum_t A_{s^*t}y_t+\eps$,
  we have that the total expenditure of $\BB(1,s)$ minus that
  of $\BB(1,s^*)$, denoted by $(\ddagger 1)$, is at least
\begin{align*}
 -\delta\cdot \alpha(\BB(1,s))-\delta\nu  &
 +\alpha(\T)\cdot \left(\sum_{t\in [n]} \nu A_{st}/2-\sum_{t\in [n]}
   \nu A_{s^*t}/2\right)+ \eps\nu\\&\quad\quad\quad\ge 
    \alpha(\T)\cdot \left(\sum_{t\in [n]} \nu A_{st}/2-\sum_{t\in [n]}
   \nu A_{s^*t}/2\right)+ \eps\nu/2
\end{align*}
using $\eps\nu \gg \delta$.
On the other hand, the budget of $\BB(1,s)$ minus that of $\BB(1,s^*)$,
  denoted $(\ddagger 2)$, is 
$$
 \sum_{t\in [n]} \nu A_{st}/2-\sum_{t\in [n]}
   \nu A_{s^*t}/2.
$$
Using $\alpha(\T)\ge 1-\gamma$ and $\gamma=1/n^7$,
  we have $(\ddagger 1)\ge (\ddagger 2)+\eps\nu /3$.
However, the total expenditure of $\BB(1,s)$ is at most her budget and
  the total expenditure of $\BB(1,s^*)$ is at least $(1-\gamma)$-fraction
  of her budget.
Given that the budget of $\BB(1,s^*)$ is $O(n^4)$,
  we also have 
$$
(\ddagger 1)\le (\ddagger 2)+\gamma\cdot O(n^4)
=(\ddagger 2)+O(1/n^3),
$$
a contradiction because $\eps\nu=\Omega(1/n^2)$.
\end{proof}

Theorem \ref{approximate_hardness}
%, one only needs to observe that the number of items $G$ is at least $4n^2$ and the number of items is at least $2n$ and invoke the result of \cite{chen2007approximation} about the 
follows from the PPAD-hardness of finding an 
  $\epsilon$-well-supported Nash equilibrium 
  in a $\{0,1\}$-cost bimatrix game \cite{WinLose}.

\section{Proof of Claim \ref{lipschitz}}\label{appendix_lipschitz}

Before stating the proof of Claim \ref{lipschitz}, we state and prove the following useful lemma.

\begin{lemma}\label{beta_lower_bound}
If $\beta \in S$ is labelled $i$, then $\beta_i \geq \min\left\{\frac{1}{n}, \frac{B_{\min}}{2nv_{\max}} \right\}$.
\end{lemma}
\begin{proof}
Without loss of generality, we will prove the lemma for $i=1$. Suppose $\beta \in S$ is labelled $1$ according to the above procedure. First, $\beta_1 > 0$ follows as a direct consequence. Furthermore, as $\max_i \beta_i \geq 1/n$ and $\sum_i\beta_i=1$, we get $t^*(\beta) = t_1 \leq n$. We consider the two possible binding cases which can define $t_1$. If $t_1 = 1/\beta_1$, then $\beta_1 \geq 1/n$, and thus the lemma holds. On the other hand, if $t_1 = \frac{B_1}{\sum_j x_{1j} p_j(\beta)}$, then $\sum_j x_{ij} p_j(\beta) > 0$ and
\begin{align*}
    B_1 = t_1 \sum_j x_{1j} p_j(\beta) \leq n \sum_{j:\ x_{ij}>0} \max_i \beta_i v_{ij} \leq n \sum_{j:\ x_{ij}>0} \frac{\beta_1 v_{1j}}{(1-\delta)} \leq n \sum_{j} \frac{\beta_1 v_{1j}}{(1-\delta)}
\end{align*}
where the second inequality follows from the definition of $(\delta, \gamma)$-approximate pacing equilibrium. Therefore, $\beta_1 \geq \min\left\{\frac{1}{n}, \frac{(1-\delta) B_1}{\sum_j v_{1j}} \right\}$.
\end{proof} 

\begin{proof}[Proof of Claim \ref{lipschitz}]
	Let $B_{\min} = \min_{i \in [n]} B_i$, $B_{\max} = \max_{i \in [n]} B_i$, $v_{\max} = \max_{i,j} v_{ij}$ and $v_{\min} = \min_{i,j: v_{ij}>0} v_{ij}$. In this proof, we will use the following facts: if $f,g$ are Lipschitz functions with Lipschitz constants $L_f, L_g$, then
	\begin{itemize}
		\item[(a)] $f + g$ is Lipschitz with constant $L_f + L_g$
		\item[(b)] $\max\{f, g\}$ is Lipschitz with constant $\max\{L_f, L_g\}$.
		\item[(c)] If $|f|, |g| \leq M$, then $fg$ is Lipschitz with constant $M(L_f + L_g)$.
	\end{itemize}

	Define $y_{ij}: S \to \mathbb{R}$ as $y_{ij}(\beta) = [\beta_i v_{ij} - (1-\delta)\max_{k} \beta_k v_{kj}]^+$. Using facts (a) and (b), we can write
	\begin{align*}
	    |y_{ij}(\beta) - y_{ij}(\beta')| \leq 2 v_{\max} \|\beta - \beta'\|_\infty
	\end{align*}
	
	Consider $\beta \in S_0$ and $i \in [n]$. As $S_0$ is panchromatic, there exists $\beta' \in S_0$ such that $T(\beta') = i$. By Lemma \ref{beta_lower_bound}, we get
	\begin{align*}
		\beta_i' \geq \min\left\{\frac{1}{n}, \frac{B_{\min}}{2nv_{\max}} \right\}	
	\end{align*}
	Then, using the definition of $\omega$, we get the following equivalent statements:
	\begin{align*}
		\beta_i \geq \frac{1}{2} \min\left\{\frac{1}{n}, \frac{B_{\min}}{2nv_{\max}} \right\}	
	    \iff \frac{1}{\beta_i} \leq U \coloneqq 2\max\left\{n, \frac{2 n v_{\max}}{B_{\min}}\right\}
	\end{align*}
	Hence, for $\beta, \beta' \in S_0$, we have
\begin{align*}
    \Biggr\lvert \frac{1}{\sum_r y_{rj}(\beta)} - \frac{1}{\sum_r y_{rj}(\beta')} \Biggr\rvert &= \Biggr\lvert \frac{ \sum_r y_{rj}(\beta') - \sum_r y_{rj}(\beta)}{\sum_r y_{rj}(\beta)\sum_r y_{rj}(\beta')} \Biggr\rvert\\
    &\leq \frac{2n v_{\max} U^2}{\delta^2 v_{\min}^2} \cdot \|\beta - \beta'\|_\infty
\end{align*} 

Using fact (c), for $\beta, \beta' \in S_0$, we can write
\begin{align*}
    |x_{ij}(\beta) - x_{ij}(\beta')| &\leq \max\left\{ v_{\max}, \frac{U}{\delta v_{\min}}\right\} \left[2 v_{\max} + \frac{2n v_{\max} U^2}{\delta^2 v_{\min}^2} \right] \cdot \|\beta - \beta'\|_\infty
\end{align*}
Set $\bar{U} = \max\left\{ v_{\max}, \frac{U}{\delta v_{\min}}\right\} \left[2 v_{\max} + \frac{2n v_{\max} U^2}{\delta^2 v_{\min}^2} \right]$. Also, note that for $\beta, \beta' \in S$,
\begin{align*}
    |p_j(\beta) - p_j(\beta')| \leq v_{\max} \|\beta - \beta'\|_\infty
\end{align*}

For $\beta, \beta' \in S_0$, combining the above Lipschitz conditions using facts (a) and (c) yields
\begin{align*}
    \biggr\lvert \sum_j x_{ij}(\beta) p_j(\beta) - \sum_j x_{ij}(\beta') p_j(\beta') \biggr\rvert \leq m  v_{\max} (\bar{U} + v_{\max}) \|\beta - \beta'\|_\infty
\end{align*}

Set $W \coloneqq m  v_{\max} (\bar{U} + v_{\max})$. Define
\begin{align*}
    P^* \coloneqq \left\{i \in [n] \biggr\lvert \exists \beta \in T \text{ s.t. } \frac{B_i}{\sum_j x_{ij}(\beta) p_j(\beta)} < \frac{1}{\beta_i} \right\}
\end{align*}

For $i \in P^*$ and $\beta \in S_0$, we can write $\frac{B_i}{\sum_j x_{ij}(\beta) p_j(\beta)} < \frac{1}{\beta_i} \leq U  \text{, which implies } \frac{1}{\sum_j x_{ij}(\beta) p_j(\beta)} \leq \frac{U}{B_{\min}}$.

Therefore, for $\beta, \beta' \in S_0$ and $i \in P^*$, we have
\begin{align*}
    \biggr \lvert \frac{B_{i}}{\sum_j x_{ij}(\beta) p_j(\beta)} - \frac{B_{i}}{\sum_j x_{ij}(\beta') p_j(\beta')} \biggr\rvert \leq  B_{\max} \frac{U^2}{B_{\min}^2} W \|\beta - \beta'\|_\infty \leq \frac{B_{\max} U^2 W }{B_{\min}^2} \|\beta - \beta'\|_\infty
\end{align*}

Also, for $\beta, \beta' \in S_0$ and $i \in [n]$, we have
\begin{align*}
    \biggr\lvert \frac{1}{\beta_i} - \frac{1}{\beta'_i} \biggr \rvert \leq U^2 \|\beta - \beta'\|_\infty
\end{align*}

Note that for $\beta \in T$, we can rewrite $t^*(\beta)$ as follows
\begin{align*}
    t^*(\beta) = \min\left\{ \min_{i \in [n]} \frac{1}{\beta_i}, \min_{i \in P^*} \min\left\{\frac{1}{\beta_i}, \frac{B_i}{\sum_j x_{ij}(\beta) p_j(\beta)}\right\} \right\}
\end{align*}

Using fact (b), for $\beta, \beta' \in T$,
\begin{align*}
    |t^*(\beta) - t^*(\beta')| \leq 2n \max\left\{U^2, \frac{B_{\max} U^2 L }{B_{\min}^2}\right\} \|\beta - \beta'\|_\infty
\end{align*}

Therefore, for $i \in [n]$, total payment made by buyer $i$ is Lipschitz for $\beta \in S_0$:
\begin{align*}
    \biggr\lvert \sum_j x_{ij}(\beta) t^*(\beta) p_j(\beta) - \sum_j x_{ij}(\beta') &t^*(\beta') p_j(\beta') \biggr\rvert\\
    &\leq \max\{n v_{\max}, n\} \left(W + 2n \max\left\{U^2, \frac{B_{\max} U^2 W }{B_{\min}^2}\right\} \right) \|\beta - \beta'\|_\infty
\end{align*}
Hence, the claim holds because
\begin{align*}
	\max\{n v_{\max}, n\} \left(W + 2n \max\left\{U^2, \frac{B_{\max} U^2 W }{B_{\min}^2}\right\} \right) \leq L = \left(\frac{2^{|G|}}{\delta}\right)^{10,000}
\end{align*}
\end{proof}
\section{Incorporating Reserve Prices}\label{appendix_reserves}

Consider the setting in which each item $j$ has a reserve price $r_j$. Now, a buyer wins a good $j$ only if her bid is the highest bid $h_j(\alpha)$ and it is greater than or equal to the reserve $r_j$. Moreover, the price of good $j$ is the maximum of the second highest bid $p_j(\alpha)$ and its reserve price $r_j$. In the presence of reserve prices, we will use $H_j(\alpha) \coloneqq \max\{h_j(\alpha), r_j\}$ to denote the winning threshold of good $j$ and $P_j(\alpha) \coloneqq \max\{p_j(\alpha), r_j\}$ to denote the price of good $j$. The next example illustrates that one needs to be careful in the way one extends the definition of pacing equilibrium (Definition~\ref{definition_exact_pacing_equilibrium}) to model the presence of reserves.

\begin{example}
    There is one buyer and one good. The buyer values the good at 4 and has a budget of 1. The goods has a reserve price of 2. If she bids strictly less than $1/2$, then she does not win any part of the good. On the other hand, if we assume that she wins the entire good upon bidding $1/2$ or higher, then she violates her budget upon doing so. This suggests that a pacing equilibrium might not even exist if we extend it naively to the setting with reserves. Instead, we will take the approach that, in a pacing equilibrium, the seller may decide to not sell a fraction of a good if the highest bid is equal to the reserve price of that good. With this new definition, we can see that a pacing equilibrium does in fact exist, namely, when the buyer has a pacing multiplier of $1/2$ and wins $1/2$ of the item.
\end{example}

Inspired by the above example, we define pacing equilibrium for the setting with reserves.

\begin{definition}[Pacing Equilibria with reserves] \label{definition_reserve_pacing_equilibrium}
Given an SPP game with reserves $G = (n, m, (v_{ij}),$ $(B_i), (r_j))$, 
we say $(\alpha, x)$ with $\alpha=(\alpha_i)\in [0,1]^n$, $x=(x_{ij}) \in [0,1]^{nm}$
  and $\sum_{i\in [n]} x_{ij}\le 1$ for all $j\in [m]$ %with $\sum_{i\in [n]}x_{ij}\le 1$
  is a \emph{pacing equilibrium} if \vspace{0.15cm}
\begin{flushleft}
\begin{itemize}
    \item[(a)] Only buyers above the winning threshold win the good: $x_{ij} > 0$ implies $\alpha_i v_{ij}= H_j(\alpha)$.\vspace{0.1cm}
    \item[(b)] Full allocation of each good for which the highest bid exceeds the reserve price:  $h_j(\alpha) > r_j$  implies $\sum_{i\in [n]} x_{ij} = 1$.\vspace{0.1cm}
    \item[(c)] Budgets are satisfied: $\sum_{j\in [m]} x_{ij} P_j(\alpha) \leq B_i$.\vspace{0.1cm}
	\item[(d)] %Complementarity/
	No unnecessary pacing: $\sum_{j\in [m]} x_{ij} P_j(\alpha) < B_i$ implies $\alpha_i=1$.\vspace{0.15cm}
\end{itemize}
\end{flushleft}
\end{definition}

Next, we extend our PPAD-membership result to the setting with reserves.

\begin{theorem}
    Finding a pacing equilibrium in a SPP game with reserves is in PPAD.
\end{theorem}
\begin{proof}
    Consider a pacing game with reserve prices $G$ and the corresponding pacing game without reserve prices $G'$. Add an auxiliary buyer $a$ to $G'$ who values good $j$ at $r_j$ for all $j \in [m]$ and has a budget large enough to ensure that her pacing multiplier is always 1 in every pacing equilibrium (this can be achieved by setting her budget to be the sum of all values $\{v_{ij}\}$ and reserve prices $\{r_j\}$). We will call this updated game $G_+'$. The theorem follows from the simple observation that if we find a pacing equilibrium $(\alpha, x)$ for $G'_+$ and disregard the terms corresponding to the auxiliary buyer, then we get a pacing equilibrium $(\alpha_{-a}, x_{-a})$ for $G$. This is because, in any pacing equilibrium of $G'_+$, the auxiliary buyer has a multiplier of 1 and hence bids $r_j$ on good $j$ for all $j \in [m]$. Moreover, any amount that the auxiliary buyer wins in $(\alpha, x)$ can be thought of as being not sold by the seller. As $(\alpha, x)$ satisfies Definition~\ref{definition_exact_pacing_equilibrium}, it is straightforward to check that $(\alpha_{-a}, x_{-a})$ satisfies Definition~\ref{definition_reserve_pacing_equilibrium}.
\end{proof}

We conclude this section by noting that our hardness results extend directly to the setting with reserves because it reduces to the setting without reserves when $r_j = 0$ for all goods $j \in [m]$.
\section{\revision{Perturbed Second-Price Pacing Games}}\label{appendix:perturbed_game}

Before stating and proving the results, we define the relevant equilibrium notions. For a perturbed pacing game $(n, m, (v_{ij}), (B_i), \delta)$, let $p'_{ij}(\alpha)$ denote the expected payment made by buyer $i$ on good $j$ when the buyers use multipliers $\alpha \in [0,1]^n$. Moreover, let $x_{ij}(\alpha)$ be the probability of buyer $i$ winning good $j$ when the buyers use the multipliers $\alpha$.

\begin{definition}
    Consider a perturbed SPP game $(n, m, (v_{ij}), (B_i), \delta)$.  Then, $\alpha \in [0,1]^n$ is a pacing equilibrium of the perturbed SPP if:
    \begin{itemize}
        \item Budgets are satisfied: $\sum_{j=1}^m p'_{ij}(\alpha) \leq B_i$
        \item No unnecessary pacing: If $\sum_{j=1}^m p'_{ij}(\alpha) < B_i$, then $\alpha_i = 1$ 
    \end{itemize}
    Moreover, $\alpha \in [0,1]^n$ is an $\gamma$-approximate pacing equilibrium of the perturbed SPP if:
    \begin{itemize}
        \item Budgets are satisfied: $\sum_{j=1}^m p'_{ij}(\alpha) \leq B_i$
        \item Not too much unnecessary pacing: If $\sum_{j=1}^m p'_{ij}(\alpha) < (1 - \gamma) B_i$, then $\alpha_i \geq (1 - \gamma) v_{ij}$ 
    \end{itemize}
\end{definition}

\begin{theorem}\label{thm:approx-perturbed}
    Computing a $\gamma$-approximate pacing equilibrium of a perturbed SPP game $(n, m, (v_{ij}),$ $ (B_i), \delta)$ is PPAD-hard when $\delta = \gamma = 1/n^8$.
\end{theorem}
\begin{proof}
    First observe that
    \begin{align*}
        (1 - \gamma)(1 - \delta) = (1 - n^{-8})^2 = 1 + n^{-16} - 2n^{-8} \geq 1 - n^{-7} 
    \end{align*}
    We will prove the theorem by reducing from the problem of computing approximate pacing equilibria of SPP games. Consider an SPP game $G = (n, m, (v_{ij}), (B_i))$. Define a perturbed SPP game $G' = (n, m, (v_{ij}), (B'_i), \delta)$ such that $B'_i = (1 - \delta) B_i$. Let $\alpha$ be a $\gamma$-approximate pacing equilibrium of the perturbed SPP game $G'$. Then, as $\epsilon_{ij} \in [1-\delta, 1]$, we get that 
    \begin{align}
         (1 - \delta)x_{ij}(\alpha) p_j(\alpha) \leq p'_{ij}(\alpha) \leq x_{ij}(\alpha)p_j(\alpha) \quad \forall\ i \in [n], j \in [m] 
         \label{eq:pspp eq 1}
    \end{align}
    where, as earlier, $p_j(\alpha)$ denotes the second highest bid in an SPP game when the buyers use multipliers $\alpha$). To complete the proof, it suffices to show that $(\alpha, x(\alpha))$ is a $(\delta, \gamma')$-approximate pacing equilibrium of the SPP game $G$ for $\gamma' = 1/n^7$. We establish the required properties below:
    \begin{itemize}
        \item[(a)] As $\epsilon_{ij} \in [1-\delta, 1]$, $x_{ij}(\alpha) > 0$ only if $\alpha_i v_{ij} \geq (1 - \delta)\max_{k \in [n]} \alpha_k v_{kj}$
        
        \item[(b)] Full allocation of each good with positive bid: This follows directly from the allocation rules of a second-price auction.
        
        \item[(c)] Budgets are satisfied: $\alpha$ being bugdet feasible for the perturbed SPP game $G$ implies $$\sum_{j=1}^m p'_{ij}(\alpha) \leq B'_i = (1 - \delta)B_i$$ for all $i \in [n]$. As $p'_{ij}(\alpha) \geq (1 - \delta)x_{ij}(\alpha) p_j(\alpha)$, we get $\sum_{j=1}^m x_{ij}(\alpha)p_j(\alpha) \leq B_i$ as required.

        \item[(d)] Not too much unnecessary pacing: Suppose $\sum_{j=1}^m x_{ij}(\alpha)p_j(\alpha) < (1 - \gamma')B_i$ for some buyer $i \in [n]$. Then, using \eqref{eq:pspp eq 1}, we get 
        \begin{align*}
            \sum_{j=1}^m p'_{ij}(\alpha) < \frac{(1 - \gamma')}{(1 - \delta)} \cdot (1 - \delta)B_i = \frac{(1 - \gamma')}{(1 - \delta)} B'_i \leq (1 - \gamma) B_i'
        \end{align*}
        where we have used $(1 - \gamma)(1 - \delta) \geq (1 - n^{-7}) = (1 - \gamma')$. Now, as $\alpha$ is a $\gamma$-approximate equilibrium of the perturbed SPP game $G'$, we get $\alpha_i \geq 1 - \gamma \geq 1 - n^{-7} = 1 - \gamma'$.
    \end{itemize}
    Hence, we have shown that $(\alpha, x(\alpha))$ is a $(\delta, \gamma')$-approximate pacing equilibrium for the SPP game $G$, where $\delta \leq n^{-7}$ and $\gamma' = n^{-7}$. As the perturbed SPP game $G'$ can be constructed from the SPP game $G$ in polynomial time, the theorem follows from Theorem~\ref{approximate_hardness}.
\end{proof}

Let the expected utility of buyer $i$ in a perturbed SPP game under multipliers $\alpha$ be denoted by $u_i(\alpha)$, i.e.,
\begin{align*}
    u_i(\alpha) = \mathbb{E}_{\{\epsilon_{ij}\}_{i,j}} \left[ \sum_{j=1}^m (v_{ij} \epsilon_{ij} - \max_{k\neq i} \alpha_k v_{kj} \epsilon_{kj}) \mathbf{1}(\alpha_i v_{ij} \epsilon_{ij} \geq \max_{k\neq i} \alpha_k v_{kj} \epsilon_{kj}) \right]
\end{align*}

\begin{definition}
    Consider a perturbed SPP game $(n, m, (v_{ij}), (B_i), \delta)$. A vector of pacing multipliers $\alpha$ is called a Nash equilibrium of this game if for each $i \in [n]$ and $\alpha'_i$ such that $\sum_{j=1}^m p'_{ij}(\alpha'_i, \alpha_{-i}) \leq B_i$, we have $u_i(\alpha_i, \alpha_{-i}) \geq u_i(\alpha'_i, \alpha_{-i}$).
\end{definition}

\begin{lemma}
    Consider a perturbed SPP game $(n, m, (v_{ij}), (B_i), \delta)$ and let $\alpha$ be a Nash equilibrium of this game. If $\sum_{j=1}^m p'_{ij}(\alpha) < B_i$ and $\alpha_i < 1$, then $\sum_{j=1}^m p'_{ij}(\alpha) =\sum_{j=1}^m p'_{ij}(1,\alpha_{-i})$.
\end{lemma}
\begin{proof}
    Suppose $\alpha$ is a Nash equilibrium of the game but not a pacing equilibrium, and buyer $i$ satisfies $\sum_{j=1}^m p'_{ij}(\alpha) < B_i$ and $\alpha_i < 1$. For contradiction, suppose $\sum_{j=1}^m p'_{ij}(\alpha) < \sum_{j=1}^m p'_{ij}(1,\alpha_{-i})$. Now, as the distribution of $\epsilon_{ij}$ is continuous, $x \mapsto p_{ij}(x, \alpha_{-i})$ is a continuous non-decreasing function. By the Intermediate Value Theorem, there exists $\alpha^*_i \in (\alpha_i, 1)$ such that
    \begin{align*}
        \sum_{j=1}^m p'_{ij}(\alpha^*_i, \alpha_{-i}) \leq B_i\,.
    \end{align*}
    Now, observe that buyer $i$ wins good $j$ if and only if
    \begin{align*}
        \alpha^*_i v_{ij} \epsilon_{ij} \geq \max_{k\neq i} \alpha_k v_{kj} \epsilon_{kj}
    \end{align*}
    Therefore, $v_{ij} \epsilon_{ij} \geq p'_{ij}(\alpha_i^*, \alpha_{-i})/\alpha^*_i$. As $\alpha^*_i < 1$, we get that
    \begin{align*}
        u_i(\alpha^*_i, \alpha_{-i}) - u_i(\alpha_i, \alpha_{-i}) \geq \frac{1}{\alpha^*_i} \cdot \left[ \sum_{j=1}^m p'_{ij}(\alpha^*_i, \alpha_{-i}) - \sum_{j=1}^m p'_{ij}(\alpha_i, \alpha_{-i}) \right] > 0
    \end{align*}
    This contradicts the fact that $\alpha$ is a Nash equilibrium. Hence, the Lemma holds.
\end{proof}

\begin{corollary}\label{cor:perturb}
    Consider a perturbed SPP game $(n, m, (v_{ij}), (B_i), \delta)$ and let $\alpha$ be a Nash equilibrium of this game. If $\sum_{j=1}^m p'_{ij}(1,\alpha_{-i}) > \sum_{j=1}^m p'_{ij}(\alpha)$, then  we have $\sum_{j=1}^m p'_{ij}(\alpha) = B_i$. Furthermore, as a consequence, if $\sum_{j=1}^m p'_{ij}(1,\alpha_{-i}) > B_i$, then $\sum_{j=1}^m p'_{ij}(\alpha) = B_i$.
\end{corollary}

\begin{theorem}
    Computing a Nash equilibrium of a perturbed SPP game $(n, m, (v_{ij}), (B_i), \delta)$ is PPAD-hard when $\delta = 1/n^8$. 
\end{theorem}
\begin{proof}
    Let $G$ be the SPP game constructed in Appendix~\ref{appendix_hardness} for the proof of Theorem~\ref{approximate_hardness}. Like the proof of Theorem~\ref{thm:approx-perturbed}, define a perturbed SPP game $G' = (n, m, (v_{ij}), (B'_i), \delta)$ such that $B'_i = (1 - \delta) B_i$. Moreover, define an auxiliary perturbed SPP game $G'' = (n+1, m +1, (v_{ij}), (B'_i), \delta)$ by adding one more buyer and one more good to $G'$. We denote the new buyer by $\T^*$ and the new good by $S$. Buyer $\T^*$ has value $1$ for good $S$, i.e., $V(\T^*, S) = 1$ and does not value any other good. She has a budget of $n^7$  (large enough to never be binding). The only other buyer who has a non-zero value for $S$ is the Threshold buyer $\T$, who has a value of 1, i.e, $V(\T, S) = 1$.

    We begin by showing that every Nash equilibrium of $G''$ is also a pacing equilibrium. Let $\alpha$ be a Nash equilibrium of $G''$. As a first step, we show that $\alpha(\T) = \alpha(\T^*) = 1$. We do so by ruling out the other cases:
    \begin{enumerate}
        \item If $\alpha(\T) < \alpha(\T^*)$, then buyer $\T$ can strictly increase her utility by setting $\alpha(\T) = 1$ as this allows her to win a strictly larger fraction of good $S$.
        \item Similarly, if $\alpha(\T^*) < \alpha(\T)$, then buyer $\T^*$ can strictly increase her utility by setting $\alpha(\T^*) = 1$ as this allows her to win a strictly larger fraction of good $S$.
        \item If $\alpha(\T) = \alpha(\T^*) < 1$,  then buyer $\T$ can strictly increase her utility by setting $\alpha(\T) = 1$ as this allows her to win a strictly larger fraction of good $S$.
    \end{enumerate}
    For every other buyer in $G''$, we use Corollary~\ref{cor:perturb} to show that they exactly spend their budget. 
    
    If $\alpha(\mathbb{C}(p,s)) \leq (1 - \delta)/2$, then the buyer $\mathbb C(p,s)$ wins no part of the threshold good $T(p,s)$ and spends strictly less than her budget because she has value at most 16 for all of the other goods and there are at most $O(n^2)$ such goods compared to her budget which is $\Omega(n^2).$ On the other hand, she can win all of the threshold good $T(p,s)$ by setting $\alpha(\mathbb C(p,s)) = 1$ and spend strictly more. Hence, by Corollary~\ref{cor:perturb}, we get that she exactly spends her budget, which is a contradiction. Therefore, $\alpha(\mathbb C(p,s)) \geq (1 - \delta)/2$.

    Consider a dummy buyer $\D(p,s)$. If we set $\alpha(\D(p,s)) = 1$, then she wins at least half of the normalization good $N(p,s)_s$ at a price of at least $\alpha(\mathbb C(p,s))$ which violates her budget of $1/(16n)$. Thus, Corollary~\ref{cor:perturb} implies that she exactly spends her budget under the Nash equilibrium $\alpha$.

    Consider buyer $\mathbb C(p,s)$. If we set $\alpha(\mathbb C(p,s)) = 1$, she she wins:\vspace{0.1 cm}
\begin{flushleft}    \begin{itemize}
        \item 
        All of 
        normalization good $N(p,s)_t$, for each $t\ne s$, by spending at least $(1-\delta)/2$ on each of them because $\alpha(\BB(p,t)) \geq (1-\delta)/2$ 
        %for all $p \in \{1,2\}$, $r \in [n]$ 
        by the earlier part of the proof.\vspace{0.15cm}
        
        \item Part of normalization good $N(p,s)_s$ by spending at least $(1-\delta) - \nu$. This is because $N(p,s)_s$ has price 
        at least $(1-\delta)$ and
        buyer $\D(p,s)$ only has budget $\nu$.
%        (Recall that $\D(p,s)$ has a budget of $\nu$).
\vspace{0.15cm}
        
        \item All of threshold good $T(p,s)$ by spending at least  $\alpha(\T)(1-\delta)n^4 = (1-\delta) n^4$.\vspace{0.15cm}
        
        \item All of expenditure good $E(p,s)_t$, for each $t\in [n]$, by spending at least $\alpha(\T)\nu A_{st}/2$ if $p=1$\\ and $\alpha(\T) \nu B_{ts}/2$ if $p=2$.\vspace{0.1 cm}
    \end{itemize}\end{flushleft}
    Hence, the total expenditure of $\BB(p,s)$ when $p=1$ is at least $$
    (1-\delta)\cdot \frac{n-1}{2} + (1-\delta)\cdot - \nu + (1-\delta ) n^4 + \sum_{t\in [n]} \nu A_{st}/2$$ 
    which is strictly higher than her budget. Similar statement holds for $p =2$. Therefore, Corollary~\ref{cor:perturb} implies that buyer $\mathbb C(p,s)$ exactly spends her budget.

    Hence, we have shown that every buyer either has her multiplier equal to 1 or exactly spends her budget, which means that $\alpha$ is a pacing equilibrium. Moreover, from our construction of $G''$ from $G'$, we get that the restriction of $\alpha$ to the buyers other than $\T^*$ is a pacing equilibrium for the game $G'$. This is because only the Threshold buyer $\T$ is affected by this change and her multipliers satisfies $\alpha(\T) = 1$ and she spends strictly less than her budget. Finally, as we showed in the proof of Theorem~\ref{thm:approx-perturbed}, $(\alpha, x(\alpha))$ is a $(\delta, \gamma)$-approximate pacing equilibrium of the SPP game $G$ where $\delta = \gamma = 1/n^7$. Invoking Theorem~\ref{approximate_hardness} completes the proof. 
\end{proof}

\end{document}